\documentclass[12pt]{article}
\usepackage{graphicx}
\usepackage[dvips,colorlinks,breaklinks]{hyperref}
\usepackage[dvips]{color}
\usepackage{amsmath}
\usepackage{amssymb}
\usepackage[export]{adjustbox}
\begin{document}
\title{MHD Turbulence}
\author{Andrey Beresnyak}
\maketitle
\thispagestyle{empty}
\tableofcontents	

\section{Introduction}
\pagenumbering{arabic}
\def\kvec{{\bf k}}
\def\small{{ }} 
\def\L{{\Lambda}}
\def\l{{\lambda}}
\def\emf{\mathcal{E}} 

Turbulence is a time-dependent, stochastic flow commonly found in fluids with
low viscosity. Shear viscosity is associated with microscopic phenomena, and the size
of the turbulent system is much larger than the viscous scale. Turbulence develops from 
laminar flow due to instabilities and has many degrees of freedom. Despite its complexity, researchers investigate turbulence because of practical importance. The effect of turbulence is not only unpredictability of each realization of the
flow, but often very important, quantifiable and predictable effects
which are attractive to scientists and engineers. For example, ideal equations of motion,
such as Euler's equation can be used, under certain conditions, to derive conservation
laws. The conservation of energy and the conservation of the velocity circulation
along the path frozen into the fluid (Kelvin's theorem) are notable examples
of these ``ideal invariants''. However, physicists realized very early on that
moving through the fluid involves drag and the loss of energy. Despite there
is always a stationary ideal flow that produces zero drag (d'Alembert paradox), in practice
such flows are not realizable due to instabilities and finite viscosity.
Turbulence research elucidated this energy loss process and argued that it could happen
for arbitrarily small viscosity due to the conserved quantity forming a ``cascade''
through scales finally dissipating on sufficiently small scales (Richardson-Kolmogorov
picture). Likewise,
Kelvin's circulation theorem is broken for flow around the wing, making possible lift force and the airplane flight.

Compared to turbulence on Earth, astrophysical turbulence is characterized by even larger scale separation between the problem size and the dissipative size, this makes turbulence in space almost unavoidable. Unlike the flows of non-conductive fluids on Earth, well-described by the Navier-Stokes equations, astrophysics deals with flows of ionized plasmas which, in most cases, can be considered perfectly conducting. These flows are described by equations with currents, magnetic fields and the
Lorentz force, namely magnetohydrodynamic (MHD) equations.

In most astrophysical environments magnetic fields are observed and often are
dynamically important. In our Galaxy, similar to other spiral galaxies, the magnetic field has a regular as well as random components. The value of the
magnetic field, around 5 $\mu G$, suggests equipartition between magnetic and kinetic
energies. In the galaxy clusters, the magnetic field is of order 1-3 $\mu G$, which is around
1/20th of the equipartition. Another example is the convective cell on the Sun. Its magnetic field is also somewhat close to the equipartition with the motions.
Our Universe would have been very boring if it had both electric and magnetic
charges, so that both electric and magnetic fields are screened out on large scales.
Fortunately, this is not the case, the magnetic field and large-scale motions result
in the acceleration of particles and the Universe if filled with non-thermal radiation
in all wavebands.

Considering the space is filled with ionizing radiation it is not surprising that
astrophysical plasmas are well-conductive. However, do they always have to be
well-magnetized as well? The process of generation or amplification of the field
is known as a dynamo, and this process seems to work sufficiently fast to do its job
everywhere. If we start with zero magnetic field in the MHD equations, this produces precisely
zero field in the future in an apparent contradiction with the ubiquity of magnetic fields.
Do we always have to rely on primordial magnetic fields or the effects beyond simple
MHD equations? In this review, among other things, we will emphasize that
the growth of magnetic energy can be described in a framework somewhat
similar to the loss of kinetic energy in the nearly ideal hydrodynamic flows.
In other words, fast dynamo is an inherent property of turbulence.

Magnetic turbulence is also the primary cause of accretion onto gravitating objects,
in particular accretion onto black holes is estimated to be the most potent source
of energy in the Universe, exceeding thermonuclear burning in stars. Thin
stationary accretion disks in a Keplerian potential are hydrodynamically stable,
so in order to generate accretion one has to rely on the excitation of the
the magnetic degree of freedom, the problem known as magnetorotational instability (MRI).
Related to MRI-unstable disks are astrophysical jets, highly collimated flows perpendicular to
the accretion disks in which magnetic field is essential in the process of launching
and collimation of the flow.

\begin{figure}[t]
\begin{center}
\includegraphics[width=0.85\columnwidth]{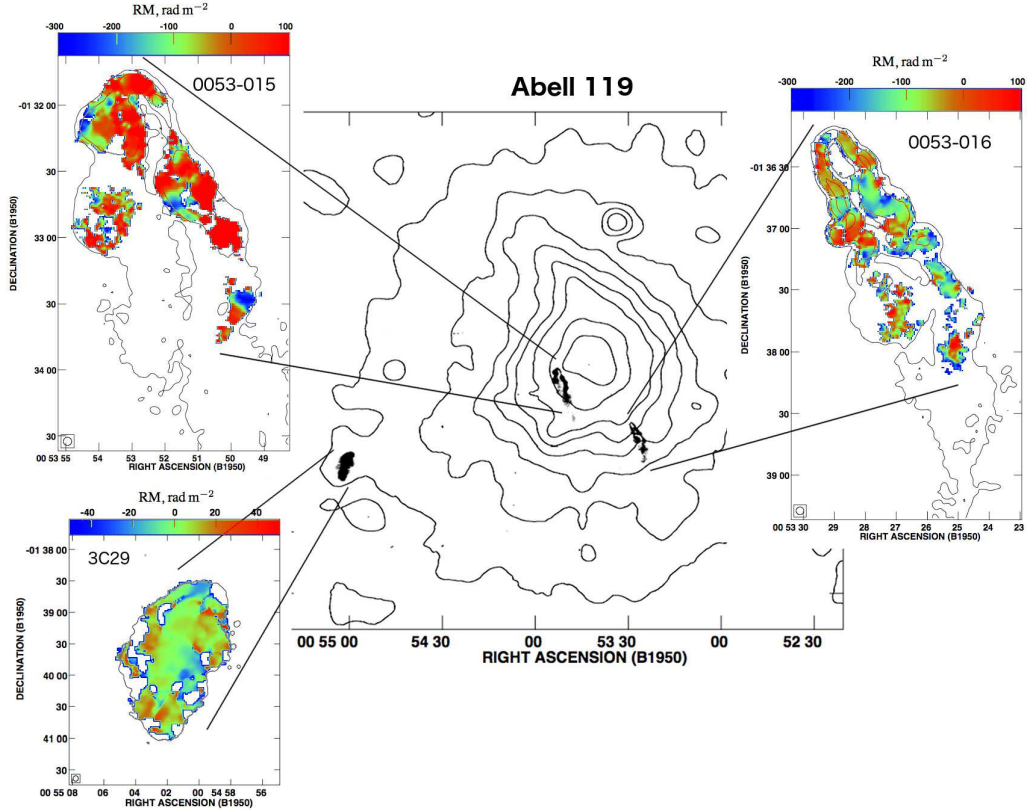}
\end{center}
\caption{Faraday rotation measure maps of radio sources within a galaxy cluster. From \cite{feretti1999,dolag2009a}. The cluster electrons act as a foreground for the radio
source. These maps indicate random magnetic fields of several $\mu$G in the cluster,
changing on scales of 10-40 kpc.}
\label{feretti}
\end{figure}

Alfven theorem of perfectly conducting magnetohydrodynamics states that magnetic field lines
are perfectly frozen into the conductive fluid, which places a severe restriction
on the process of the so-called magnetic reconnection -- the change of topology of the magnetic
configuration by magnetic field lines crossing and moving through magnetic null.
By way of restricting such change, Alfven theorem also precludes fast release of magnetic
energy in highly conductive environments. This is in gross contradiction with high-energy
phenomena above the solar surface known as X-ray flares. Again, turbulence comes
to the rescue and allows for the radical breaking of the Alfven theorem, even in near-perfectly
conducting fluids, very much like breaking of Kelvin's theorem of hydrodynamics.

\begin{figure}[!t]
\begin{center}
\includegraphics[width=0.5\columnwidth]{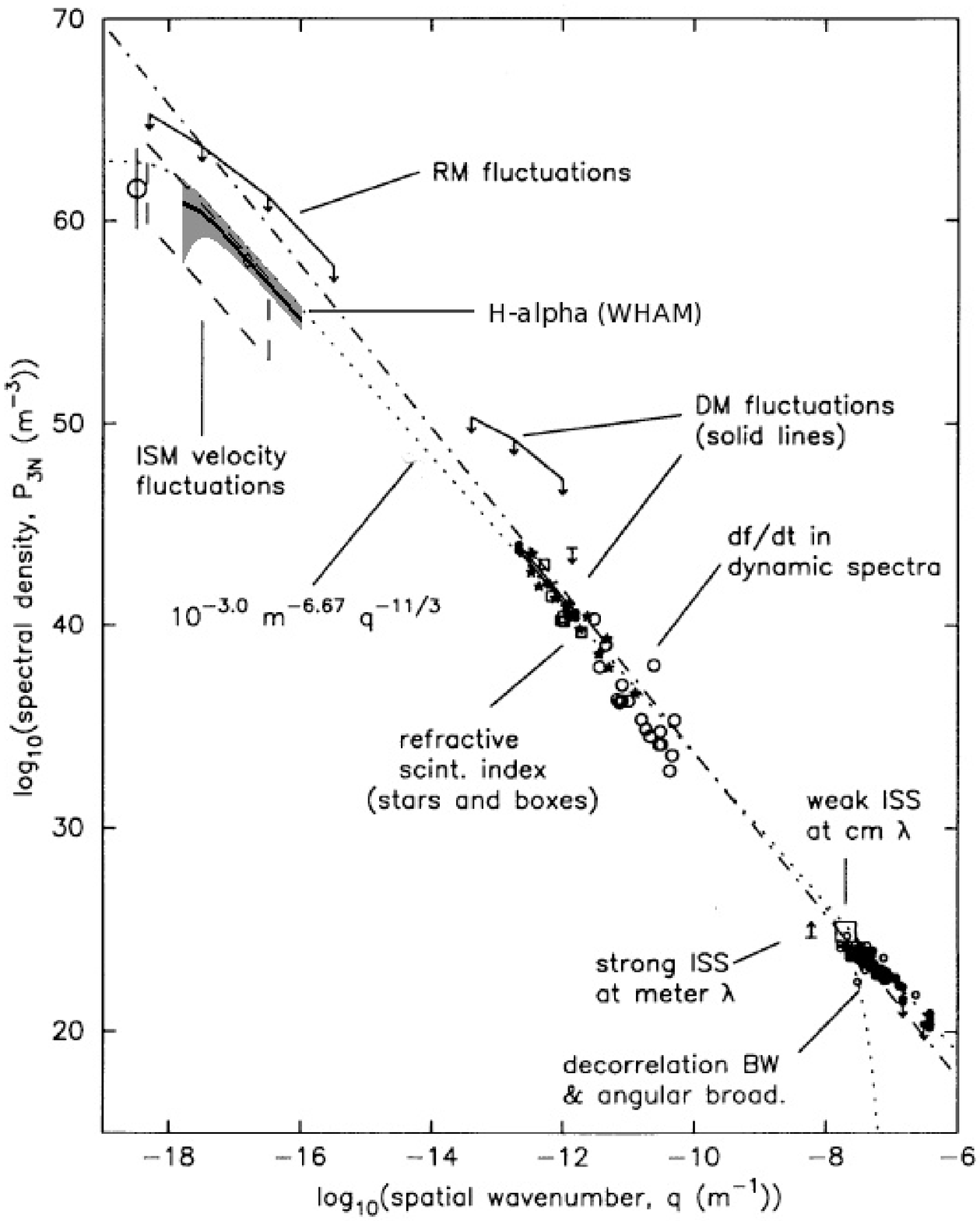}
\end{center}
\caption{Fluctuation of density in the interstellar medium (ISM), ``big power law in the sky'' from \cite{armstrong1995, Chepurnov2010}. This 3D spectral density with slope of $-11/3$ correspond to Kolmogorov's 1D spectrum with slope of $-5/3$. Whether this power
law, obtained from a variety of observations of different ISM components, is part
of a single turbulent cascade is still an open question.}
\label{bpl}
\end{figure}

In recent years the two traditional pillars of physics -- the theory and the experiment has been complemented by a new method, numerical simulations. Numerics is valuable because it covers the
gap between the real world, experimental data, and the idealized theory, a gap often being too
great and precluding discovery. Numerics solves idealized equations directly, in this aspect it is similar to theory. On the other hand, numerics may be referred to as ``numerical experiment'', measurement of physical quantities without invoking much of the assumptions or prerequisites.
Compared to the real-life experiment, in numerics, it is easier to study idealized cases such as statistically homogeneous or statistically stationary turbulence for which the theory has something to say
At the same time numerics reduces almost infinite space of theoretical
ideas by weeding out theories which are incompatible with numerical measurements.
In the studies of turbulence, the strength of numerics is manifested in the high statistical
accuracy of the results, especially on small scales. Compared to the experimental
measurements or observations which have high statistical and systematic uncertainties
this helps to discriminate between theories and make quicker progress. 

\begin{figure}[t]
\begin{center}
\includegraphics[width=0.6\textwidth]{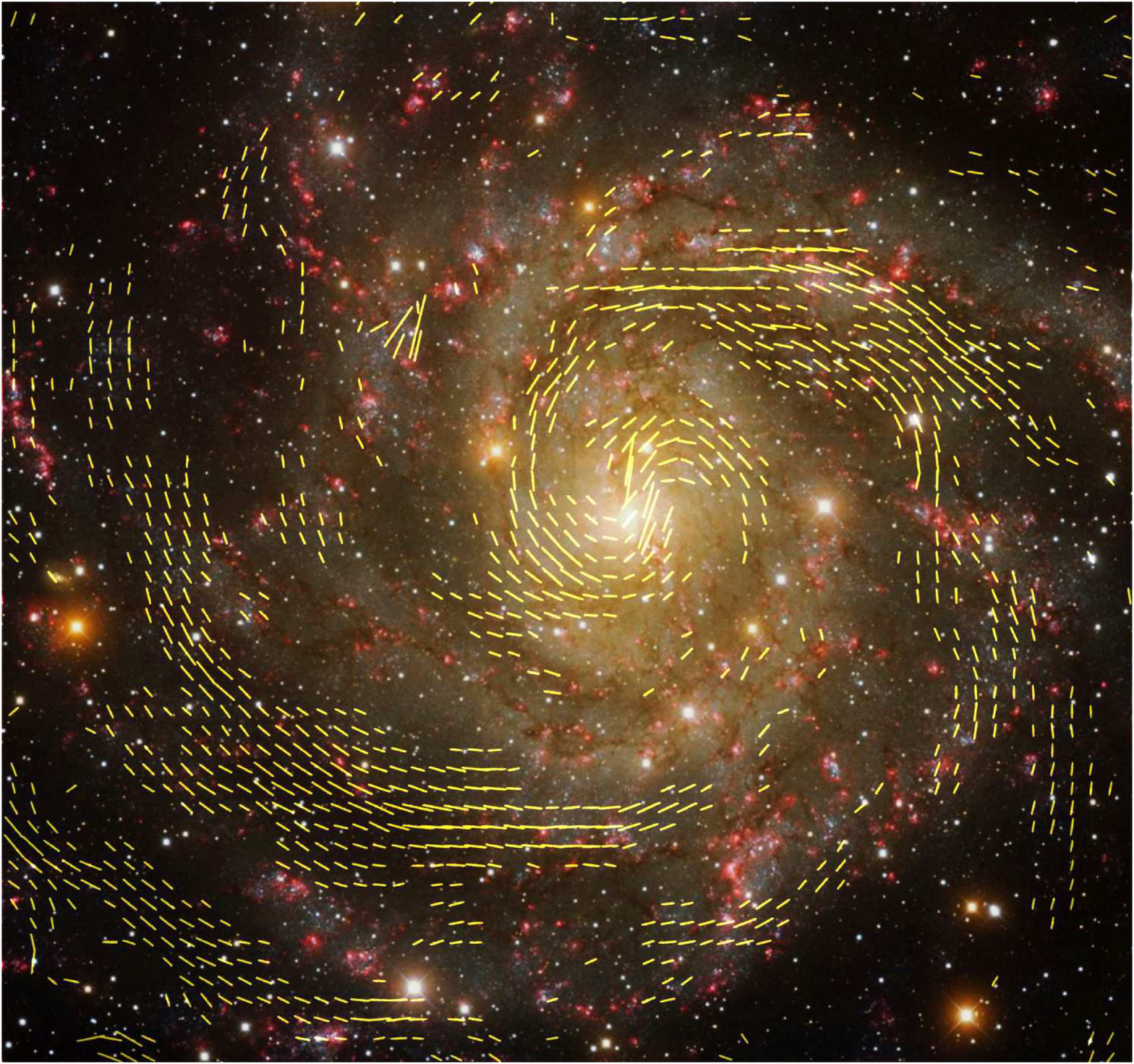}
\caption{B-vectors in spiral galaxy IC~342 observed at 6 cm from \cite{Beck2015a} (VLA and Effelsberg) overlaid on the Kitt Peak Observatory image (credit: T.A. Rector, University of Alaska Anchorage, and H. Schweiker, WIYN and NOAO/AURA/NSF).}
\label{kitt}
\end{center}
\end{figure}

One type of numerics, direct numerical simulations (DNS) will be highlighted in this review.
DNS refers to ``fully resolved'' numerical experiment, where numerics is very accurate and
faithfully reproduce solutions of the original equations on all scales. On the other end, there are Implicit Large Eddy Simulations (ILES), calculations aiming to get
the large-scale features of the flow correct without caring about the details
of the dissipation in small-scale turbulence or shocks. These are very common is astrophysics,
allowing to simulate large objects which are indeed out of reach of DNS,
however as we will show in the dynamo section this should be used with caution.

This, primarily theoretical, review mostly deals with homogeneous (although usually anisotropic) turbulence. Homogeneous models benefit from the opportunity to average quantities over volume, instead of averaging over ensemble (see more in Section~\ref{statistical}).
Few examples of important physical problems involving inhomogeneous turbulence: 1) large-scale dynamo, where inhomogeneity is required to break the statistical symmetry of turbulence, typically mirror symmetry, to produce large-scale magnetic fields, see also Section~\ref{dynamo}; 2) generation of imbalanced turbulence with a localized source of perturbations, see also Section~\ref{imbal}; 3) large-scale dynamics of expanding solar wind; 4) magnetic shear as a driver of turbulence, see Section~\ref{mag_shear}; 5) MRI, mentioned above. Very often inhomogeneous problems are treated with the scale-separation technique, where turbulence is described is some sort of ``local box'' approximation, within the box it is assumed homogeneous, but have overall driving, for example, shear boundary condition, as in the case of MRI. 

\section{MHD turbulence in astrophysics}
Turbulence results from instabilities of large-scale fluid motions experiencing low friction forces. Dimensionless Reynolds number characterizes the relative importance of viscosity 
\begin{equation}
{\rm Re}=LV/\nu, 
\end{equation}
where $L$ is the characteristic scale of the flow, often called ``outer scale,'' e.g., the diameter of a jet, $V$ is its velocity, and $\nu$ is fluid kinematic viscosity (in units of $[L]^2/[T]$). Likewise, one
can introduce similar magnetic Reynolds number 
\begin{equation}
{\rm Re}_m=LV/\eta,	
\end{equation}
 where $\eta=c^2/4\pi\sigma$ is magnetic diffusivity, $c$ is a speed of light, and $\sigma$ is a conductivity and Lundquist number 
\begin{equation}
S=Lv_A/\eta,	
\end{equation}
where 
\begin{equation}
v_A=B/\sqrt{4\pi\rho}
\end{equation}
 is Alfven speed, in units of velocity. Re, ${\rm Re}_m$ and S
are typically very large in astrophysics, meaning that viscous and resistive effects should be very small, the numbers of order $10^{10}$ or larger are common. \textit{\small A notable caveat of this simple picture
is that astrophysical plasmas are very often collisionless and the rigorous derivation of simple diffusive transport coefficients, such as Chapman-Enskog expansion, simply fails. So, our transport
coefficients refer to some ``effective'' diffusivities, the physical meaning of which can be understood as follows. In molecular physics, the kinematic viscosity can be estimated as $\bar u l$, a product of thermal speed and the mean free path. It is then clear that the Reynolds number is the ratio of the product of velocity and scale corresponding to macro- and micro-scales.
A suitably chosen ``effective'' mean free path will allow estimating ${\rm Re}$ and the scale at which fluid motion transitions into the dissipative or dispersive regime, for example, the Kolmogorov scale that we introduce in Section~3. Such trickery works extremely well for fluid flows with turbulence or shocks and also the logic behind ILES. Throughout this review, the reader will see many examples
of the so-called scale locality of turbulence in action, in particular, large-scale properties of the flow are insensitive to the diffusivities.}

\begin{figure}[t]
\begin{center}
\includegraphics[width=0.5\textwidth]{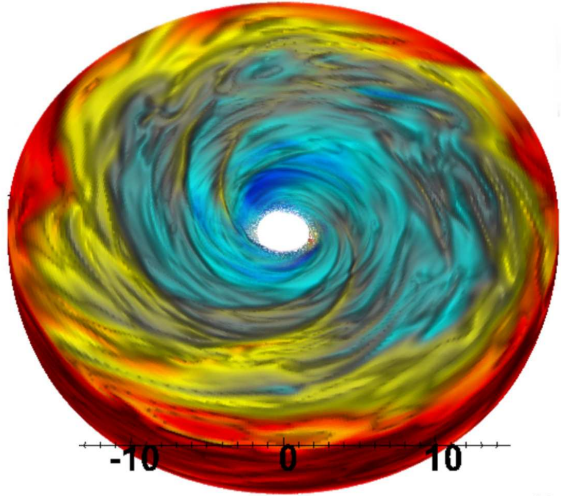}
\caption{Density, shown in color, in a simulated accretion disk around a black hole, subject to MRI.  The length unit is two gravitational radii of the black hole. From \cite{Jiang2017}.}
\label{jiang}
\end{center}
\end{figure}

One way to understand astrophysical turbulence is to understand the source of energy
and how it is converted to turbulent motions. The biggest source of energy in the Universe
is gravity. Turbulence can be driven by cosmological flows when gravity amplifies initially small
density perturbations and cause structure formation. This is a very slow process, however,
and typically dynamical times of voids are less then unity, in units of the age of the Universe, the dynamical times of filaments (superclusters) are of order unity, while dynamical times in the
intra-cluster medium (ICM) of the galaxy clusters are of order 20, meaning they are expected to be turbulent. The size of a typical large galaxy cluster is of the order of several megaparsecs (Mpc), and the main source of turbulence is the infalls of large chunks of matter into it, so-called major mergers. While the direct evidence of kinetic turbulence motions in clusters is still scarce, the evidence of magnetic fields produced
by dynamo action is available, see, e.g., Fig.~\ref{feretti}.
\begin{figure}[t]
\begin{center}
\includegraphics[width=0.5\columnwidth]{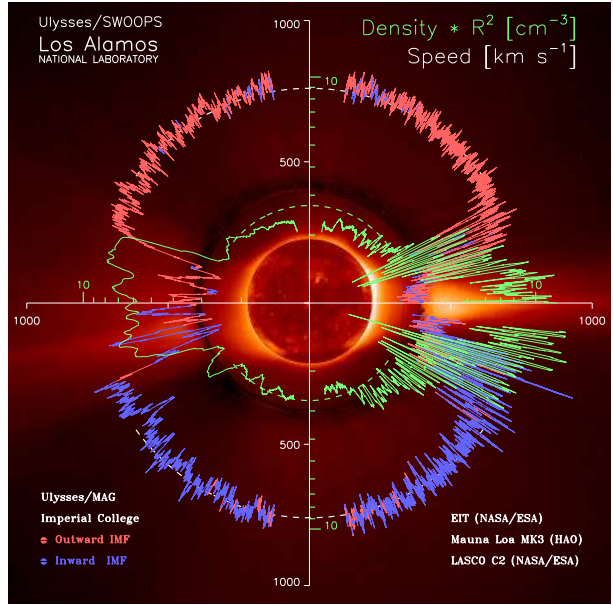}
\end{center}
\caption{Density and speed of the solar wind recorded by ULYSSES/SWOOPS. From \cite{mccomas2000}}
\label{ulysses}
\end{figure}

The source of turbulence in the plasma of ordinary galaxies, called interstellar medium (ISM), is likely multiple. At the present time, collisions with other galaxies are fairly rare. Several other mechanisms of driving turbulence can be identified, however: a) galactic disk is conductive and as such is subject to MRI, b) supernova explosions expand to the scale of several parsecs colliding with density inhomogeneities
of the ISM, producing large-scale irregular motions,
c) ISM turbulence is subject to Parker's instability when volumes of the ISM which are more magnetized and filled with more cosmic rays (CRs) are more buoyant compared to volumes
poorly magnetized and scarce in CRs, so convective instability against the gravity of the disk ensues,
d) CRs and starlight heat ISM at the same time cools itself by atomic and molecular emission on lower frequencies, and this produces thermal instability in the gas,
e) jets and winds from young stars collide with ISM inhomogeneities producing irregular motions, f) at high redshifts also accretion and merger.
The complexity of the ISM turbulence is rather overwhelming, and we refer to \cite{maclow2004,mckee2007} for further reading. The evidence of turbulence present in the
ISM was compiled from different sources by \cite{armstrong1995} and sometimes is referred to as a ``Big power law in the sky'', see Fig.~\ref{bpl}.
One amusing property of ISM turbulence
is the large-scale dynamo, which produces the magnetic field on the scales of the disk, tens of kpc,
while the outer scale of turbulence is only 10-100 pc. 
The evidence of large-scale magnetic
fields in other galaxies is abundant, see, e.g. Fig.~\ref{kitt}, while the evidence
of fluctuating component of the magnetic field is mostly limited to our own Galaxy, due to the limited resolution of the observations. 
\begin{figure}[t]
\includegraphics[width=0.49\textwidth]{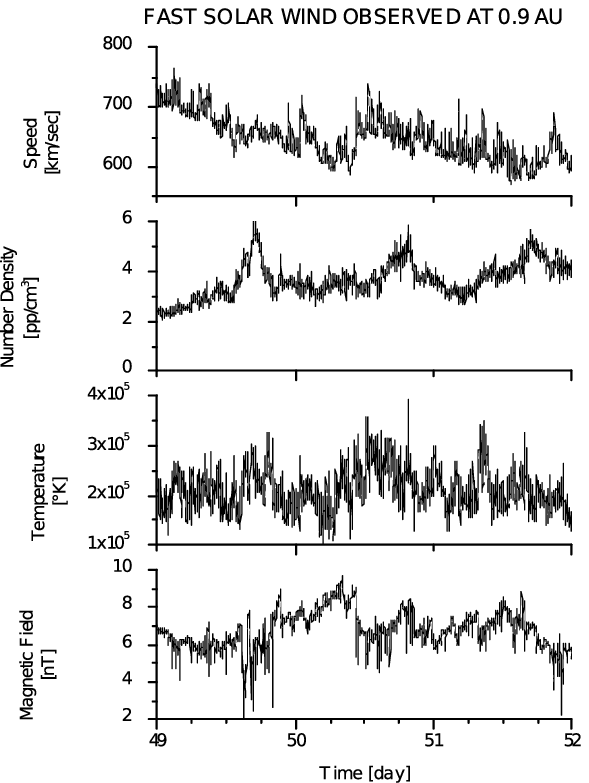}
\includegraphics[width=0.49\textwidth]{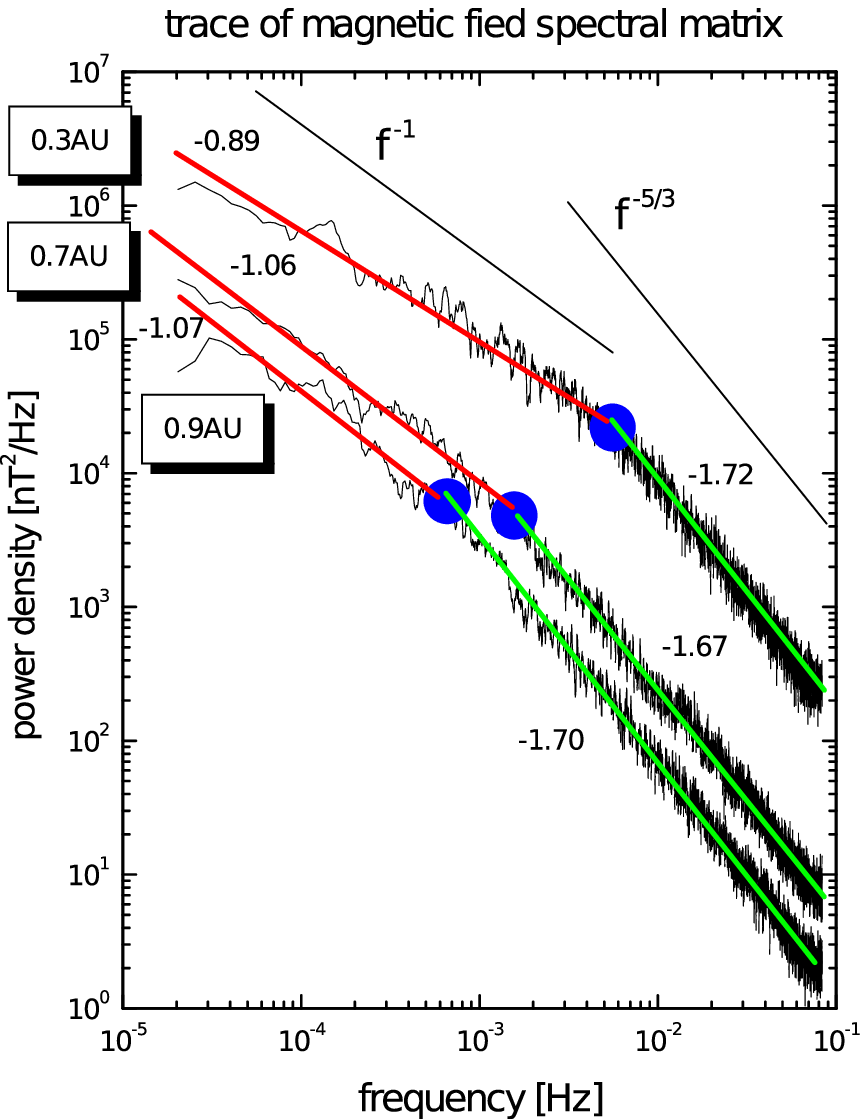}
\caption{Left: A sample of fast solar wind at a distance of 0.9 AU measured by the Helios 2 spacecraft.
Right: Power density spectra of magnetic field fluctuations observed by Helios 2 between 0.3 and
1 AU. From \cite{bruno2005}.}
\label{helios}
\end{figure}

Another source of kinetic energy derived from gravity is jets from disks around black holes.
This process is actually more efficient than thermonuclear burning in stars, including explosive burning in supernovae. The disk is conductive and unstable to MRI, turbulence in the disk creates rather large-scale structures, see, e.g., Fig.~\ref{jiang} and magnetic fields. The jet
may be driven from the rotating black hole directly or may be driven centrifugally at larger distances by gas escaping along open magnetic field lines.

In the solar system, we can make in-situ measurements in the solar wind, the flow of tenuous magnetized plasma emitted from the Sun at speeds 400-800 km/s and propagating outwards to the boundaries of the solar system. Such direct measurements of the solar wind parameters and fluctuations in different regions from 0.3 to 5 AU distance to the Sun are especially valuable
because they convey much more precise information about turbulent fluctuations compared to
astrophysical observations of ISM and ICM mired by limited resolution and projection effects.
Ion and electron counters on the satellite provide information about the flow, while magnetometers measure magnetic fields. Solar wind properties widely vary depending on the flow angle with respect to the ecliptic, see Fig.~\ref{ulysses}. The measurements by a single spacecraft represent time-sequence, demonstrating fluctuations on timescales from days to seconds, which can be Fourier-analyzed.
The velocity of the solar wind is much larger than the local Alfv\'en speed of around 30 km/s so that
the measurement can be interpreted as the spacial spectrum, see, e.g., Fig.~\ref{helios} for spectra obtained from measurement by Helios 2 spacecraft. The $f^{-1}$ part of the spectrum corresponds to the shot-noise statistics of features emitted by the Sun, while the $f^{-5/3}$ part is the evidence of well dynamically evolved turbulence, the characteristic timescales on these
scales are indeed shorter than the time of flight from the Sun.

\begin{figure}[t]
\begin{center}
\includegraphics[width=0.7\columnwidth]{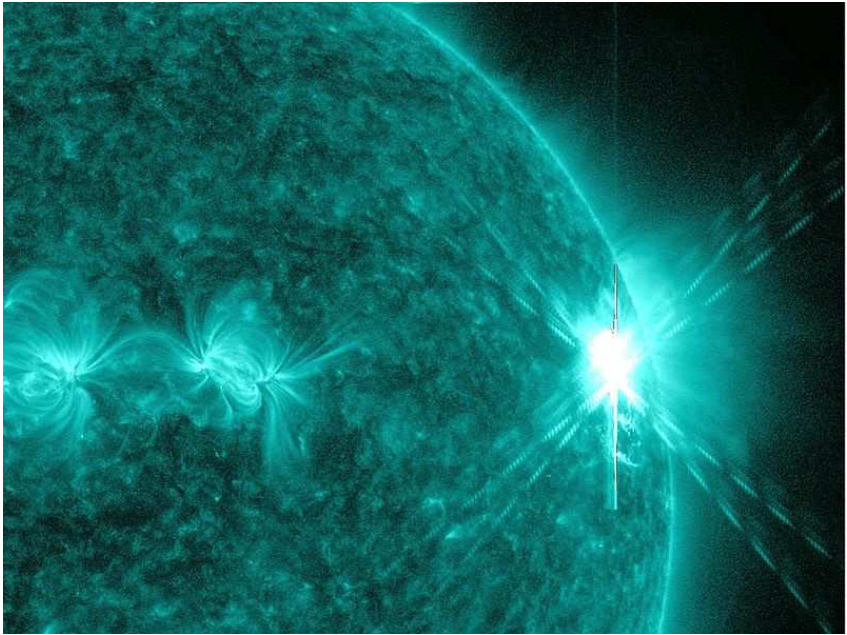}
\end{center}
\caption{X-class solar flare caused by reconnection of large arcs of magnetic field above
the solar surface (NASA/SDO).}
\label{x_class}
\end{figure}

The Sun's outer envelope transports energy to the surface by convection, also generating magnetic
fields in the process. The magnetic field is distributed extremely unevenly on the surface reaching
several kilogauss in sunspots. Sunspots are connected by magnetic arcs visualized by structure
because hot plasma has high thermal conductivity along the field and low conductivity
perpendicular to it. Interaction of strong magnetic flux tubes above the solar surface
leads to magnetic reconnection which results in two spectacular phenomena: X-ray flares (see Fig.~\ref{x_class}) and coronal mass ejections (CME). It is conjectured that reconnection
and the release of magnetic energy are due to the thin current sheet at the intersection of flux tubes becomes unstable and generate turbulence, see Section~\ref{mag_shear}.

\section{Statistical description of turbulence}
\label{statistical}
\begin{figure}
\begin{center}
\includegraphics[width=0.4\columnwidth]{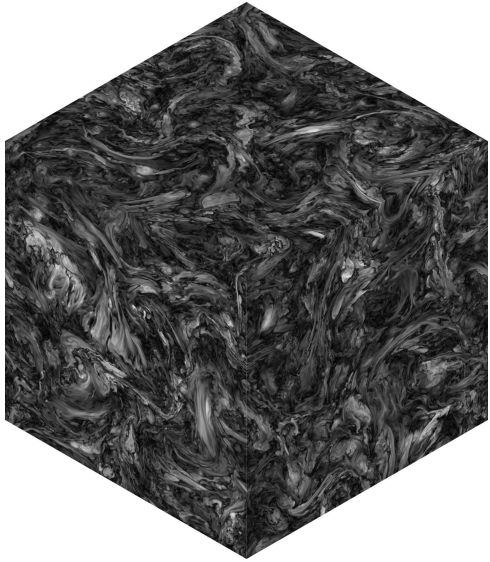}
\end{center}
\caption{Simulated MHD turbulence visualized by magnetic field magnitude shown in grayscale. This statistically homogeneous, isotropic turbulence with zero net magnetic flux was driven by volumetric force to statistically stationary state, one snapshot of which is shown on the picture.}
\label{driven_turb}
\end{figure}

In this section, we will briefly introduce a statistical description. For a more in-depth review
of this subject, we highly recommend the monograph by Monin and Yaglom \cite{monin1975}.
While single realizations of turbulent flow are chaotic and unpredictable, there's some order.
This order can mostly be described statistically, at the same time engineers and scientists
are not interested in individual realizations, but rather in averaged quantities, such as
an averaged lift of an airfoil. Turbulence is a volume-filling and persistent process, 
its realizations filling configuration space densely so that 
the statistical ensemble measurements sometimes can be replaced with time- and volume-averaging (ergodic hypothesis.) The theory relies typically on ensemble averaging, but numerical experiments
mostly use volume and time averaging, see Fig.~\ref{driven_turb}. We will designate averaging $<>$ without specifying whether its statistical, time or volume averaging. 

The spatial variability of a physical variable, e.g., ${\bf v}({\bf r})$, over some scale $l$
can be described as some function of the difference of ${\bf v}$ between points separated by a distance $l$. Second-order statistics can be related to energy, for example
second order structure function (SF) of velocity,
\begin{equation}
 {\rm SF}^2({\bf l})=\langle(v({\bf r}-{\bf l})-v({\bf r}))^2\rangle.
\end{equation}
In the limit of large $l$ this equals to four times kinetic energy, while for smaller
$l$ it is four times ``characteristic energy'' on this scale and all smaller scales.
Fourier-transformed SF can be related to ``energy spectrum'' $E(k)$ (see below). The energy spectrum is the energy distributed in wavenumber space, with $dE=E(k)dk$ being the energy at a particular
wavenumber and $\int E(k)dk$ being total energy. When turbulence is statistically self-similar
we expect a power-law scaling of statistical quantities, e.g., $E(k)$.

The SF above represents the sum of the longitudinal and transverse components of the velocity with respect to direction perpendicular and parallel to ${\bf l}$. Naturally, longitudinal
and transverse functions can be calculated separately. The longitudinal SF
is historically important in experimental research of hydrodynamic turbulence due to being
the primary quantity measured by the heated wire technique. \textit{\small The large-scale flow ${\bf v}_0$ around the wire carries smaller fluctuations which cause fluctuations of the absolute value of ${\bf v}$,
which is what is measured by the changing resistance of the wire, however, fluctuations in $|v|$ are mostly due to fluctuations parallel to the average ${\bf v}_0$.
The Taylor hypothesis assumes that time variations correspond to variations in space, i.e.
measurements separated by $t$ correspond to ${\bf l}= {\bf v}_0 \delta t$. Thus we measure
only the component parallel to ${\bf l}$.} In the solar wind measurements, all
three vector components are recovered so that the transverse, longitudinal and full structure functions can be calculated.

In the case of isotropic turbulence ${\rm SF}({\bf l})$ is only a function of $l$, MHD turbulence
is not isotropic, however, so there is a wider variety of structure functions
that we can measure. However, as we show below, in the reduced MHD limit there is a particular structure function which plays the similar role as the isotropic SF in hydrodynamics, the perpendicular SF 
\begin{equation}
{\rm SF}^2_\perp (l)=\langle(w^\pm({\bf r}-l{\bf n})-w^\pm({\bf r})  )^2\rangle_{\bf r},
\end{equation}
where ${\bf n}$ is a vector perpendicular to the magnetic field.

\begin{figure}[t]
\begin{center}
\includegraphics[width=0.65\textwidth]{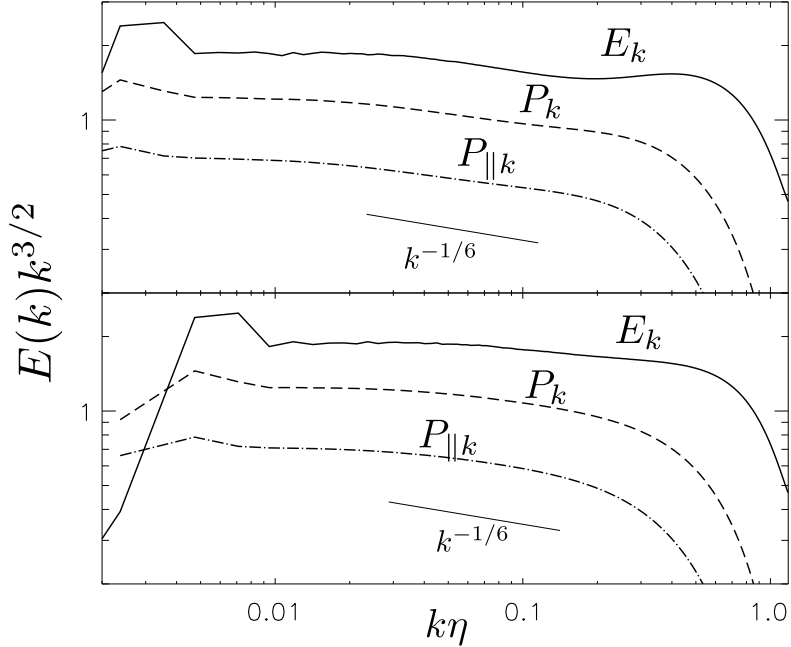}
\end{center}
\caption{Three types of spectra from MHD numerical simulation. $E(k)$ -- solid, $E_1(k)$ -- dashed,
$E_{\|}(k)$ -- dash-dotted. The upper and lower simulations differed in the shape of the elementary cell and the influence of the numerical error. Note the different shapes of the three types of spectra, despite all three spectra convey essentially the same information about energy content of turbulence on different scales.}
\label{spec1d3d}
\end{figure}

The turbulent quantity $u({\bf r})$ can be Fourier-transformed:
\begin{equation}
u({\bf r})=\int e^{i{\bf k}x} d\hat u({\bf k}),
\end{equation}
with the square of the transform called power spectrum:
\begin{equation}
F(\kvec)d \kvec = <|d\hat u(\kvec)|^2>.
\end{equation}
This function can be integrated over the sphere in k-space, e.g. if $F({\bf k})$ depends only on the magnitude of k
we have $E(k)=4 \pi k^2 F(k)$, the resulting quantity we will call three-dimensional spectrum.
Similar procedure is possible when sampling the field along the line, i.e. in one dimension,
this quantity will be called a one-dimensional spectrum $E_1(k)=2 F_1(k)$. \textit{\small Note that 2 comes from F(k) being defined for positive and negative wavenumbers.} $E(k)$, $E_1(k)$ 
can be related in isotropic case by 
\begin{equation}
E_1(k)=\int^\infty_k E(k_1) \frac{dk_1}{k_1},
\end{equation}
while the above mentioned parallel one-dimensional spectrum, which we designate as $E_{\|}(k)$ where only parallel component of velocity is used, for a solenoidal isotropic field:
\begin{equation}
E_{\|}(k)=\int^\infty_k E(k_1)\left(1-\frac{k^2}{k_1^2} \right)\frac{dk_1}{k_1}, 
\end{equation}
so that if $E(k)$ is a power law $E(k) \sim k^\gamma$ then $E(k)=-\gamma E_1(k)$ and 
$E(k)=\gamma(\gamma-2) E_{\|}(k)$. In practice spectras are never exact power laws
so the shape of these spectra are different Fig.~\ref{spec1d3d} shows three types
of spectra from a simulation of MHD turbulence.

Spectra and structure functions
have one-to-one correspondence by Fourier transforms:
\begin{equation}
{\rm SF}^2(r)=2 \int^\infty_0 (1-\frac{\sin kr}{kr}) E(k) dk,
\end{equation}
\begin{equation}
{\rm SF}^2(r)=2 \int^\infty_0 (1-\cos kr) E_1(k) dk.
\end{equation}
If the spectrum has a power-law dependence $k^\alpha$, then by substitution $k=x/r$
we obtain 
\begin{equation}
{\rm SF}^2(r) \sim r^{-1-\alpha},
\label{sf_power}
\end{equation}
provided that remaining dimensionless integral converge. This relation is satisfied for $\alpha$ between $-3$ and $-1$.

From statistical viewpoint turbulence \emph{self-similarity}, i.e. the assumption that
turbulence have a single-fractal structure would mean that for structure functions of arbitrary orders $n$ and $m$ one can write:
\begin{equation}
({\rm SF}^n(r))^{1/n} \sim ({\rm SF}^m(r))^{1/m}
\label{self-similarity}
\end{equation}

Some exact relations for structure functions in turbulence are known for hydrodynamics and MHD,
which helps to test numerics. In the subsequent Section, we explain in more detail the concept
of the \emph{inertial range} -- a range of scales where energy is being overall conserved and is being transferred from one scale to another. From the dynamical viewpoint, these are scales
at which dissipation term can be ignored, and the energy is only injected from large-scale motions
but not from an external force.

The Kolmogorov $-4/5$ law relates a parallel signed structure function
for velocity in the inertial range with the turbulent dissipation rate:
\begin{equation}
{\rm SF}^3_{\|h}(l)=\langle(\delta v_{l\|})^3 \rangle=-\frac{4}{5}\epsilon l.
\end{equation}

Another exact relation, similar to the Yaglom's -4/3 law for incompressible hydro exists for axially symmetric MHD turbulence:
\begin{equation}
{\rm SF}^3_{\|}(l)=\langle\delta w^{\mp}_{l\|} (\delta w^{\pm}_{l})^2 \rangle=-2\epsilon l,\label{PP}
\end{equation}
where $l$ is taken perpendicular to the axis of statistical symmetry -- the direction of the mean magnetic field ${\bf B}$ (\cite{biskamp2003,politano1998}).

The testing of numerics involves measuring SFs and comparing them with theoretical
predictions which help to establish which part of the spectrum is the inertial range,
and which scales are dissipative and driving scales. 
The inertial range in a simulation is often defined as a range
of scales where $-{\rm SF}_{3\|}/l$ is closest to its theoretical value, i.e.,  where the
influence of energy injection from driving and energy dissipation from the viscous
term is minimized. 

\begin{figure}[t]
\includegraphics[width=0.49\textwidth]{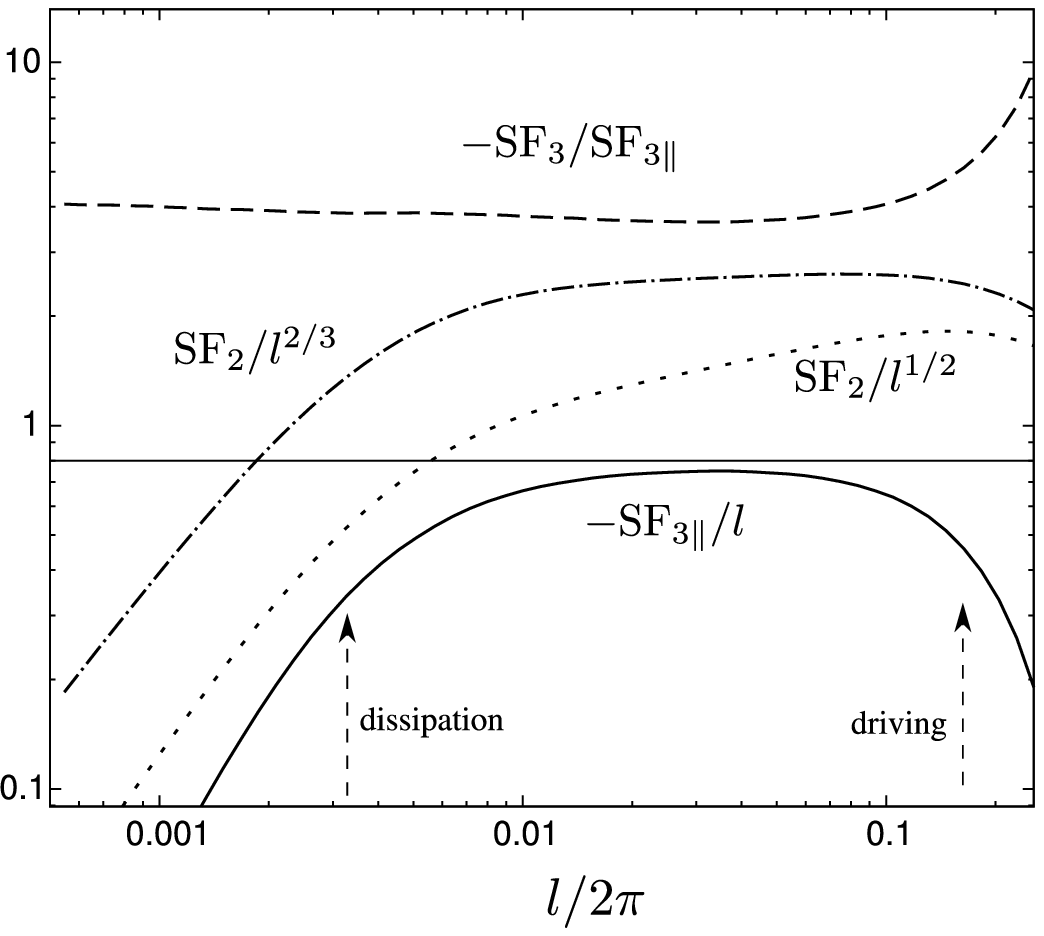}
\includegraphics[width=0.49\textwidth]{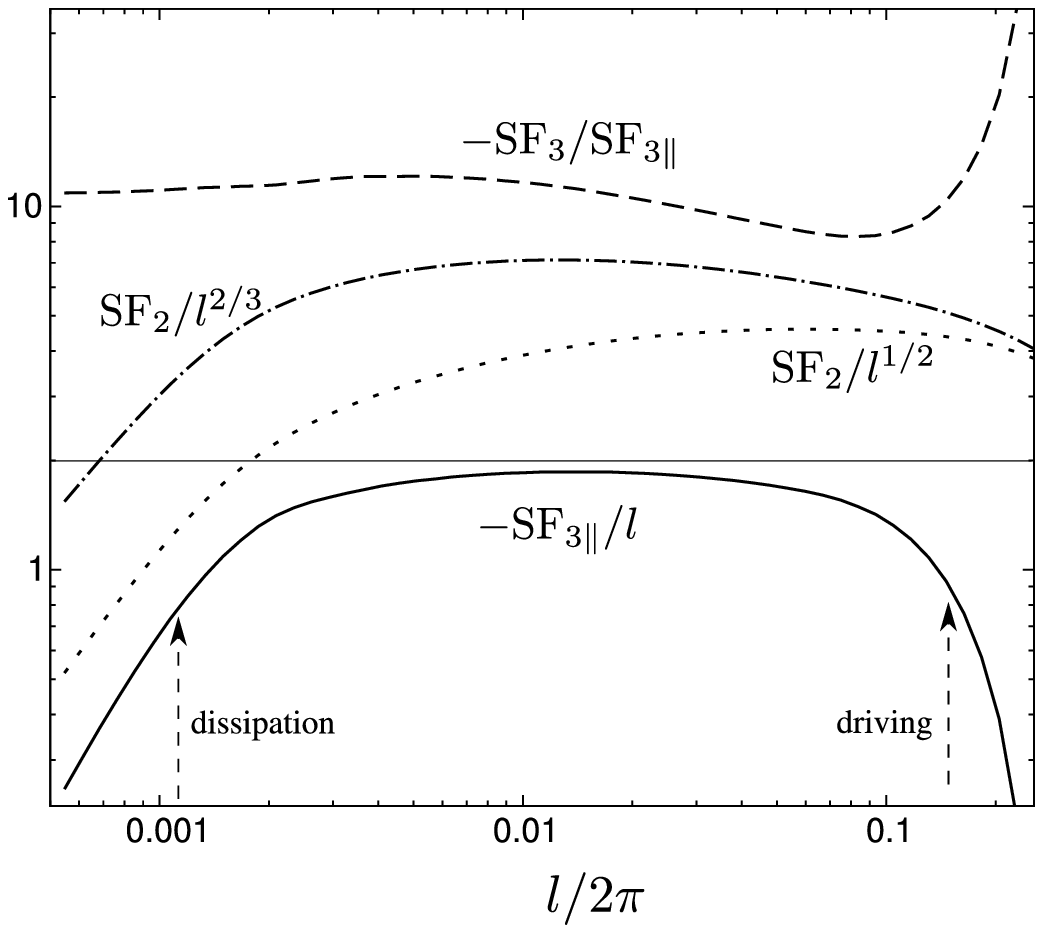}
\caption{Different structure functions vs the distance $l$,
measured in hydrodynamic (left) and MHD (right) simulations.
Solid lines show $-{\rm SF}_{3\|}/l\epsilon$. The influence of driving and dissipation
is minimized in the point where $-{\rm SF}_{3\|}/l\epsilon$ is closer to its theoretical value.
The dashed line indicates the ratio of the third order signed and unsigned SFs as a test for self-similarity. Dotted and dash-dotted lines indicate second-order structure functions,
compensated by $l^{1/2}$ and $l^{2/3}$ correspondingly, in arbitrary units.
Here $l^{2/3}$ is the Richardson-Kolmogorov scaling and $l^{1/2}$ is the scaling that appears in Kraichnan DIA model for hydrodynamics or Iroshnikov-Kraichnan model for MHD.}
\label{sfs}
\end{figure}

Fig.~\ref{sfs} shows several structure functions, compensated by
various powers of $l$. 
The ratio
of different structure functions can test turbulence self-similarity. If this ratio is dimensionless, it is supposed to
be constant through scales. 
For the test of self-similarity on Fig.~\ref{sfs} we show the ratio
of parallel third order structure function and full third
order SF, ${\rm SF}_3=\langle|{\bf v}({\bf r}-{\bf l})-{\bf v}({\bf r})|^3\rangle$.
Fig.~\ref{sfs} shows that hydrodynamic
turbulence is rather self-similar at the same time the scaling of the second-order
structure function in the inertial range is around $l^{0.7}$,
i.e., close to the Kolmogorov scaling (see next section). 

\section{Kolmogorov cascade model}
\label{kolm_cascade}
In hydrodynamic turbulence, a useful starting point is the Kolmogorov model\cite{kolm41} for incompressible turbulence. The incompressible case has constant density so that the energy
dissipation can be defined per unit mass and assumed statistically homogeneous as well.
This quantity, $\epsilon$ have units of ${\rm cm}^2/{\rm s}^3$ and plays a crucial role in many
situations and will be used plenty through this review.
The Kolmogorov model assumes that the statistical properties of turbulence are uniquely determined by the amount of energy available in this stationary homogeneous system,
i.e., by the $\epsilon$ alone. Furthermore, it is argued that the energy self-similarly cascades
through the series of scales known as the \emph{inertial range}. Cascade means that the energy is being transferred from one scale to another without dissipation.

The dimensional derivation of Kolmogorov scaling involves noting that the spectrum defined
in the previous section have units of ${\rm cm}^3/{\rm s}^2$ and the wavenumber have units
of ${\rm cm}^{-1}$, so that 

\begin{equation}
E(k)=C_K \epsilon^{2/3} k^{-5/3},
\label{kolm53}
\end{equation}

where $C_K$ is a dimensionless Kolmogorov constant. From the previous Section we also know
that the 3D spectrum $F(k)\sim E(k)k^{-2} \sim k^{-11/3}.$

The hand-waving derivation involves introducing ``characteristic velocity on scale $l$'' $u_l$
and imagining that the energy rate is constant for all scales:

\begin{equation}
u_l^2/t_{\rm casc}=\epsilon,
\label{cascading}
\end{equation}

where $t_{\rm casc}$ is the ``cascading timescale'' the time it takes for nonlinearity to remove
energy from scale $l$ and transfer it to smaller scales. It is further assumed that in the hydrodynamic cascade $t_{\rm casc}$ is a dynamical time on each particular scale, i.e. $t_{\rm casc}\approx l/u_l$, which results in 

\begin{equation}
u_l^3/l \sim \epsilon.
\label{u_l_law}
\end{equation}

\begin{equation}
u_l \sim (\epsilon l)^{1/3} \sim \epsilon^{1/3} k^{-1/3}.
\label{u_l_law2}
\end{equation}

From the definition of the spectrum and its relation to the SF, we can argue that $E(k)k \sim u_l^2$ so that the two formula for Kolmogorov scaling agree.

A compilation of experimental results for hydrodynamic turbulence \cite{sreenivasan1995} suggests that a Kolmogorov constant $C_K$ is universal for a wide variety of flows. High-resolution numerical simulations of isotropic incompressible hydrodynamic turbulence, see Fig.~\ref{hydro} and  \cite{gotoh2002} suggest the value around 1.6. See Fig.~\ref{hydro_spec} for a spectrum from $1024^3$ simulation from \cite{B11}.

The expression Eq.~\ref{u_l_law} can be written for the largest scale in the system,
the outer scale:
\begin{equation}
\epsilon=C'_K \delta v^3/L,
\label{eps_zeroth_law}
\end{equation}

This can be regarded as scaling with the outer scale velocity $v$, and/or the scale
of the system $L$. Three different things can be done to study this law, known
as the zeroth law of turbulence, empirically. One can scale an experimental
apparatus from $L$ to $L'$, one can increase or decrease velocity, and one can change
the fluid to vary viscosity. From the symmetries of the hydrodynamic equations, we know
that the only real change would be a change in Reynolds number. The same type of turbulent flow results in approximately the same dimensionless coefficient of $C'_K$. 
In the systems that generate turbulence easily, e.g., flow past the grid the above expression
is reasonably precise for ${\rm Re}>200$. Note that in the statistically stationary case $\epsilon$
is also the energy dissipation rate, which happens on small scales due to viscosity.
This fact illustrates that the outer scale $L$ the dissipative scale only
know each other through $\epsilon$. 

Outer scale is sometimes formally defined through integral over the spectrum, e.g. $L=3\pi/4E\int_0^\infty k^{-1}E(k)\,dk.$ \label{outer_scale}. Usually, this is around
a scale where energy is injected into the system. \textit{\small Inverse cascade of energy in two-dimensional hydrodynamic turbulence is one counter-example of this.}.

The energy ``cascades'' down to smaller scales until it hits the so-called Kolmogorov
scale, where dissipative processes overcome nonlinear transfer of energy.
The Kolmogorov scale can be expressed as a combination of viscosity/diffusivity and energy dissipation rate, which gives a unit of length.
\begin{equation}
 \eta=(\nu_n^3/\epsilon)^{1/(3n-2)},
 \label{kolm_scale}
\end{equation}
where $n$ is the order of the viscosity, e.g. n=2 for classic molecular viscosity,
$\nu_n$ is the value of the diffusivity, so that we obtain Navier-Stokes equation by
adding to the RHS of the Euler equation the dissipation operator $-\nu_n(-\nabla^2)^{n/2}$.

Dimensionless ratio $L/\eta$ could serve as a ``length of the inertial range'', although in practice spectrum is around an order of magnitude shorter. 

Criticism of the Kolmogorov model points to the fact that the assumption
of self-similarity is quite arbitrary and points to the examples of turbulence
which are notably not self-similar. In the three-dimensional hydrodynamic turbulence
deviations from self-similarity in the second-order measurements, such as energy spectrum,
are fairly small, however. The more precise formula can be obtained by multiplying RHS of Eq.~\ref{kolm53} by the ``intermittency correction'' $(kL)^\alpha$, where $\alpha\approx 0.035$.
For more details see \cite{frisch1995}.

\subsection{Lagrangian spectrum}
\label{lagrangian}
Lagrangian measurements are performed by following a fluid element. Lagrangian viewpoint
offers a simpler conceptual picture as the model above is conceptually simpler in Lagrangian formulation. The Euler's equation is a third Newton's law for the fluid element:
\begin{equation}
\frac{D \bf v}{Dt} = - \frac{\nabla P}{\rho}.
\label{Euler_lagr},
\end{equation}
Here $D/Dt$ is the advective (Lagrangian) derivative corresponding to changes of a fluid element's properties over time, $D/Dt = \partial/\partial_t+{\bf v}\cdot{\bf \nabla}$.
The work per unit mass, done upon a fluid element by pressure of surrounding fluid elements will
be expressed, therefore as ${\bf v} \cdot {d \bf v}/dt$. The Kolmogorov theory would therefore
assume that, given a characteristic time interval $\tau$, the work done per unit mass upon a fluid
element during this interval, $\delta \bf{v}_\tau \cdot \delta \bf{v}_\tau/\tau$, will be constant
when $\tau$ corresponds to inertial-range timescales and equal to the turbulence energy cascade
rate per unit mass $\epsilon$. Formally, in stationary turbulence the second-order Lagrangian structure function of velocity should satisfy:
\begin{equation}
{\rm SF}(\tau)=\langle ({\bf v}(t+\tau)-{\bf v}(t))^2 \rangle \approx \epsilon \tau
\end{equation}
in the inertial range, where ${\bf v}(t)$ is a velocity as a function of time for a given fluid element. This time structure function will correspond to the frequency spectrum of
\begin{equation}
E(\omega) \approx \epsilon \omega^{-2},
\end{equation}
see Eq.~(\ref{sf_power}).
This first appeared in the texbook \cite{ll06a} and also in \cite{corrsin1963,tennekes1972}.
The scaling $\omega^{-2}$ and the fact that the energy \emph{spectrum} is proportional to \emph{energy injection rate} $\epsilon$ appear to be conceptually simpler than $\epsilon^{2/3}$ scaling of the standard Eulerian Kolmogorov scaling.

This spectrum has a dissipation \emph{timescale} associated with the lifetime of critically damped eddies, also called the Kolmogorov timescale (for n=2):
\begin{equation}
\tau_\eta=(\nu/\epsilon)^{1/2},
\label{tau_eta}
\end{equation}
this is the location of the dissipative cutoff in Lagrangian spectrum.
The direct measurement of the Lagrangian frequency spectrum is fairly challenging, however,
as the probe has to be embedded in the flow. Temporal measurement of spectra from a wind
tunnel or channel flow (e.g. \cite{Grant1962}) does not correspond to the Lagrangian spectrum but can be connected, by Taylor hypothesis, to spatial spectrum (see Section~\ref{statistical}). Below, in Section~\ref{anis_phenom} we explain how a parallel
spectrum in MHD turbulence can act as a surrogate of the Lagrangian spectrum.

\subsection{More general Kolmogorov phenomenology}
\label{kolm_gen_sec}
More general phenomenology is possible assuming cascading time scale
relating to the dynamical timescale as 
\begin{equation}
t_{\rm casc}=\frac{l}{u_l} \left(\frac{l}{L}\right)^{-\alpha},
\end{equation}
in which case using Eq.~\ref{cascading} we get
\begin{equation}
u_l \sim \epsilon^{1/3} l^{(1-\alpha)/3} L^{\alpha/3},
\label{kolm_gen}
\end{equation}
so that, assuming self-similarity,
the spectral slope will be $-1-2(1-\alpha)/3=-5/3+2\alpha/3$ (Eq.~\ref{sf_power}).
This will also result in a different Kolmogorov scale which we get
by equating cascading time above and the viscous time $l^n/\nu_n$:
\begin{equation}
 \eta_\alpha^{3n-2+2\alpha}=L^{2\alpha}\frac{\nu_n^3}{\epsilon}.
 \label{eta_gen}
\end{equation}
This reduces to Eq.~\ref{kolm_scale} for $\alpha=0$.
Alternatively, one can also assume that interaction is reduced by $u_l/c_s$:
\begin{equation}
t_{\rm casc}=\frac{lc_s}{u_l^2},
\end{equation}
where $c_s$ is the sound speed. This model is called acoustic/wave turbulence and gives
the $-3/2$ spectral slope.

\subsection{Scaling convergence in turbulence: numerics and experiments}
\label{convergence}
\begin{figure}[t]
\begin{center}
\includegraphics[width=0.5\columnwidth]{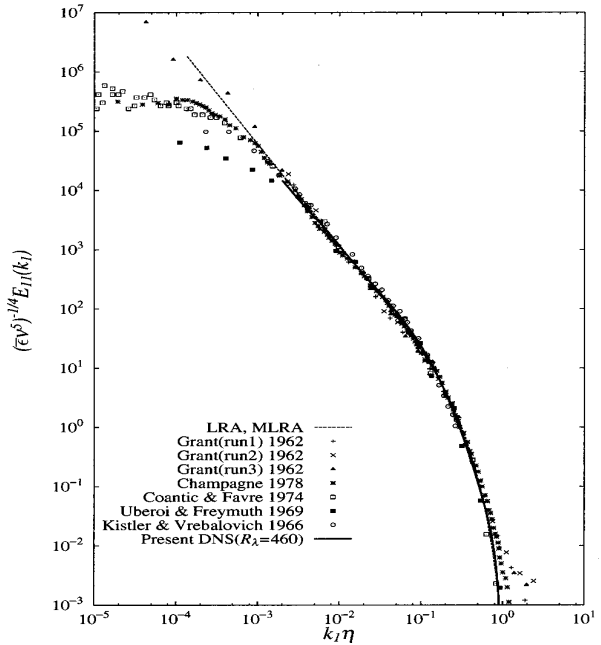}
\end{center}
\caption{Spectra of hydrodynamic turbulence from numerics (solid) and experiments (dots)
in dimensionless units, from \cite{gotoh2002}.}
\label{hydro}
\end{figure}

\begin{figure}
\begin{center}
\includegraphics[width=0.6\columnwidth]{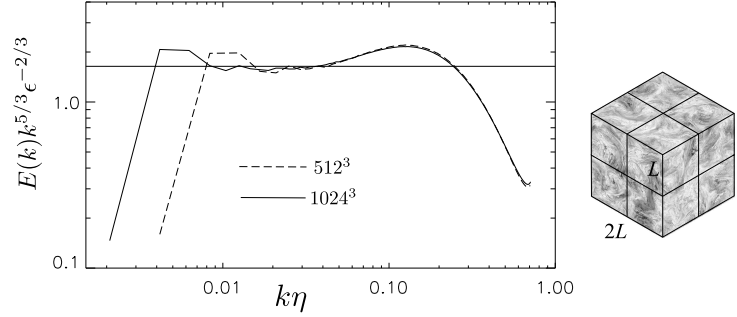}
\end{center}
\caption{Spectrum of hydrodynamic turbulence compensated by Kolmogorov scaling to give approximate constant function vs. wavenumber. Statistics from smaller datacube should largely repeat
statistics from larger datacubes, as we see on collapsing spectra on the left and visually
represented on the right.}
\label{hydro_spec}
\end{figure}

Inertial range in the 3D numerics is not as big as in nature.
In numerics, we use a rigorous quantitative argument to elucidate
asymptotic inertial-range scaling. Imagine we performed several simulations
with different Reynolds numbers. If we believe that turbulence is universal,
and the separation of scales between forcing scale and dissipation scale
is large enough, the properties of small scales should not depend on
how turbulence was driven and also on the scale separation itself. 
This is because MHD or hydrodynamic equations do not explicitly contain
any designated scale, so the simulation with a smaller dissipation scale could be considered, because of the symmetry from equations,
as a simulation with the same dissipation scale, but larger driving scale.
For example, the small-scale statistics in a $1024^3$ simulation will look similar to small-scale statistics in $512^3$ simulation, if we keep physical sizes of the grid cell and the
dissipation scale the same as on Fig.~\ref{hydro_spec}.
Another example is the convergence of experimental data as well as numerics
onto the same curve on Fig.~\ref{hydro}. Note how x- and y-axis units
were made dimensionless using the Kolmogorov length scale and the
Kolmogorov velocity scale.

Scaling convergence can be used to compare numerics with measurements, in which can numerics should faithfully reproduce dynamics on all relevant scales, including dissipation scales, e.g. numerics should be ``well resolved''. However, 
in the case when several numerical experiments with different Re
are compared using scaling convergence, this condition can be somewhat relaxed. 
Instead, the condition is that discretized formulation works similarly in the
compared experiments. The discretization error and other numerical inaccuracies
of statistically averaged quantities should depend only on the ratio of
Kolmogorov scale to the grid scale provided that the timestep is also
determined by the grid scale. So we need to keep the grid scale as the fixed
fraction of the Kolmogorov scale. Keep in mind, that Kolmogorov scale itself
is determined based on particular phenomenology of the cascade (see Section~\ref{kolm_gen_sec}) and may not be known apriori. In this case, rigorous
scaling convergence would require going through available hypotheses and checking each
in turn.

We express the spectra of several simulations in dimensionless units
corresponding to the expected scaling, for example, a $E(k)k^{5/3}\epsilon^{-2/3}$
for the Kolmogorov model and plot it versus dimensionless wavenumber $k\eta$,
where dissipation scale $\eta$ again, corresponds to the same phenomenology.
On the plot, the two spectra should collapse onto the same curve on the viscous scales, as long as the model works. The method has been used extensively in hydrodynamics
\cite{yeung1997,gotoh2002,kaneda2003} with great success. In numerics, it is especially
efficient since, while experimental data may suffer from systematic uncertainties,
numerics does not, and it collects tremendously large statistics on small scales,
driving statistical error virtually to zero. Let us understand why this is the case.
If we refer to the Kolmogorov cascade picture, described above, the energy cascade
is local in scale and the only information that is being transferred from large scales
to small scales is the local cascade rate $\epsilon$. Now, assuming that at each scale, each eddy is independently created, its energy content on this scale should only depend
on $\epsilon$ as $\epsilon^{2/3}$. So if we normalize the measurement by $\epsilon^{-2/3}$,
each eddy will represent, presumably, independent estimate of such normalized energy
content at each scale. Given a characteristic eddy scale $l$, the number of eddies in a datacube goes as $l^{-3}$, while the number of correlation timescales for strong turbulence goes as $l^{-2/3}$, so the statistical error due to volume and time-averaging should decrease as $l^{-11/6}$. The plotted normalized spectrum $I(x)=I(k\eta)=E(k)\epsilon^{-2/3}k^{5/3}$ should be ``pinned'' on the dissipation scale, because it should satisfy 
\begin{equation}
\int^\infty_0 I(x) x^{1/3} dx = \frac12.
 \label{pinned}
\end{equation}
The precision of the convergence method
was demonstrated in \cite{kaneda2003}, where $4096^3$ simulations allowed
to capture the intermittency correction, which is a correction of $-0.04$ to the $-5/3$
spectral slope.

\section{MHD equations, modes}
\label{mhd_eq}
Below we write ideal MHD equations that describe perfectly conducting, inviscid fluid. It should be kept in mind that solving ideal equations often require including special treatment
of shocks and turbulence.

\begin{eqnarray}
\partial_t \rho+{\bf \nabla\cdot}(\rho{\bf v})=0,\\
\rho(\partial_t+{\bf v}\cdot\nabla){\bf v}=-\nabla P+{\bf j\times B},\\
{\bf \nabla\cdot B}=0,\\
\partial_t{\bf B}=\nabla\times({\bf v \times B}),\label{mhd0}\\
P=P(\rho,s)
\end{eqnarray}
with current ${\bf j}={\bf \nabla \times B}$ and vorticity ${\bf \omega}={\bf \nabla \times v}$,
$P(\rho,s)$ is an equation of state. 

In the ideal case, specific entropy $s$ is decoupled from the rest of the equations and can be described as a passive scalar. \textit{\small Here we use Heaviside units, redefining electric charge with a factor of $1/4\pi$ and getting rid of $4\pi$ factors in Maxwell's equations.}

Introducing sound speed $c_s^2=\partial P/\partial \rho$, linearized MHD equations reveal four perturbation modes: 

1) Alfv\'en mode -- transverse waves with
v and B perturbations along ${\bf k\times B}$ and 
 dispersion relation $\omega={\bf (v_A \cdot k)}$, where ${\bf v_A=B}/\sqrt{4\pi\rho}$,
so-called Alfv\'en velocity = magnetic field ${\bf B}$ in velocity units introduced earlier.
The phase velocity of Alfv\'en mode is 

\begin{equation}
u_A=\omega/k=\pm {\bf (v_A \cdot \hat k)}= \pm v_A \cos\theta, 
\label{u_A}
\end{equation}

while its group velocity $\partial\omega/\partial {\bf k}=\pm {\bf v_A}$, hence the term Alfv\'en velocity.  

2,3) Fast and slow modes -- compressible waves with perturbations in the ${\bf k,B}$ plane propagating correspondingly faster  and slower than $v_A$, with the dispersion relation

\begin{equation}
u_{f,s}^2=\omega^2/k^2= \frac{1}{2} \left[ (v_A^2 +c_s^2) \pm 
               \sqrt{ (v_A^2 +c_s^2)^2 - 4 v_A^2 c_s^2 \cos^2\theta } \right].
\label{u_fs}			   
\end{equation}

4) Entropy mode -- non-propagating passive scalar perturbations of specific entropy \cite{ll08,biskamp2003}. 

In this section, we will skip eigenvectors for the three modes for brevity and write them out
in Section~\ref{compressible}.

\section{Numerical methods to simulate MHD turbulence}
Several tools are available to simulate turbulence numerically:

\subsection{Pseudospectral codes}
The pseudospectral code solves MHD equations as a set of ordinary differential equations in time
for each spacial Fourier harmonic. The coupling of harmonics is through the nonlinear term
which is calculated in real space (hence ``pseudo'') and then converted
back to Fourier space. Since the pseudospectral method approximates derivatives non-locally,
using all data points, it does not suffer from dispersion error. Also, if some care
is taken with timestep integration, for example, a symplectic integrator is used, it also
preserves energy, i.e., does not suffer from dissipation error. The explicit dissipation, e.g.
viscosity or resistivity are done with simple algebraic operations in Fourier space and can be made
unconditionally stable, irrespective of the time step with no numerical expense.
A typical pseudospectral code have symplectic integrator, corrects for aliasing error
and has explicit dissipation in the form $a_{t+\Delta t}=a_t \exp(-\nu k^2 \Delta t)$. Aliasing error
comes from frequencies above $2/3$ or below $-2/3$ of the Nyquist frequency if the nonlinear term
is second order. E.g., if the Nyquist frequency is $\pi$, and keeping frequencies within
 $[-2/3\pi,2/3\pi]$, the sum or difference will still be within the interval modulo $2\pi$
(e.g. $2/3\pi+2/3\pi-2\pi=-2/3\pi$). Another advantage of pseudospectral code
for the incompressible case is that divergence-free condition for velocity and magnetic
field can be done with simple algebraic operations in Fourier space.
The spacial reconstruction uses all Fourier harmonics. As a result, the method's precision
increases exponentially with the number of points in one dimension. This makes it practical
to do ``fully resolved'' simulations, i.e., when the viscosity and magnetic diffusivities are explicit, and all scales of interests are represented with reasonable precision.
The usual rule of thumb for well-resolved simulation in a periodic box with size $2\pi$ and
number of points $N$ in one direction, when the wavenumbers are represented by integers $[-N/2+1,...,0,1,...,N/2]$ is $k_{\rm max} \eta \gtrsim 1$, where $k_{\rm max}=N/3$ for
a 2/3 dealiased code.
The main disadvantage is difficulty in introducing arbitrary boundary conditions. This method's periodic box comes naturally when we try to simulate homogeneous isotropic turbulence, however.
One example of a publicly available pseudospectral code is Snoopy:
\url{http://ipag.osug.fr/~lesurg/snoopy.html}.

\subsection{Finite difference codes}
Finite difference codes estimate derivatives by finite differencing. The precision of the code
increases, typically, as a power law with the number of points, the index of the power law is
the ``order'' of the code. The main advantage is simplicity and numerical speed. Disadvantages include special treatment of shocks with ``shock viscosity''. High order finite difference codes with explicit diffusivities can be rather precise in simulating turbulence.
The divergence-free condition can be kept with ``divergence cleaning'' or with equations formulated in terms of magnetic potential.
Pencil code is popular publicly avalable high order finite difference 
code: \url{https://github.com/pencil-code}

\subsection{Finite volume codes with Riemann solvers}
Also known as Godunov codes. Finite volume codes keep values of cell averages
then \textit{reconstruct} (interpolate) values on the interface of the cells both from the right and from the left. Thus the interface value is discontinuous and may be evolved for a short time
as a ``Riemann problem'' -- initial value problem with a single discontinuity. The time-average fluxes of conserved quantities through the interface are then computed from, typically, the approximate solution of the Riemann problem by the ``Riemann solver''. Finally, fluxes
are used to advance cell averages in time.
The inherent ability to describe discontinuities makes Godunov codes very robust and a code
of choice to simulate supersonic turbulence.
See also documentation of the publicly available code Athena++: 

\url{https://github.com/PrincetonUniversity/athena-public-version}

\subsection{Lagrangian codes}
Lagrangian codes refer to codes which use grid or material elements that are moving with the fluid.
These include N-body codes (e.g., collisionless particles moving under the action of gravity), smooth particle hydrodynamics (SPH) codes -- particle codes simulating hydrodynamics, moving mesh codes, among which purely Lagrangian (mesh moving with the fluid) or arbitrary Lagrangian-Eulerian (ALE, mesh moving in an arbitrary way). These codes are often used to simulate collapse under gravitational forces, but less common to simulate turbulence.

Publicly avalable code Gadget2: \url{https://wwwmpa.mpa-garching.mpg.de/gadget/}

\section{Theory of Alfvenic Turbulence}
\label{alfvenic}
In Section ~\ref{kolm_cascade} we introduced the standard Kolmogorov description
of the inertial range of incompressible hydrodynamic turbulence. It is clear, however,
that this picture is not applicable to MHD. Turbulence spectra are typically steeper
than $k^{-1}$ meaning that RMS fields are dominated by large scales. In hydrodynamics,
however, large-scale velocity can be
nullified by an appropriate choice of the reference frame. In MHD large-scale magnetic field
cannot be nullified and will be dynamically important on all scales, including very small scales.
This combination of sizable RMS large-scale field and small-scale fluctuations of the fields is the main difference from hydrodynamics. Also known as a ``strong field limit'', it was pointed out by Iroshnikov and Kraichnan \cite{iroshnikov1964, kraichnan} and it was suggested that inertial-range MHD turbulence is weak turbulence. Here weak turbulence refers to the picture of wave turbulence where wave packets propagate almost freely, and collision between waves leads to the small perturbation in their structure so that the perturbation theory is applicable \cite{zakharov1992}. The interaction of wave packets in MHD, however, is very different from the collision of sound waves. Introducing wavevector components parallel and perpendicular to the mean field, $k_\|$ and $k_\perp$ we see that the wave frequency $\omega=k_\| v_A$ depends
only on $k_\|$. This anisotropic dispersion relation results in anisotropic turbulence. 

The subsequent analytic work demonstrated that MHD turbulence tends to become stronger and not weaker during the cascade \cite{galtier2000}, as we will show below. We will also show that the Alfvenic part of MHD perturbations governs this highly anisotropic turbulence, hence the terms ``Alfvenic Turbulence''.

The rationale of working with simplified incompressible equations is similar to hydrodynamics.
Assuming that a) turbulence have no shocks, b) no sizable energy is carried by sound waves
(in MHD case, fast MHD mode), c) the Mach number $M_s=V_L/c_s$ is small, we can argue that a scale-wise Mach number $M_s=\delta v/c_s$ should also be small and decrease
with scale. The fluid compressibility will, therefore, be small in the inertial range.

Incompressible MHD equations consist of two dynamical equations and two constraints:
\begin{eqnarray}
\partial_t{\bf v}&=&-\nabla P'/\rho - 
(\nabla \times {\bf v}) \times {\bf v}+(\nabla \times {\bf b}) \times {\bf b},\\
\partial_t{\bf b}&=&\nabla\times({\bf v \times b}),\label{mhd01}\\
{\bf \nabla\cdot v}&=&0,\\
{\bf \nabla\cdot b}&=&0.
\end{eqnarray}
Here we normalized magnetic field to velocity units in the same manner as in previous
Sections, i.e., ${\bf b}={\bf v_A}={\bf B}/\sqrt{4\pi\rho}$.
The dynamical equations are known as the momentum equation and the induction equation.
The induction equation honors the divergence-free constraint for the magnetic field, this effectively results in no new constraint, if the initial condition is chosen divergence-free.
The divergence-free constraint for velocity is satisfied by the appropriate choice
of the scalar function $P'=P+\rho v^2/2$. The pressure, therefore, is a dummy variable.

Introducing solenoidal projection
$\hat S=(1-\nabla\Delta^{-1}\nabla)$ we can rewrite the equations without explicit constraints:
\begin{eqnarray}
\partial_t{\bf v}&=&\hat S(-(\nabla \times {\bf v}) \times {\bf v}
+(\nabla \times {\bf b}) \times {\bf b}),\\
\partial_t{\bf b}&=&\nabla\times({\bf v \times b}).\label{mhd1}
\end{eqnarray}

Very useful change of variables to the Els\"asser variables ${\bf w^\pm=v\pm b}$
makes these equations even more compact: 
\begin{equation}
\partial_t{\bf w^\pm}+\hat S ({\bf w^\mp}\cdot\nabla){\bf w^\pm}=0.\label{mhd2}
\end{equation}

We introduce total energy, $1/2 \int (v^2+b^2)\, d\bf r$ and cross-helicity $\int {\bf v\cdot b} \, d\bf r$ which are conserved in this incompressible formulation.
By taking the sum and difference of these quantities, we obtain conservation of each of Elsasser
energies $1/2 \int (w^\pm)^2\, d\bf r$.

\subsection{From weak to strong turbulence}
\label{weak_turb}
Keeping in mind the above argument of a sizable mean field let us explicitly
write it down as the constant field ${\bf v_A}$ in a given volume,
and perturbations as $\delta{\bf w^\pm=w\pm v}_A$:
\begin{equation}
\partial_t{\bf \delta w^\pm}\mp({\bf v_A}\cdot\nabla){\delta \bf w^\pm}
+\hat S ({\delta \bf w^\mp}\cdot\nabla){\delta \bf w^\pm}=0\label{mhd3}.
\end{equation}

Let us denote $\|$ and $\perp$ as directions parallel and perpendicular to ${\bf v}_A$ and the subscript to the vector means projection to ${\bf v}_A$ or the perpendicular plane, respectively.

In the limit of small $\delta w$'s they represent perturbations, propagating along $\bf B$ or in the opposite direction, with the nonlinear term describing their interaction. Note that ``self-interaction'' of $\delta w^+$ or $\delta w^-$ is absent, both being an exact solution in the absence of another. The dominant nonlinear interaction is a three-wave process,
so writing the dispersion relation and the conservation laws for energy and momentum, 
\begin{eqnarray}
\omega_n=k_{\|n} v_A,\\
\pm \omega_1=\pm \omega_2 \pm \omega_3,\\
k_{\|1}= k_{\|2}+k_{\|3},\\
k_{\perp 1}= k_{\perp 2}+k_{\perp 3}
\end{eqnarray}
we see that one of the $\omega_n$ must be zero. Let us choose $\omega_3=0$, 
this means $|k_{\|1}|= |k_{\|2}|$, but there's no restrictions on $k_{\perp 1,2}$. 
The cascade preserves frequencies and goes forward by increasing only $k_{\perp}$. 

In wave turbulence theory the interaction strength $\xi$ is the ratio of the nonlinear shear rate
$k_\perp \delta w$ to the wave frequency $k_\| v_A$, it describes a fractional perturbation
during one wave period:
\begin{equation}
\xi=k_\perp \delta w /k_\| v_A.
\end{equation}
It is also the estimate of the ratio of the nonlinear term to the mean-field term in Eq.~\ref{mhd3}.
In MHD turbulence the dynamical timescale $\tau_{\rm dyn}=1/k_\perp \delta w$ does not
have to be proportional to the cascade timescale as in hydrodynamic turbulence. Instead, $\tau_{\rm casc}$ is increased by a factor of $1/\xi$. This can also be understood in terms of perturbations of a wave packet being a random walk. Each individual perturbation is $\xi$ strong, so it takes
$(1/\xi)^2$ steps to destroy the wavepacket completely:
\begin{equation}
\tau_{\rm casc}=(1/k_\| v_A)(1/\xi)^2=k_\| v_A/(k_\perp \delta w)^2.
\label{casc_weak}
\end{equation}

The energy cascade rate is the energy on each scale divided by the cascade time on this scale. This rate is expected to be constant through scales and we designate it $\epsilon$:
\begin{equation}
\delta w^2 \frac{(\delta w k_\perp)^2}{v_A k_\|}= \epsilon.
\label{weak_casc}
\end{equation}

Note $k_\|$ here is constant, so the phenomenological cascade spectrum is determined by
$\delta w^2 \sim k_\perp^{-1}$, which corresponds to one-dimensional perpendicular
spectrum $E(k_\perp) \sim \delta w^2 k_\perp^{-1} \sim k_\perp^{-2}$. This argument
can be followed rigorously by perturbation collision integral approach, used in
wave turbulence\cite{zakharov1992} and solved exactly by Zakharov transformation,
which was accomplished in \cite{galtier2000,galtier2002}.

One consequence of this solution is that turbulence grows anisotropic,
with $k_\perp/k_\| \sim k_\perp$. Interestingly, it becomes stronger
and not weaker on smaller scales, in other words, $\xi$ is an increasing function of $k_\perp$.
Indeed, if we maintain $k_\|$ constant, this will result in
\begin{equation}
\xi=\frac{\delta w k_\perp}{v_A k_\|} \sim k_\perp^{1/2} \xrightarrow[k_\perp\to \infty]{} \infty.
\end{equation}
Our two conclusions from this simple perturbation theory is that: 
a) the resonance condition results in a ``perpendicular cascade'', making MHD turbulence
anisotropic, b) turbulence becomes stronger along the cascade until $\xi\sim 1$.

One can wonder if weak MHD turbulence is ever realized in nature. We can hypothesize
that this is the case in astrophysical objects where the strong magnetic field is anchored
in a heavy object, i.e., a star and is extended into the magnetosphere where perturbations of the field are much smaller than this anchored field. The empirical evidence for this
case and specifically for the $k^{-2}$ perpendicular spectrum is lacking, however.
One can argue that large-scale dynamo (which we consider in subsequent Sections) can generate a mean field which is much stronger than perturbations, but empirically
we know from the ISM observations that they are of the same order. This results in MHD turbulence being strong on the outer scale.

\subsection{Reduced MHD approximation}
\label{sec:rmhd}

Equation \ref{mhd3} can be further simplified assuming anisotropy $k_\perp \gg k_\|$ and the fact that $\delta w \ll v_A$. This allows to neglect parallel gradients
in the nonlinear term, indeed, the mean field term with the parallel gradient
$(v_A \nabla_\|) \delta w^\pm$ is always much larger than similar contribution
from the nonlinear term,
$(\delta w^\mp_\| \nabla_\|)\delta w^\pm$ and the latter could be ignored.
So the three vector components of equation~\ref{mhd3} are split into interdependent equations
for the scalar $\delta w^\pm_\|$ and vector $\bf \delta w^\pm_\perp$:
\begin{eqnarray}
\partial_t{\bf \delta w^\pm_\|}\mp({\bf v_A}\cdot\nabla_\|){\delta \bf w^\pm_\|}
+\hat S ({\delta \bf w^\mp_\perp}\cdot\nabla_\perp){\delta \bf w^\pm_\|}=0,\label{mhd4}\\
\partial_t{\bf \delta w^\pm_\perp}\mp({\bf v_A}\cdot\nabla_\|){\delta \bf w^\pm_\perp}
+\hat S ({\delta \bf w^\mp_\perp}\cdot\nabla_\perp){\delta \bf w^\pm_\perp}=0,\label{mhd5}
\end{eqnarray}
Note that Equation~\ref{mhd4} depends on Eq.~\ref{mhd5}, but not vice-versa.
Since Equation~\ref{mhd4} represent passive dynamics and does not have essential nonlinearity,
the nonlinear cascade is completely governed by Eq.~\ref{mhd5}. This latter equation
is known as reduced MHD. In this anisotropic limit, the ${\bf \delta w^\pm_\perp}$ is purely
the Alfv\'en mode, and ${\delta w^\pm_\|}$ is the amplitude of the slow mode. 
Turbulence in Eq.~\ref{mhd5} is called Alfv\'enic turbulence.

Slow mode for ${\delta w^+_\|}$ is a passive scalar to ${\bf \delta w^-_\perp}$
and vice versa. If Alfv\'en and slow modes will be injected similarly from large scales, they will have the same statistics. In practice, the slow mode content can be determined from
numerics. 

It turns out, reduced MHD is more general than incompressible MHD and can be used beyond collisional fluid description. Alfv\'enic perturbations are transverse and rely only
on the tension of the magnetic field line as a restoring force,
the charged particles tied to this magnetic field line provide inertia.
The $[\bf E\times B]$ drift waves with wavelengths much smaller than the ion skin depth are
indeed just Alfv\'en waves, and they exist regardless of the collisionality of the plasma \cite{Schekochihin2009}, which is useful for the description of the collisionless solar wind.
The anisotropy of MHD turbulence has been known empirically for a while, and RMHD had been formulated for perturbations in plasma in strongly magnetized case some time ago \cite{kadomtsev1974,strauss1976}. Since RMHD motions do not require plasma pressure we
assume the results that we find for Alfvenic turbulence in this section do not depend
on the ratio of plasma pressure to magnetic pressure ``$\beta$'', despite we started the derivation assuming infinite $\beta$.

Introducing parallel length $\Lambda=2\pi/k_\|$ and perpendicular length $\lambda=2\pi/k_\perp$
we see that reduced MHD has a two-parametric symmetry:
\begin{equation}
{\bf w} \to {\bf w}A,\ \lambda \to \lambda B,\ t \to t B/A,\ \Lambda \to \Lambda B/A.
\end{equation}
$A$ and $B$ are arbitrary parameters of the transformation. This is the same symmetry as in hydrodynamics, except for the parallel scale $\Lambda$ transforms similar to time, not to length. 
$\Lambda$ being similar to time is very important and leads to analogies
between dynamics in time and parallel structure in space as we show below.
MHD equations do not have such symmetry, so Kolmogorov self-similarity arguments, technically,
can not be applied to the MHD case. In practice, this regime for MHD can be achieved within
the inertial range where $\delta w \ll v_A$ condition and anisotropy condition are satisfied. In numerics, it is challenging to reach these universal dynamics directly from
isotropic scales with $\delta w \sim v_A$. Instead, one can directly solve RMHD equations.
As a practical comment, the statistics from the full MHD with $\delta w^\pm \sim 0.1 v_A$ is very
close of that one of RMHD, see \cite{BL09a}.

Another symmetry is evident in RMHD, related to the value of $v_A$,
The equations are unchanged under transformation $v_A \to v_A A,\ \Lambda \to \Lambda A$. 
The parallel scale and the Alfv\'en speed can be rescaled simultaneously without changing the dynamics.

\subsection{Strong MHD turbulence}
\label{anis_phenom}
As we demonstrated above in Section~\ref{weak_turb}, the perpendicular cascade will result in the growth of $\xi$ and will naturally lead to strong turbulence, with $\xi \sim 1$. Goldreich and Sridhar \cite{GS95} proposed that the growth of $\xi$ will be limited by the uncertainty relation between the cascading timescale and the wave-packet frequency, namely that the cascade time cannot be shorter than the wave period: $\tau_{\rm casc}\omega \geq 1$. Using Eq.~\ref{casc_weak} we get $\xi \leq 1$. This will make $\xi$ to be stuck around unity, which was termed as ``critical balance'' by Goldreich and Sridhar. As far as $\xi \sim 1$ we have $\tau_{\rm dyn} \sim \tau_{\rm casc} \sim 1/\omega$,
we can regard turbulence as ``strong'' and apply Kolmogorov phenomenology.
For the cascade of the two Elsasser energies: 
\begin{equation}
\epsilon^\pm=\frac{(\delta w^\pm_\lambda)^2\delta w^\mp_\lambda}{\lambda},
\label{fluxes}
\end{equation}
These are two independent cascades, but in a theory with $\epsilon^+=\epsilon^-$ this becomes
standard Kolmogorov phenomenology in the $k_\perp$ direction. 
\begin{equation}
E(k)=C_K \epsilon^{2/3} k_\perp^{-5/3},\label{spec53}
\end{equation}
We will return to the more general ``imbalanced'' case with $\epsilon^+ \neq \epsilon^-$ in the 
next section, but briefly note that such theory is non-trivial since it is impossible to maintain 
critical balance for resonant waves with different amplitude.

The assumption of critical balance $\xi \sim 1$ allow us to estimate perturbation anisotropy directly. The ``wavevector anisotropy'' relates two wavevectors
at which the one-dimensional spectrum along the field and perpendicular to the field
have the same power. A similar relation can be obtained between parallel and perpendicular scales
$\Lambda$ and $\lambda$ vis SFs. Using $\xi = 1$, and the $-5/3$ scaling $\delta w \sim \lambda^{1/3}$ we obtain $k_\| \sim k_\perp^{2/3}$, which is known as GS95 anisotropy.

There is a different argument, however, that is sufficient to obtain this anisotropy. This argument
is based on RMHD symetry $\Lambda \sim v_A$ we discussed in Section \ref{sec:rmhd} and dimensional grounds. Indeed, if we have $\Lambda \sim v_A$, the rest of the expression for $\Lambda$ must have units of time, which is uniquely obtained
from $\lambda$ and $\epsilon$ as $\lambda^{2/3} \epsilon^{-1/3}$:

\begin{equation}
\Lambda=C_A v_A \lambda^{2/3} \epsilon^{-1/3}\label{anis},
\end{equation}
where we introduced a dimensionless ``anisotropy constant'' $C_A$.

The perpendicular SF which correspond to $k_\perp^{-5/3}$ spectrum will have the scaling 
${\rm SF}_\perp \sim \lambda^{2/3}$ (Eq.~\ref{sf_power}), while inserting $\Lambda \sim \lambda^{2/3}$, we get parallel structure function as ${\rm SF}_\| \sim \lambda^{2/3} \sim \Lambda$.
\begin{figure}[t]
\begin{center}
\includegraphics[width=1.00\columnwidth]{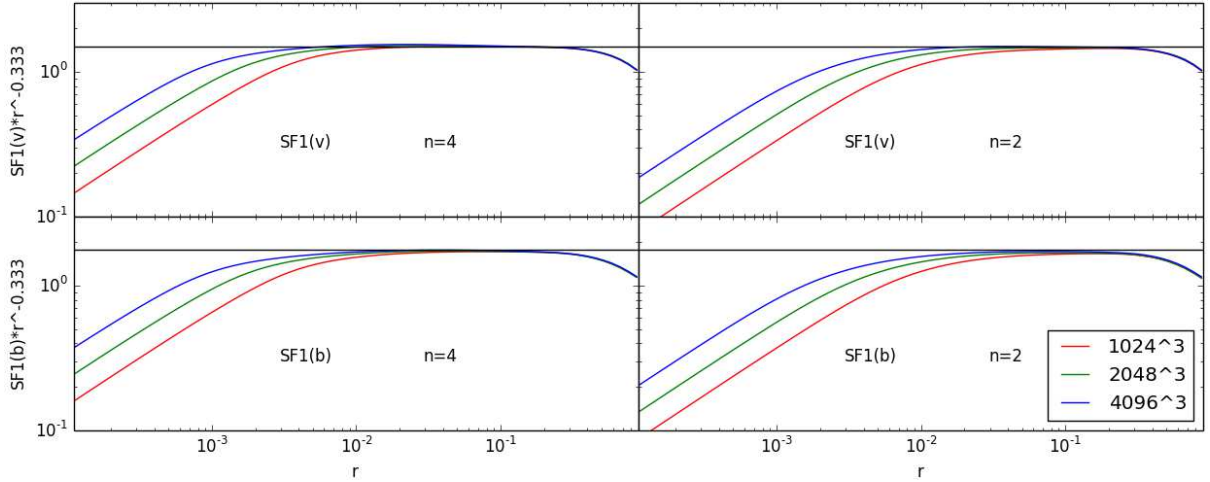}
\end{center}
\caption{First order SFs of velocity (top) and magnetic field (bottom) compensated by $r^{1/3}$ (Kolmogorov scaling). Left: M1-3H, right: M1-3}
\label{SF1_13}
\end{figure}

The parallel spectrum, which corresponds to such SF is $E(k_\|) \sim k_\|^{-2}$ and from dimensional arguments we recover the prefactor as
\begin{equation}
E(k_\|)= C_\| \epsilon v_A^{-1} k_\|^{-2},
\label{par_spec}
\end{equation}
where we introduced dimensionless constant $C_\|$.

Equations ~\ref{spec53} and ~\ref{anis} (or, alternatively ~\ref{par_spec}) describe the spectrum and anisotropy of MHD turbulence, which may still be corrected for intermittency. 

A modification which leads to a shallower and not steeper spectrum was proposed
in \cite{boldyrev2005,boldyrev2006}, henceforth B06 suggesting that GS95 scalings
are modified by a scale-dependent factor that decreases the strength
of the interaction, effectively, the theory described in Sec.~\ref{kolm_gen_sec} with
$\alpha=1/4$. Different arguments to the same effect were proposed in \cite{gogoberidze2007}. 
In this case the spectrum will be expressed as $E(k)=C_{K2} \epsilon^{2/3} k^{-3/2}L^{1/6}$,
see Eq.~\ref{kolm_gen}, the factor $\xi$ is modified by $(l/L)^{1/4}$, so that anisotropy follows modified critical balance with $k_\| \sim k_\perp{^1/2}$. The Kolmogorov scale of B06 model is 
obtained from Eq.~\ref{eta_gen}: $\eta_{1/4}=(\nu_n^3/\epsilon)^{1/(3n-1.5)}L^{0.5/(3n-1.5)}$.
\begin{figure}[t]
\includegraphics[width=0.49\textwidth]{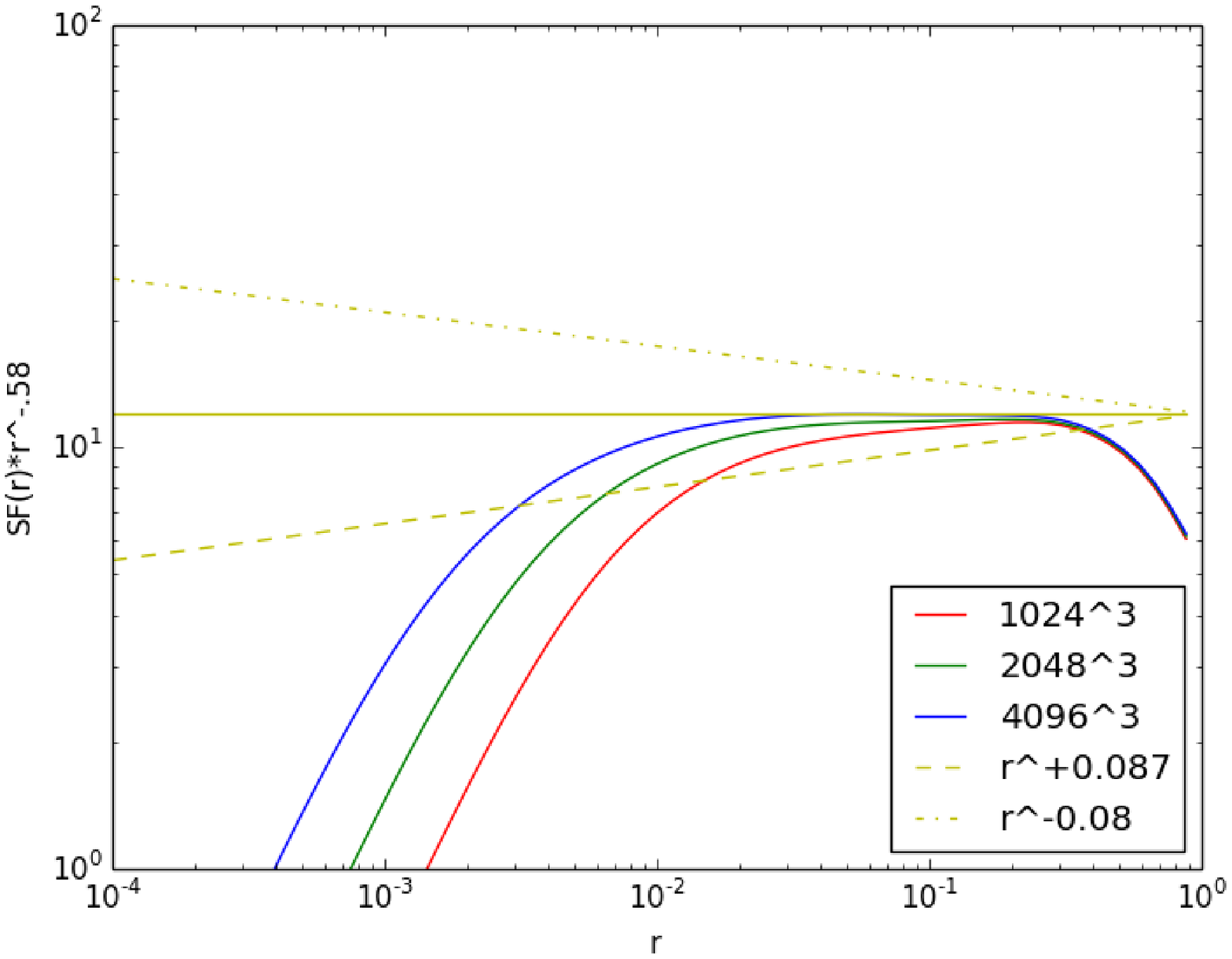}
\includegraphics[width=0.49\textwidth]{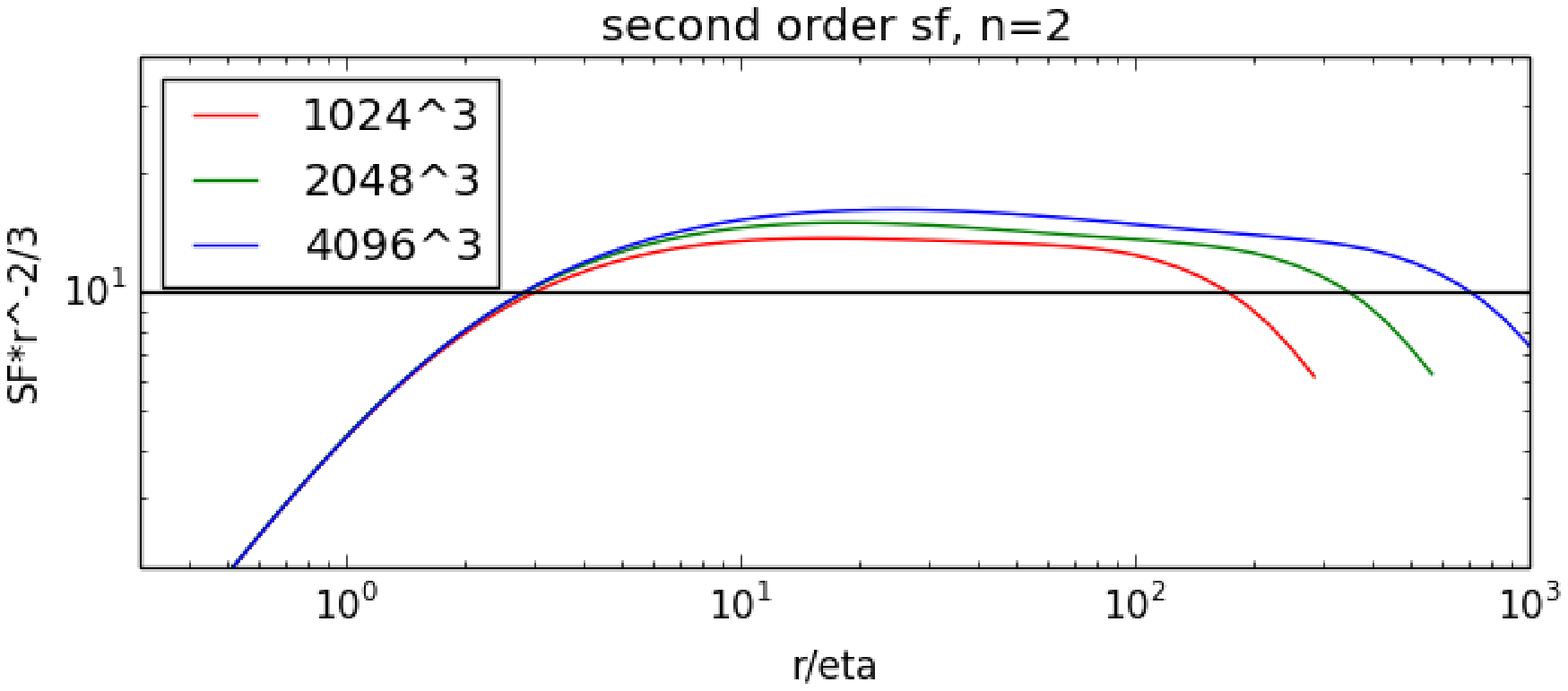}
\caption{Second order SF scaling for M1-3. Left: SF plotted vs dimensional distance $r$ compensated by $r^{-0.58}$. Right: scaling convergence study for the $r^{2/3}$ (Kolmogorov) scaling, described in Sec.~\ref{convergence}, both axes are dimensionless. We see convergence, i.e., the overall scaling is $r^{2/3}$.}
\label{sf_n}
\end{figure}

It turns out the anisotropy can be argued from the Lagrangian frequency spectrum without
postulating critical balance or involving uncertainty relations.
In the incompressible MHD all modes propagate with the same speed, the Els\"asser components propagate either along or against the local magnetic direction, i.e., along with the magnetic field line. This propagation will be described by the functional form $f(s \mp v_A t)$, where $s$ is a distance along the field line. The nonlinear interaction will contribute to the slower time evolution
of $f$ and the trajectory $s =\pm v_A t$ will be analogous to following hydrodynamic fluid element in the Lagrangian formulation. Let us record ${\bf w^+}$ and
${\bf w^-}$ along the field line in a fixed time. The positive direction $s$ will be equivalent to following the evolution of ${\bf w^+}$ backward in time and ${\bf w^-}$ forward in time.
In measuring the frequency spectrum, the sign of time will be unimportant.
So the measurement of power spectrum along the field line will be analogous to Lagrangian
frequency spectrum with frequency $\omega$ replaced by the wavenumber $k_\|=\omega/v_A$ \cite{B14a}:

\begin{equation}
E(k_\|) =  C_\| E(\omega) \frac{d\omega}{dk_\|}=  C_\| \epsilon (v_A k_\|)^{-2} v_A = C_\| \epsilon v_A^{-1} k_\|^{-2} \label{par_spec2}
\end{equation}
This is the same expression as obtained in Section \ref{anis_phenom} from phenomenological
considerations. The parallel structure function ${\rm SF}_\|(l) \sim \epsilon l v_A^{-1}$
The dimensional argument involving Alfv\'en symmetry of reduced MHD arrive at the same result \cite{B12a}. This symmetry allows $E(k_\|) dk_\|$ to depend only on $k_\| v_A$, which
will require that $E(k_\|)\sim v_A^{-1}$. The rest of the expression can be obtained from units.

\subsection{Numerics: perpendicular spectrum}
\begin{table}
  \caption{Three-dimensional MHD and RMHD simulations.\label{4096exp}}	
\begin{center}
  \begin{tabular*}{1.00\columnwidth}{@{\extracolsep{\fill}}c c c c c c c c}
    \hline\hline
Run  & $N^3$ & Dissipation & $v_A$ &   $\epsilon$ & $\eta$ &  $k_{\rm max}\eta$ & $v_A \tau_\eta$ \\
\hline
MHD1 & $1536^3$ & $-5\cdot10^{-10}k^4$   &0.73& 0.091 &  0.0021  & 1.08 &  0.026     \\
MHD2 & $1536^3$ & $-6.2\cdot10^{-10}k^4$ &1.53& 0.728 &  0.0018  & 0.92 &  0.025 \\
   \hline
M1 & $1024^3$ & $-1.75\cdot10^{-4}k^2$   &1&    0.06  &  0.0031  & 1.05 &  0.044 \\
M2 & $2048^3$ & $-7\cdot10^{-5}k^2$      &1&    0.06  &  0.00155 & 1.06 &  0.028 \\
M3 & $4096^3$ & $-2.78\cdot10^{-5}k^2$   &1&    0.06  &  0.00077 & 1.06 &  0.017 \\
\hline
M1H & $1024^3$ & $-1.6\cdot10^{-9}k^4$   &1&    0.06  &  0.0030  & 1.04 &  0.045 \\
M2H & $2048^3$ & $-1.6\cdot10^{-10}k^4$  &1&    0.06  &  0.00152 & 1.04 &  0.029 \\
M3H & $4096^3$ & $-1.6\cdot10^{-11}k^4$ &1&     0.06  &  0.00076 & 1.04 &  0.018  \\
   \hline 
\end{tabular*}
\end{center}
\end{table}

\begin{figure}[t]
\begin{center}
\includegraphics[width=0.9\columnwidth]{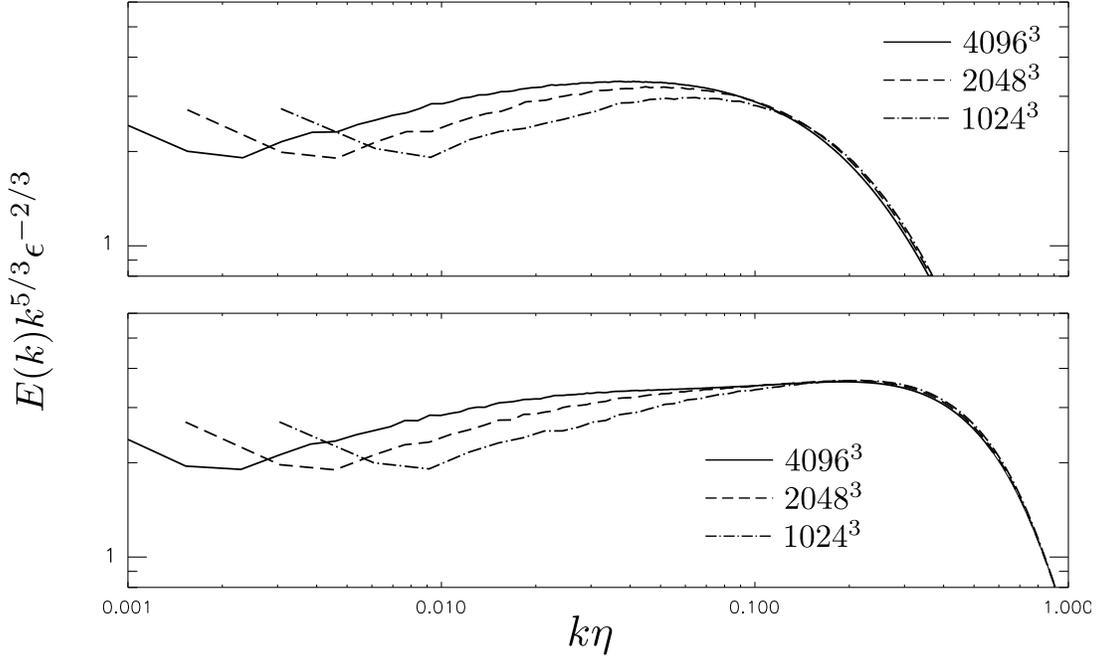}
\end{center}
\caption{Checking $-5/3$ hypothesis with the scaling convergence study (Sec.~\ref{convergence}) of perpendicular spectrum.
Solid, dashed and dash-dotted lines are the spectra from $4096^3$,  $2048^3$ and $1024^3$ simulation correspondingly. The upper plot shows normal diffusion M1-3 simulations, and the lower plot shows hyperdiffusive M1-3H simulations. The convergence is reasonable around the dissipation scale. The scaling that achieves the best convergence is $\approx -1.70$. Applying the same method to the $-3/2$ slope model results in the lack of convergence.
From \cite{B14a}.}
\label{energy17}
\end{figure}

\begin{figure}[t]
\begin{center}
\includegraphics[width=0.9\columnwidth]{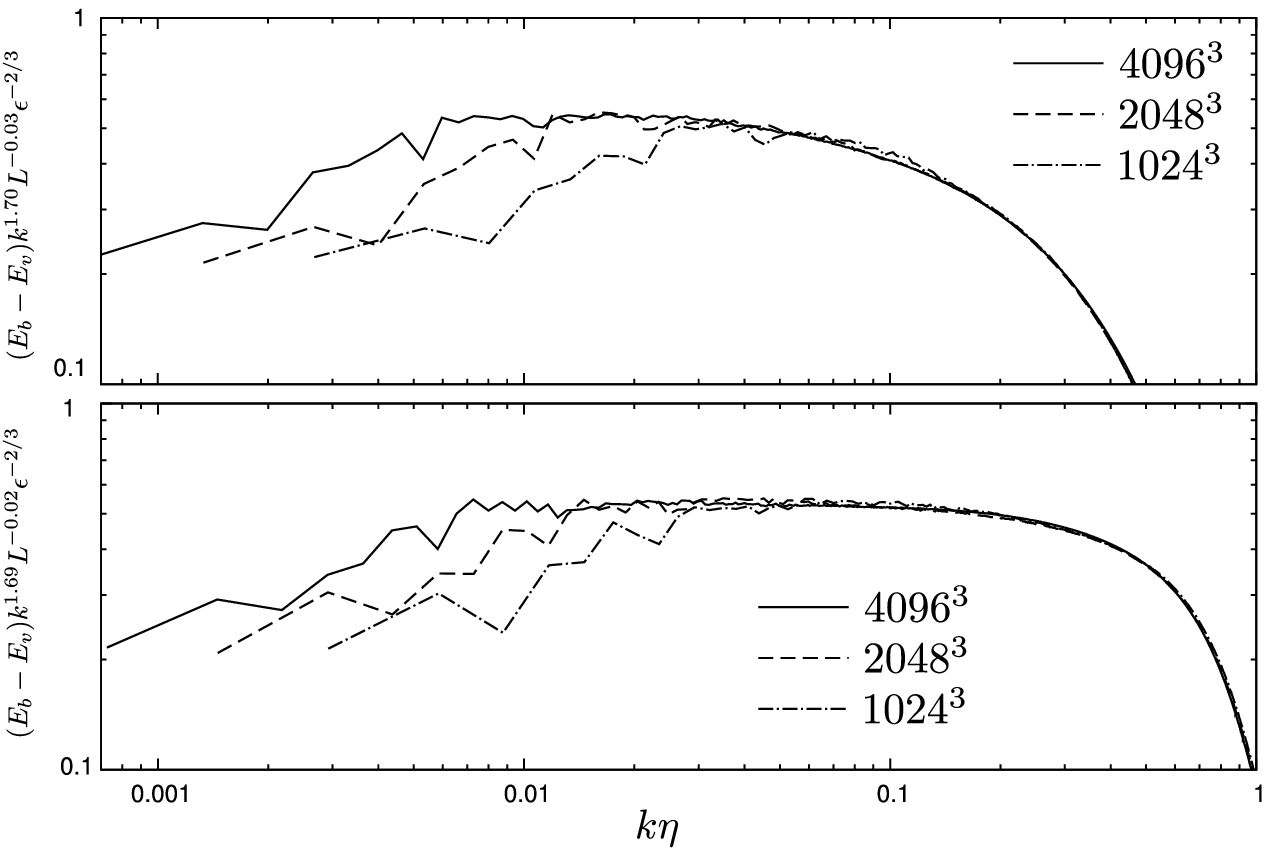}
\end{center}
\caption{Residual energy convergence. Best convergence is $k^{-1.70}$ scaling for M1-3 and $k^{-1.69}$ scaling for M1-3H. From \cite{B14a}}
\label{residual}
\end{figure}

Table~\ref{4096exp} presents parameters of DNS strong MHD and RMHD turbulence (first presented in \cite{B14a}, averaged statistics are publicly available at\\ \url{https://sites.google.com/site/andreyberesnyak/simulations/big3}.), these are a well-resolved driven statistically stationary simulations intended to precisely calculate averaged quantities. MHD cases labeled MHD1-2 have no mean field, $B_0=0$ so that $v_A$ is defined only locally with the Table listing RMS values of $v_A$.                     
Two series of RMHD driven simulations are described in Table~\ref{4096exp} as M1-3 and M1-3H.
These have a strong mean field we denote $B_0$, RMS fields $v_{\rm rms}\approx B_{\rm rms} \approx 1$, perpendicular box size of $2\pi$ and parallel box size of $2\pi B_0$. The driving was anisotropic with anisotropy $B_0/B_{\rm rms}$ so that turbulence starts being strong from the outer scale. Technically, $B_0$ is arbitrary. However, the RMHD limit is only applicable to very large $B_0$ as we showed above.

In simulations, we see a rapid decrease of parallel correlation length right after the driving scale, which indicates the efficiency of nonlinear interaction and the regime of strong turbulence. The correlation timescale for $v$ and $B$ was around $\tau\approx 0.97$, so the box contained around 6.5 parallel correlation lengths. Each simulation was started from long-evolved low-resolution simulation and was subsequently evolved for $\Delta t=13.5$ in code units in high resolution, and we used the last $7$ dynamical times for averaging. In earlier work \cite{B11,B12b} it was found that averaging over $\sim 7$ correlation timescales gives a reasonably good statistic on the outer scale and very good statistics on smaller scales.

Numerically, we used $k_{\rm max} \eta>1$ resolution criterion, with $\eta$ being classic Kolmogorov scale. Additionally, we checked the precision of the spectra by performing a resolution study on lower resolutions. In particular, we saw spectral error lower than $8\times 10^{-3}$, up to $k \eta =0.5$ when increasing resolution from $576^3$ to $960^3$ and the spectral error lower than $3\times 10^{-3}$ when we increased parallel resolution in a $1152^3$ simulation by a factor of two. We presume this error is a mostly systematic error, associated with grid effects because the statistical error is likely to be vanishingly small, see the end of Section~\ref{convergence}.
We did not use any data above $k \eta =0.5$ for fitting as the spectrum sharply declines after this point and contains negligible energy. 

On Fig.~\ref{SF1_13} we plotted first order perpendicular structure functions of velocity and magnetic field. These seem to scale with the Kolmogorov power of $r^{1/3}$.
On Fig.~\ref{sf_n} we show second-order SFs for M1-3. On the left of this figure are SFs vs. distance. We see that the scaling is not obvious, with higher-resolution SF having the shallower slope
of the flat part, the $4096^3$ seemingly having $r^{0.58}$ scaling not expected from theory.
On the right of Fig.~\ref{sf_n} we use rigorous scaling convergence study (Sec.~\ref{convergence}) to show that the overall scaling is $r^{2/3}$ (Kolmogorov).

Fig.~\ref{energy17} presents a convergence test of the perpendicular 3D spectrum for the $-5/3$ model, and the convergence is reasonable, while the best convergence is reached at the -1.7 scaling. Fig.~\ref{residual} shows a convergence study of the residual energy spectrum (magnetic energy minus kinetic energy). The best convergence is, again, near -1.7 slope. In all cases
the convergence is consistent across two simulation groups with different dissipation prescriptions, M1-3 and M1-3H.
 
\begin{figure}[t]
\begin{center}
\includegraphics[width=0.6\columnwidth]{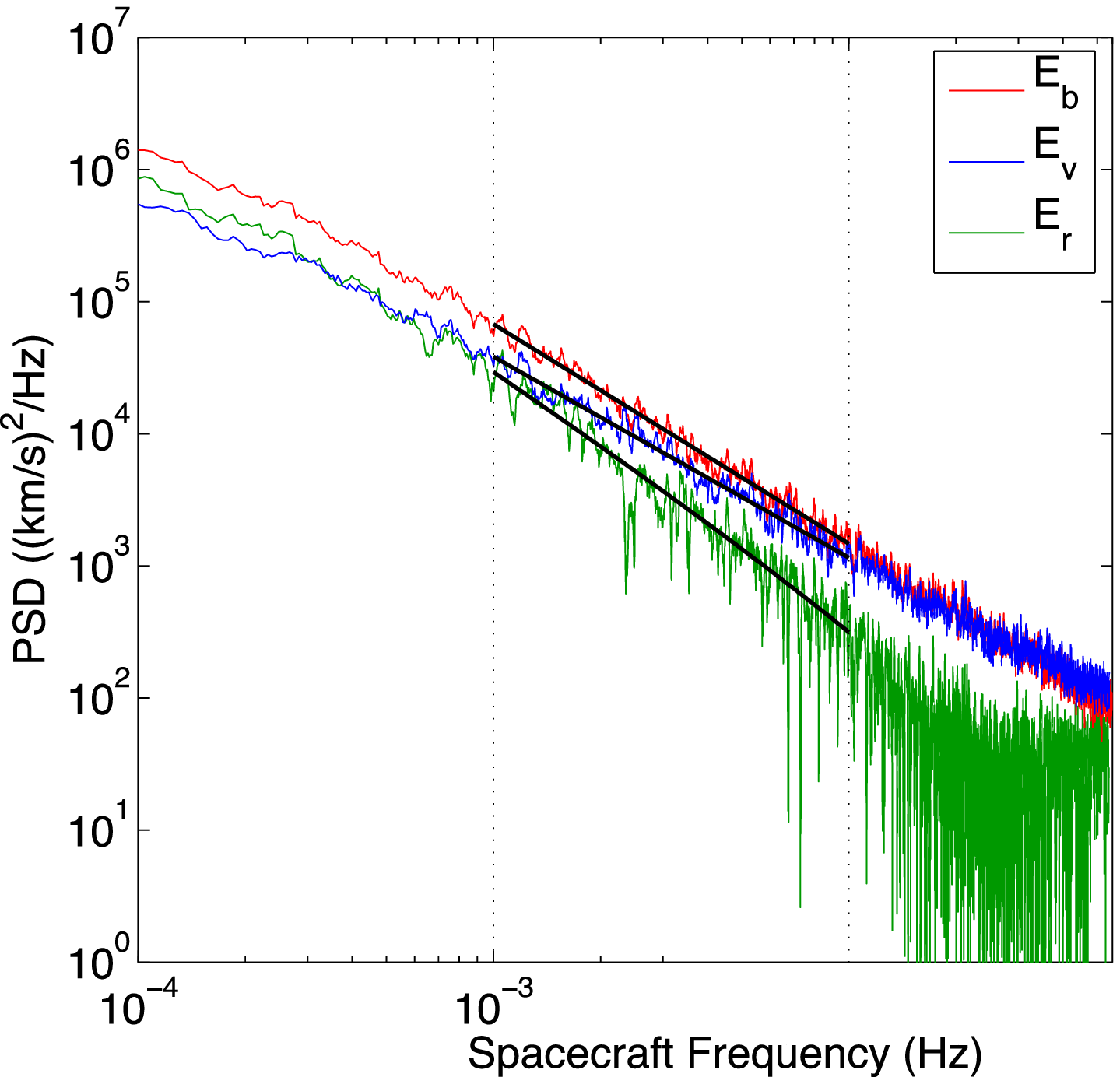}
\end{center}
\caption{Power spectra of magnetic field, velocity and residual
energy measured in the solar wind. Alfven ratio was
strongly fluctuating, and the average was around 0.71.
From \cite{chen2013}}
\label{sw_power}
\end{figure}

\label{resolsub}

The flat part of the normalized spectrum can be used to obtain a Kolmogorov constant of $C_{KA}=3.3\pm 0.1$, which was first reported in \cite{B11}. The total Kolmogorov constant
for both Alfv\'en and slow mode in the above paper
was estimated as $C_K=4.2\pm0.2$ for isotropically
driven turbulence with zero mean field. This was obtained
using an empirical energy ratio between slow and Alfv\'en mode,
$C_s$ which is between 1 and 1.3. This larger
value of Kolmogorov constant, $C_K=C_{KA}(1+C_s)^{1/3}$ is due to slow mode being passively
advected and not contributing to nonlinearity. 

We also from these simulations that the residual energy, $E_B-E_v$ have the same spectral slope
as the total energy, i.e., there is a constant fraction of residual energy in
the inertial range. The results in Fig.~\ref{residual} show that residual energy scaling is the same as for total energy so that residual energy is a constant fraction of the total energy. Our best estimate for this fraction is $\sigma_r=0.15\pm0.03$. More commonly used in the solar wind community, Alfven ratio $r_A=E_v/E_B=(1-\sigma_r)/(1+\sigma_r) \approx 0.74$. Residual energy and its scale-dependence has been discussed
in the past and has recently been associated with the so-called called alignment measures in simulations \cite{BL09b} and in the solar wind measurements \cite{wicks2013,chen2013}. Explaining previously reported $-2$ scaling \cite{muller2005} for the residual energy is challenging from the theoretical standpoint. Assuming particular residual energy on the outer scale, and the $-2$ scaling, its value in the inertial range will depend on the scale separation. This would mean a nonlocal character of residual energy. Our simulations, showing that the residual energy is just a fraction of the total energy in the inertial range, resolve this conceptual difficulty and make theories suggesting different scalings for magnetic and kinetic energies obsolete.

The solar wind spectra often feature different kinetic and magnetic scalings, see Fig.~\ref{sw_power}. The amount of residual energy changes from measurement to measurement and is different for the fast and the slow solar wind \cite{wicks2013,chen2013}. These deviations are not observed in numerics and will be the subject of future study. We optimistically believe that RMHD is valid for large-scale solar wind fluctuations. However, the disagreement between simulations and the measurements could also be due to the solar wind being inhomogeneous, expanding and accelerating \cite{Tenerani2017}, anisotropic with respect to the sunward direction \cite{Grappin1996}, and having the large number of discontinuities \cite{Borovsky2010}. 

\subsection{Numerics: parallel spectrum}

\begin{figure}[t]
\begin{center}
\includegraphics[width=0.7\columnwidth]{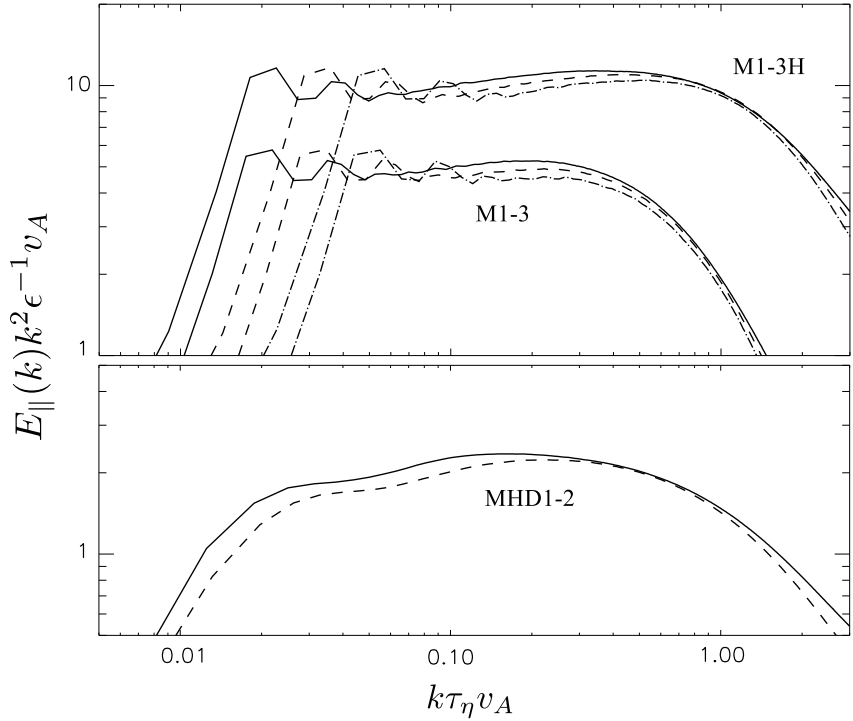}
\end{center}

\caption{Energy spectrum along the magnetic field line $E(k_\|)$ compensated by the theoretical scaling $\epsilon k_\|^{-2}$ (\ref{par_spec}).
Upper plot: Solid, dashed and dash-dotted are spectra from $4096^3$,  $2048^3$ and $1024^3$ RMHD simulations.
The M1-3H has been multiplied by a factor of two to separate the curves.
Lower plot: dashed and solid are MHD1 and MHD2. From \cite{B15b}}
\label{par_sp}
\end{figure}

We plotted the parallel spectrum $E(k_\|)$ vs dimensionless wavenumber $k v_A \tau_\eta$,
compensated by $k^2\epsilon^{-1} v_A$ to see how the scaling is consistent with (\ref{par_spec}). This measurement is presented on Fig.~\ref{par_sp}. For the RMHD case the spectra collapsed, meaning the overall scaling of $k^{-2}$.

Reduced MHD can be performed with different the mean field strength, which in practice requires a particular choice of $\epsilon$ to generate strong turbulence from the outer scale.
The Alfven symmetry of numerical RMHD formulation ensures that $E(k_\|)$ scale precisely linearly $\epsilon$. However, statistically isotropic MHD simulations with $B_0=0$ MHD1-2 do not have this symmetry, and the inertial range scaling (\ref{par_spec}) cannot be rigorously argued based on units. Our test of Eq.~(\ref{par_spec}) is the test not only of the Lagrangian spectrum idea but also the Kraichnan hypothesis of dominant local $v_A$. We substituted the RMS field instead of $v_A$ in Eq.~(\ref{par_spec}). Fig.~\ref{par_sp} demonstrates that there's convergence
to $\epsilon k^{-2}$ in this zero mean field case as well.

\begin{figure}[t]
\begin{center}
\includegraphics[width=0.8\columnwidth]{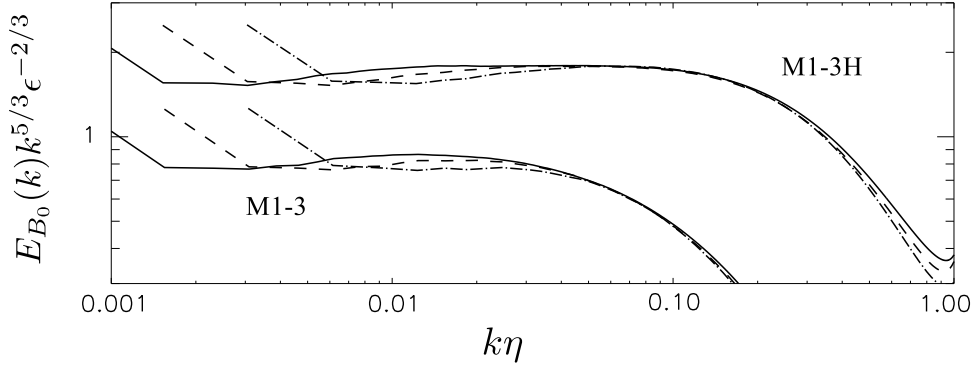}\label{b0spec:fig}
\end{center}

\caption{The spectra along the global mean field in M1-3, M1-3H. The M1-3H spectra have been shifted by a factor of two. The energy spectrum scales as $k^{-5/3}$, i.e., in the same way as the perpendicular scaling. From \cite{B15b}.}
\label{b0_sp}
\end{figure}

Another spectral measurement is along the direction of the global mean field in M1-3, M1-3H.
We expect these scalings to be the same as the perpendicular scalings, i.e., Kolmogorov
because while Alfv\'en waves propagate along the local field direction which
deviates by an angle of $\delta B_L/B_0$ from ${\bf B_0}$, the angular anisotropy in this frame is $\delta B_l/B_0$, with inertial range values of $\delta B_l$ much smaller than the outer scale value of $\delta B_L$. It follows that the anisotropy will be washed
out. \label{wander:argument}
Fig.~\ref{b0_sp} presents a measurement
of the spectrum along the global mean field direction, which is
consistent with $-5/3$.

Worth noting that the application of the critical balance fails in the imbalanced turbulence (more on this in Section \ref{imbal}). A more rigorous Lagrangian argument does not have this problem. We can imagine that the energy cascade is manifested both in space and time domains, with the parallel direction is equivalent to the time domain, the anisotropy relation $k_\| \sim k_\perp^{2/3}$ being the correspondence between space domain (Eulerian) and frequency domain (Lagrangian) spectra. Observational data from the solar wind points to the $k^{-2}$ parallel spectrum, e.g. \cite{horbury2008}. Numerical studies overwhelmingly
support $k^{-2}$, as long as the measurements was along the local field direction see, e.g., \cite{Cho2000,Maron2001,BL09a,BL09b,B12b}, while the measurements in the global frame usually
demonstrate scale-independent anisotropy, see, e.g. \cite{Grappin2010}.

\subsection{Numerics: anisotropy}

\begin{figure}[t]
\begin{center}
\includegraphics[width=1.0\columnwidth]{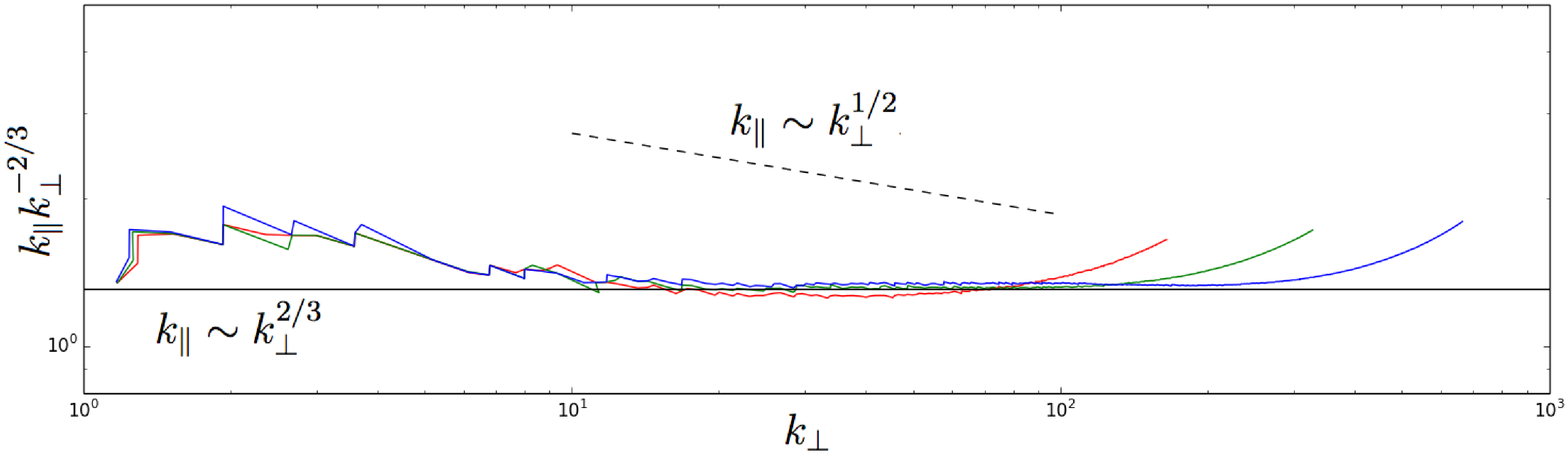}
\includegraphics[width=1.0\columnwidth]{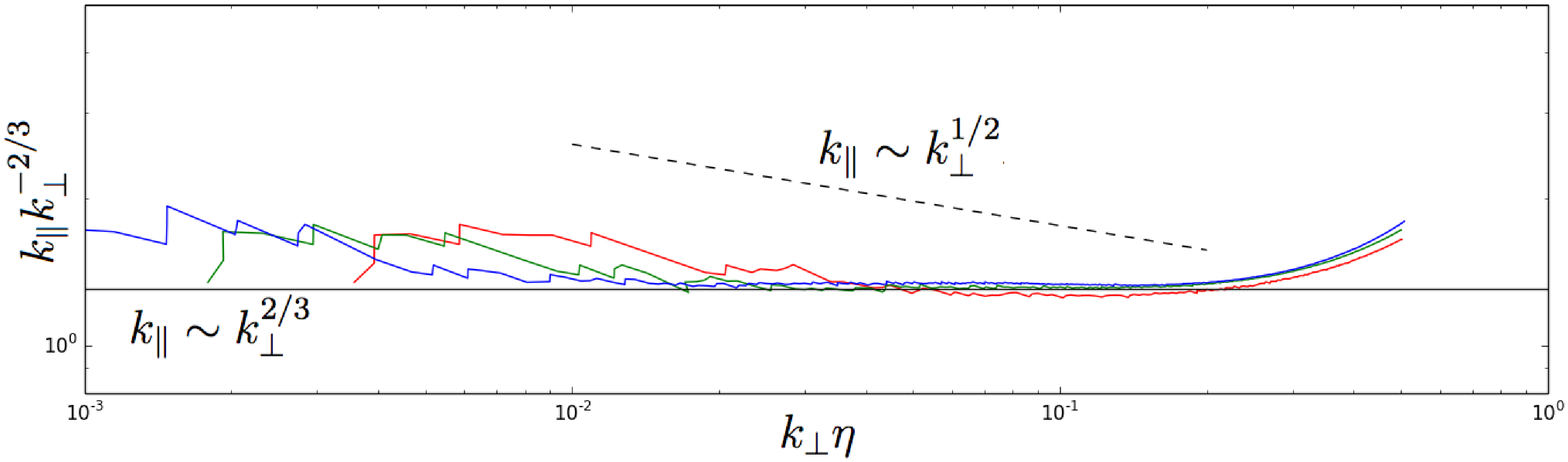}
\end{center}
\caption{The relation between parallel and perpendicular wavenumber in Alfv\'enic turbulence simulations M1-3H. The $k_\|$ is compensated by $k_\perp^{2/3}$, which is expected from theory and
represented by the solid line. The dashed line corresponds to the theory with 3/2 spectral scaling.
Top: x-axis is a dimensional wavenumber, bottom: x-axis is a dimensionless $k_\perp\eta$, so this plot corresponds to scaling study for anisotropy.}
\label{spec_anisotropy}
\end{figure}

\begin{figure}[t]
\begin{center}
\includegraphics[width=0.8\columnwidth]{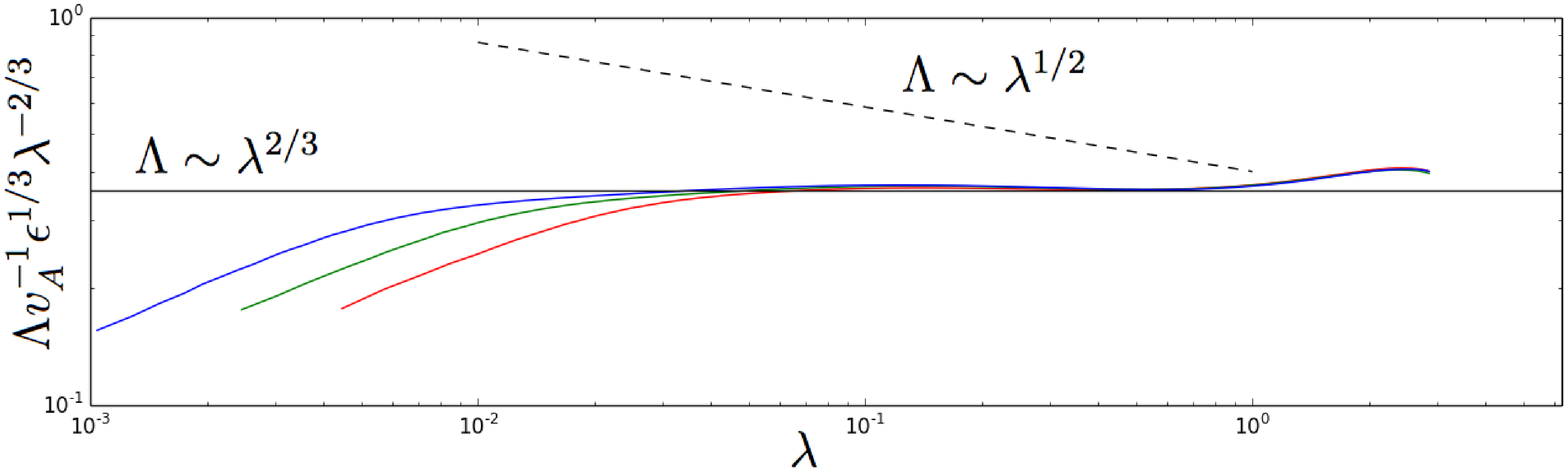}
\includegraphics[width=0.8\columnwidth]{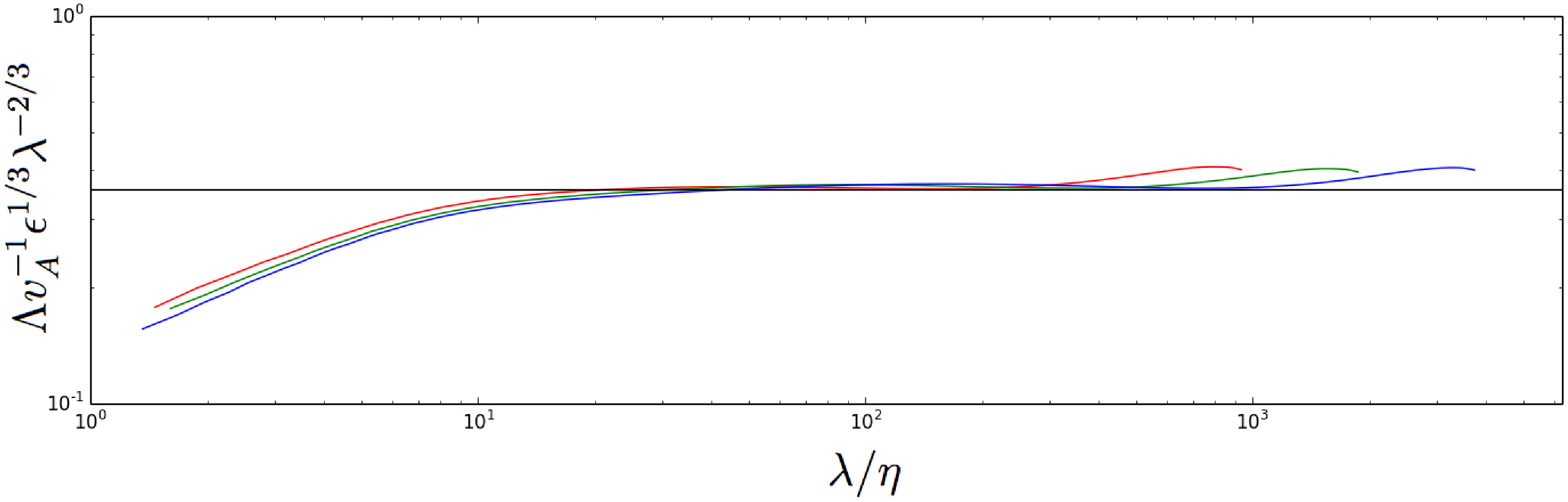}
\includegraphics[width=0.8\columnwidth]{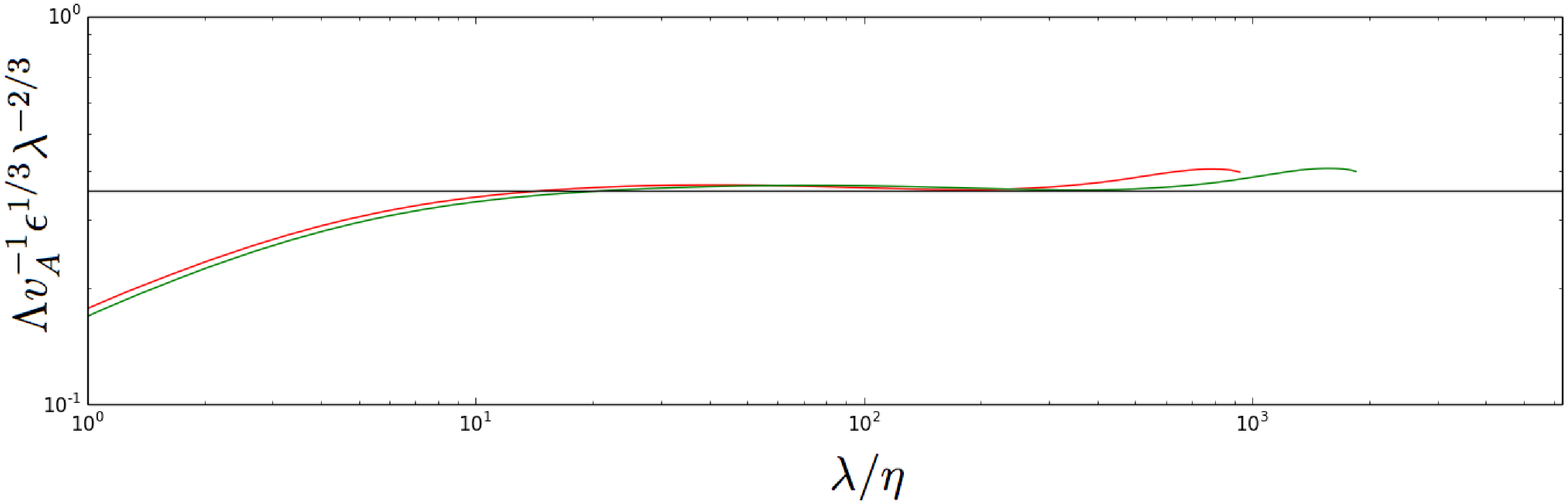}
\end{center}
\caption{Anisotropy in real space, the relation between perpendicular scale $\lambda$
and parallel scale $\Lambda$, compensated in this plot by the prefactor $v_A \lambda^{-2/3} \epsilon^{1/3}$, making it dimensionless. On the top plot we use $\lambda$ on x-axis and plot
results from M1-3H. On the middle we use dimensionless $\lambda/\eta$ on x-axis and again plot
results from M1-3H. On the bottom we use results from M1-3.}
\label{anisotropy}
\end{figure}

We alluded above that the anisotropy should be universal in the inertial
range due to the relation between Lagrangian and Eulerian spectra.
We expect the relation between parallel and perpendicular
scales to follow Eq.~\ref{anis}. Both Alfv\'enic and slow modes
are expected to have the same anisotropy. Similar relation is expected
to hold between parallel and perpendicular wavenumber:
\begin{equation}
k_\|=(2\pi)^{1/3}C_A^{-1} v_A^{-1} k_\perp^{2/3} \epsilon^{1/3}\label{anis_k},
\end{equation}

On Fig.~\ref{spec_anisotropy} we plotted wavevector anisotropy. We determine anisotropy
relation by solving the equation for $k_\|$, given a range of $k_\perp$:
\begin{equation}
E_\|(k_\|)= E(k_\perp).
\end{equation}
A similar procedure is done with parallel and perpendicular second order structure functions to obtain the relation between $\Lambda$ and $\lambda$ of Fig.~\ref{anisotropy}. We also did a convergence
study in the same spirit that was done for spectra in previous sections, which are on the bottom parts of each two figures above. We see that the theoretical scalings are followed fairly well.
To summarize the above three sections, we showed that the spectrum and anisotropy
of Alfvenic turbulence follows the two relations below:

In this Section, we argued that the properties of Alfv\'en and slow components of incompressible MHD turbulence in the inertial range would be determined only by the Alfv\'en speed $v_A$, dissipation rate $\epsilon$ and the scale of interest $\lambda$. The energy spectrum and anisotropy of Alfv\'en mode will be expressed as
\begin{eqnarray}
E(k)=C_K \epsilon^{2/3} k^{-5/3},\\
\Lambda/\lambda=C_A v_A (\lambda \epsilon)^{-1/3}.
\end{eqnarray}
Also, we found numerically that the ratio of kinetic to magnetic energies in the inertial range is constant, $r_A=E_v/E_B \approx 0.74$.

\subsection{Dynamic alignment models}
\label{align}

\begin{figure}[t]
\begin{center}
\includegraphics[width=0.9\columnwidth]{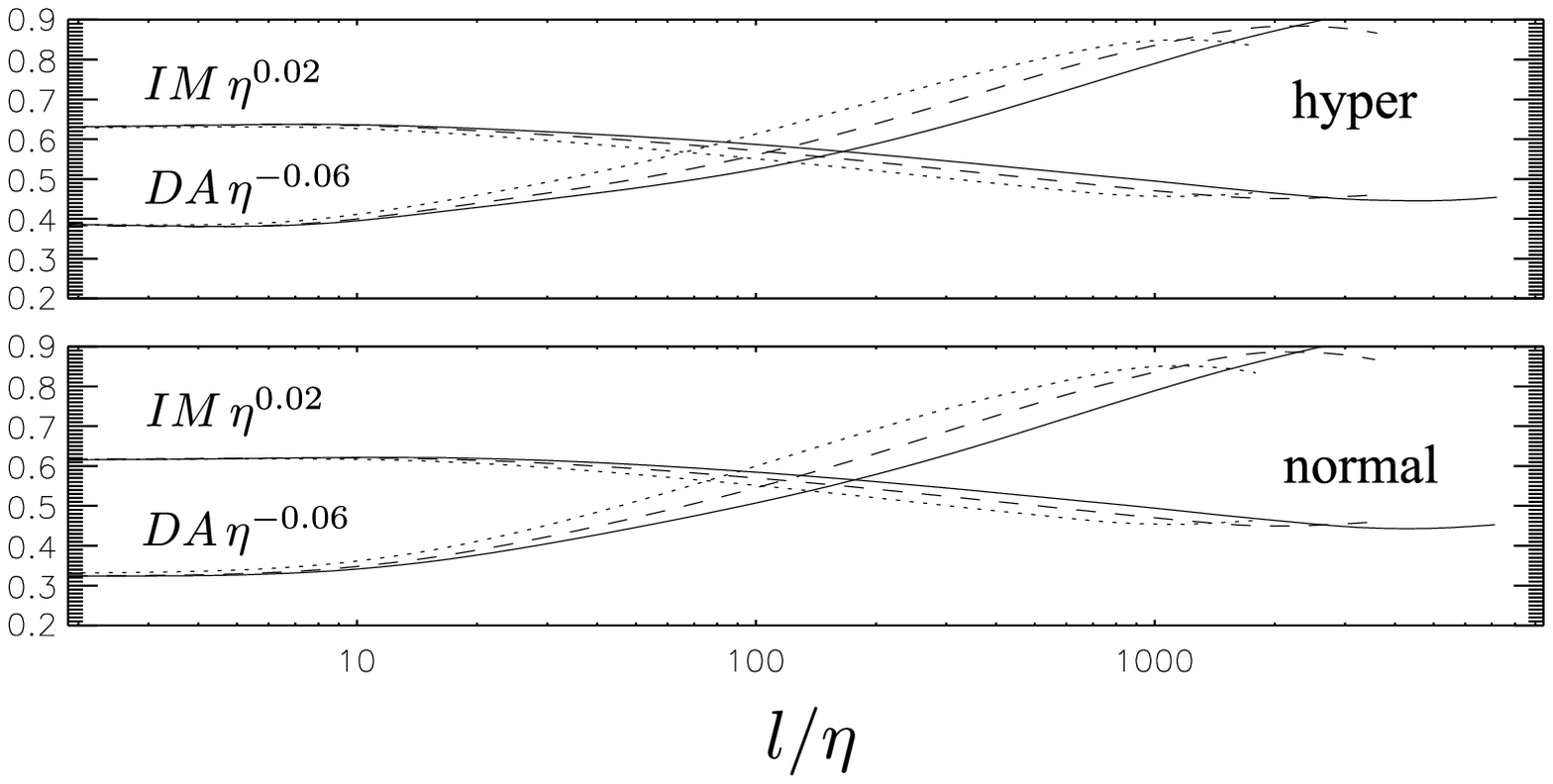}
\end{center}
\caption{Scaling study of alignment measures ${\rm DA}=\langle
|\delta {\bf v} {\bf \times} \delta {\bf b} |\rangle /\langle |\delta v \delta
b|\rangle$ and ${\rm IM}=\langle |\delta (w^+)^2- \delta (w^-)^2|\rangle /\langle \delta
(w^+)^2+ \delta (w^-)^2\rangle$ from M1-3H (top) and M1-3 (bottom). The alignment slopes converge to relatively small values,
e.g., 0.06 for ${\rm DA}$ which is smaller than 1/4, predicted in alignment theories.
From \cite{B14a}, see also \cite{BL09b,B11,B12b}.}
\label{align_conv}
\end{figure}
MHD has more degrees of freedom than hydro, which results, in first-order measures, in two independent dimensionless quantities (four degrees of freedom of $w^\pm_\perp$ minus rotational freedom minus normalization). One example of this is the fraction of residual energy, introduced earlier. These dimensionless quantities may, in principle, have a non-trivial scaling in the inertial range. 

After spectral scaling of $k^{-3/2}$ has been found in \cite{muller2005}, a number of models have been proposed suggesting that strong turbulence phenomenology have to be modified along the lines of Sec.~\ref{kolm_gen_sec} to become consistent with this scaling. Among these models are \cite{boldyrev2005} and \cite{gogoberidze2007}. A sizable confusion ensued, however to which alignment measure represent the scale-dependent weakening of interaction
more accurately. The original \cite{boldyrev2005} idea was analyzed  
in \cite{BL06} and no significant alignment was found for the averaged
angle between ${\bf w}^+$ and ${\bf w}^-$,
${\rm AA}=\langle|\delta {\bf w}^+_\lambda\times \delta {\bf w}^-_\lambda|/|\delta
{\bf w}^+_\lambda||\delta {\bf w}^-_\lambda|\rangle$, but when this angle
was weighted with the amplitude ${\rm PI}=\langle|\delta {\bf
  w}^+_\lambda\times \delta {\bf w}^-_\lambda|\rangle/\langle|\delta {\bf
  w}^+_\lambda||\delta {\bf w}^-_\lambda|\rangle$, some scale-dependent alignment was found.
Later \cite{boldyrev2006} proposed the alignment between ${\bf v}$
and ${\bf b}$ and subsequently \cite{mason2006} suggested a particular
amplitude-weighted measure, ${\rm DA}=\langle|\delta {\bf
  v}_\lambda\times \delta {\bf b}_\lambda|\rangle/\langle|\delta {\bf
  v}_\lambda||\delta {\bf b}_\lambda|\rangle$, that was claimed to
  depend of perpendicular scale as $\lambda^{1/4}$. 
  In a sense, ${\rm DA}$ is very similar to ${\rm PI}$ but uses ${\bf B}$ and ${\bf v}$ instead of ${\bf w}^\pm$. The measure for local imbalance was introduced in \cite{BL09b} as 
${\rm IM}=\langle |\delta (w^+_\lambda)^2- \delta (w^-_\lambda)^2|\rangle /\langle \delta
(w^+_\lambda)^2+ \delta (w^-_\lambda)^2\rangle$. While reevaluating the logic in \cite{boldyrev2006}
it becomes clear that the choice of ${\rm DA}$ as a measure exclusively responsible for the interaction weakening is arbitrary, at the same time the argument that ${\rm DA}$ scales as $l^{1/4}$ due
to field line wandering, is invalid keeping in mind the Alfven symmetry of RMHD equations.
The argument in \cite{boldyrev2006} suggests that ${\rm DA}$ will tend to increase, but
will be bounded by field wandering, i.e., the alignment on each scale will be
created independently of other scales and will be proportional to the relative
perturbation amplitude $\delta B/B_0$. However, this violates Alfv\'en
symmetry of RMHD equations (see Section~\ref{sec:rmhd}), which requires that
$B_0$ can be factored out of the dynamics and appear only in combination with $k_\|$.
The most apparent contradiction we find while following \cite{boldyrev2006} is that
a perfectly aligned state, e.g., with $\delta {\bf w}^-=0$ is a precise solution of
MHD equations and it is not destroyed by its field wandering. Empirically
we know that alignment measures showed very little or no dependence on $\delta B_L/B_0$ (see, e.g., \cite{BL09b}).

Fig.~\ref{align_conv} studies scale-dependency of ${\rm DA}$ and ${\rm IM}$ by the method of scaling study.
The asymptotic scale-dependency slope for ${\rm DA}$ for our data is found around $0.06$, which is way below its value of $1/4$ suggested in \cite{boldyrev2006}. From this figure it is evident that the scale-dependency of ${\rm DA}$ seems is tied to the outer scale, i.e., it is non-universal.

\section{Imbalanced MHD turbulence}
\label{imbal}
\begin{figure}[t]
\begin{center}
\includegraphics[width=0.5\columnwidth]{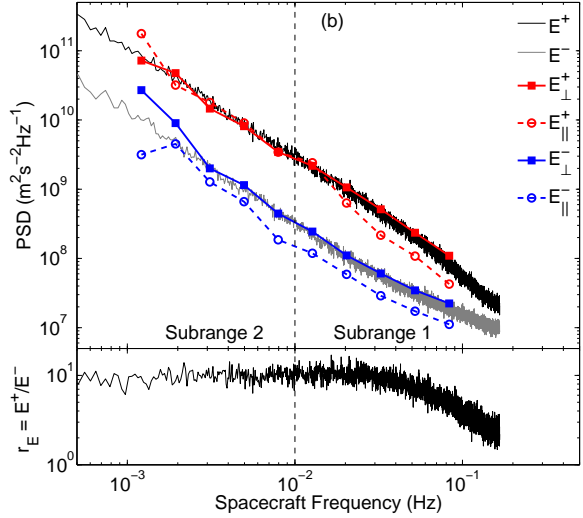}
\end{center}
\caption{Spectra of imbalanced turbulence measured in the solar wind.
Trace of the Fourier and wavelet power spectra of $(w^+)^2$ (black line and red symbols)
and $(w^-)^2$ (gray line and blue symbols) parallel and perpendicular
to the local magnetic field. The bottom panel shows the ratio $(w^+/w^-)^2$. From \cite{wicks2011}.}
\label{wicks2011_fig}
\end{figure}

While hydrodynamic turbulence have only one energy cascade, the
incompressible MHD turbulence has two, due to the exact conservation of
the Els\"asser (oppositely going wave packets') ``energies''.
The situation of zero total cross-helicity, which we considered in previous sections
has been called ``balanced'' turbulence as the amount of oppositely moving
wavepackets balance each other, the alternative being ``imbalanced''
turbulence. Imbalanced turbulence, however, is very common,
it is enough to have a mean magnetic field (like in the ISM) and
a localized source of perturbations. Another example
is the solar wind, where the dominant perturbation component propagate
away from the Sun, see Fig.~\ref{wicks2011_fig}. Likewise, we expect
similar phenomena in active galactic nuclei (AGN), where the jet
has a mean magnetic field component and the perturbations
will propagate primarily away from the central engine.

\subsection{Theoretical considerations}
A conceptual difficulty in the imbalanced case arises from the application of the critical
balance idea. If $\delta w^+_l$ critical balance depends on $\delta  w^-_l$ amplitude,
their parallel scales and frequencies are mismatched, and the cascade cannot proceed
in a normal manner. Below we describe three models that tried to deal with this issue,
references \cite{LGS07,BL08,PB09}, Lithwick, Goldreich \& Sridhar (2007) that
we designate LGS07, Beresnyak \& Lazarian (2008) that we designate BL08 and 
Perez \& Boldyrev (2009) model that we designate PB09. In short, only BL08 model
is consistent with all numerical evidence, taking into account cascading rates, spectra and anisotropy. Below we shortly describe these theories.
\begin{figure}[t]
\begin{center}
\includegraphics[width=0.6\textwidth]{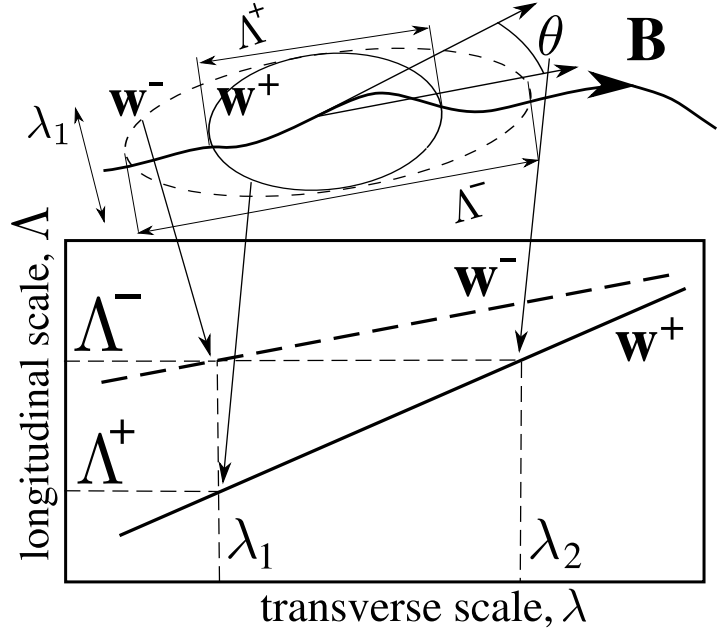}
\end{center}
\caption{Upper: a ${\bf w}^+$ wavepacket, produced
by cascading by ${\bf w}^-$ wavepacket is aligned with respect
to ${\bf w}^-$ wavepacket, but misaligned with respect
to the local mean field on scale $\lambda_1$, by the angle $\theta$.
Lower: the longitudinal scale $\L$ of the wavepackets,
as a function of their transverse scale, $\l$; $\L^+$, $\L^-$, $\l_1$, $\l_2$
are the notations used in this review. From \cite{BL08}.}
\label{anis_cartoon}
\end{figure}

PB09 employs dynamic alignment which depends on the scale as $l^{1/4}$, this alignment,
however, is acting differently on $w^+$ and $w^-$ so that it effectively results
in the same nonlinear timescales for both components. It could be rephrased that
PB09 predicts turbulent viscosity on each scale which is equal for both components.
this results in an expression for the ratios of energies 
\begin{equation}
(w^+)^2/(w^-)^2=\epsilon^+/\epsilon^- \ \ \ {\rm for (PB09)}.
\end{equation}
It is not clear how this is consistent with the limit of large imbalances, where the weak
component will not be able to produce any sizable interaction.

LGS07 the authors proposed that the parallel scale for both components is determined by
the shear rate of the stronger component, despite the cascading timescale is different
for both components. The energy cascade is still a strong cascade:
\begin{equation}
\epsilon^\mp=\frac{(w^\mp(\l))^2 w^\pm(\l)}{\l} \ \ \ {\rm for (LGS07)}.
\end{equation}
This results in the prediction 
\begin{equation}
w^+/w^-=\epsilon^+/\epsilon^- \ \ \  {\rm for (LGS07)}.
\end{equation}
Both PB09 and LGS07 predict the same anisotropy for both components.

In BL08 the authors proposed a new formulation of critical balance
for the stronger component. This model is described in greater detail below.
BL08 relaxes the assumption of local cascading for the strong component $w^+$,
while saying the $w^-$ classic critical balance and local cascading. In BL08 picture
the waves have different anisotropies (see Fig.~\ref{anis_cartoon}) and the $w^+$ wave
have smaller anisotropy than $w^-$, which is opposite to what
a naive application of critical balance would predict.
The anisotropies of the waves are determined by
\begin{equation}
w^+(\l_1)\L^-(\l_1)=v_A\l_1,
\end{equation}
\begin{equation}
w^+(\l_2)\L^+(\l^*)=v_A\l_1,
\end{equation}
where $\l^*=\sqrt{\l_1\l_2}$, and the energy cascading
is determined by weak cascading of the dominant wave
and strong cascading of the subdominant wave:
\begin{equation}
\epsilon^+=\frac{(w^+(\l_2))^2 w^-(\l_1)}{\l_1}\cdot\frac{w^-(\l_1) \L^-(\l_1)}{v_A\l_1}\cdot f(\l_1/\l_2),
\end{equation}
\begin{equation}
\epsilon^-=\frac{(w^-(\l_1))^2 w^+(\l_1)}{\l_1}.
\end{equation}

\begin{figure}[t]
\begin{center}
\includegraphics[width=0.85\columnwidth]{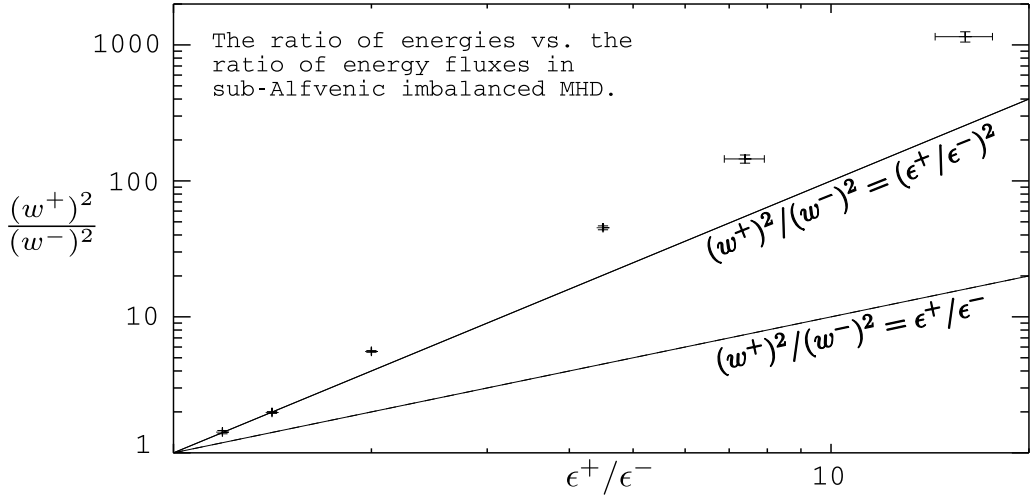}
\end{center}
\caption{Elsasser energy ratio plotted versus dissipation rate ratio in simulations
I2, I4, I5, I6 (Table~\ref{imb_experiments}) and A7 and A5 from \cite{BL09a}.
The solid line is the LGS07 prediction, and the dashed line is a PB09 prediction, also a
prediction for purely viscous/non-turbulent dissipation of eddies.}
\label{dissip}
\end{figure}

BL08 model, unlike LGS07, does not produce self-similar (power-law) solutions when turbulence is driven with the same anisotropy for $w^+$ and $w^-$ on the outer scale. BL08, however, claims that, on sufficiently
small scales, the initial non-power-law solution will transit into
asymptotic power law solution that has
$\Lambda^-_0/\Lambda^+_0=\epsilon^+/\epsilon^-$
and $\lambda_2/\lambda_1=(\epsilon^+/\epsilon^-)^{3/2}$.
The larger imbalance will require larger transition to this asymptotic regime.

\subsection{Numerics}

Table~\ref{imb_experiments} summarizes RMHD simulations with imbalanced driving.
In these simulations, we kept the energy injection rate constant.
All experiments were evolved into the stationary state.
The imbalanced runs have to be were evolved for a longer time to achieve
stationary state due to longer cascading timescales
for the stronger component.

\begin{table}[t]
\caption{Three-Dimensional RMHD Imbalanced Simulations}
\begin{tabular*}{0.99\columnwidth}{@{\extracolsep{\fill}}l c c c c r}
    \hline\hline
Run  & Resolution & $f$ & Dissipation & $\epsilon^+/\epsilon^-$ & $(w^+)^2/(w^-)^2$   \\
   \hline
I1 &  $512\cdot 1024^2$ & $w^\pm$ & $-1.9\cdot10^{-4}k^2$ & 1.187 &  $1.35\pm 0.04$ \\
I2 &  $768^3$ & $w^\pm$ & $-6.8\cdot10^{-14}k^6$ & 1.187 &  $1.42 \pm 0.04$   \\
I3 &  $512\cdot 1024^2$ &$w^\pm$ & $-1.9\cdot10^{-4}k^2$ & 1.412 &  $1.88\pm 0.04$   \\
I4 &  $768^3$ & $w^\pm$ & $-6.8\cdot10^{-14}k^6$ &  1.412 & $1.98\pm 0.03$ \\
I5 &  $1024\cdot 1536^2$ & $w^\pm$ & $-1.5\cdot10^{-15}k^6$ &   2   &    $5.57\pm 0.08$  \\
I6 &  $1024\cdot 1536^2$ & $w^\pm$ & $-1.5\cdot10^{-15}k^6$ & 4.5   &  $45.2\pm1.5$   \\
   \hline
\end{tabular*}
  \label{imb_experiments}
\end{table}

Compared to spectral slopes, dissipation rates are robust quantities that require
much smaller dynamical range and resolution to converge. Fig.~\ref{dissip} shows
energy imbalance $(w^+)^2/(w^-)^2$ versus dissipation rate imbalance $\epsilon^+/\epsilon^-$ for simulations I2, I4, I5 and I6. We also use two data points from
earlier simulations with large imbalances, A7 and A5 from \cite{BL09a}.
I1 and I3 are simulations with the normal viscosity similar to I2 and I4.
They show slightly less energy imbalances than I2 and I4.
We see that most data points are above the prediction of LGS07, which is consistent with BL08.
In other words, numerics strongly suggest that 
\begin{equation}
\frac{(w^+)^2}{(w^-)^2}\geq \left(\frac{\epsilon^+}{\epsilon^-}\right)^2.
\label{imbal_ratios}
\end{equation} 
Although there is a tentative correspondence between LGS07 and the data for small degrees of imbalance, the
deviations for large imbalances are significant. As to PB09 prediction, it is inconsistent with data for all degrees of imbalance including those with small imbalances.

The approximate equality in Eq.~\ref{imbal_ratios} for very small imbalances, however,
is an excellent test of the expression in Eq.~\ref{fluxes} that we assumed in the balanced case.
So in some sense, empirical study of the imbalanced case validated the theory of the balanced case as well.

\begin{figure}[t]
\begin{center}
\includegraphics[width=\columnwidth]{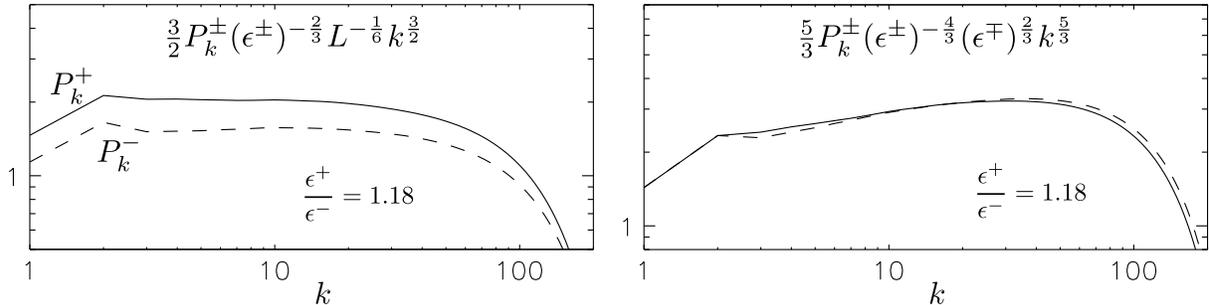}
\end{center}
\caption{Energy spectra for $w^+$ (solid) and $w^-$ (dashed) from simulation I2, compensated by
factors that correspond to PB09 (left) and LGS07 (right). Spectra collapse on the right, but not on the left.}
\label{imb2models}
\end{figure}

Fig.~\ref{imb2models} shows spectra from low-imbalance simulation I2,
compensated by the predictions of PB09 and LGS07.
We see that the collapse of two curves for $w^+$ and $w^-$ is much better for the LGS07 model.

\begin{figure}[t]
\includegraphics[width=0.9\columnwidth]{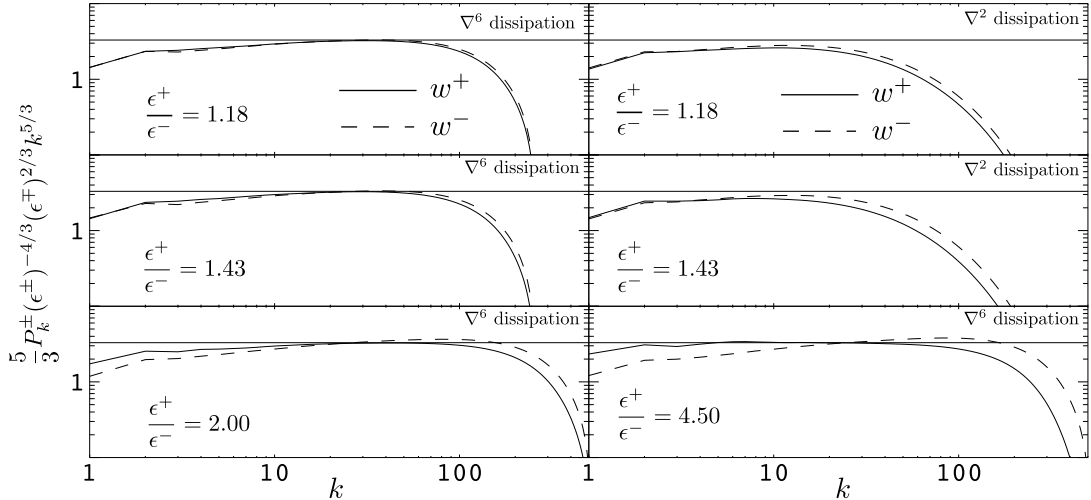}
\caption{Energy spectra for $w^+$ (solid) and $w^-$ (dashed) for simulations I1-I6,
compensated by factors that correspond to LGS07. The thin solid line corresponds to Kolmogorov
constant for Alfv\'enic turbulence $C_{KA}=3.27$.}
\label{imb_spectra}
\end{figure}

Fig.~\ref{imb_spectra} shows spectra from all I1-6 simulations, compensated by the prediction of LGS07.
For lower imbalances, the collapse is reasonably good and become progressively worse for
larger imbalances. This deviation, however, does not entirely follow the prediction of the
asymptotic power-law solutions from BL08, which will predict that the solid curve will
go above $C_{KA}$ and the dashed curve -- below it. This is possibly explained by the fact
that asymptotic power-law solutions were not reached in these limited resolution
experiments, this is also observed for anisotropies.

We measured parallel and perpendicular structure functions in simulations I1-I6, in order to quantify the anisotropies of eddies. 

\begin{figure}[t]
\begin{center}
\includegraphics[width=0.9\columnwidth]{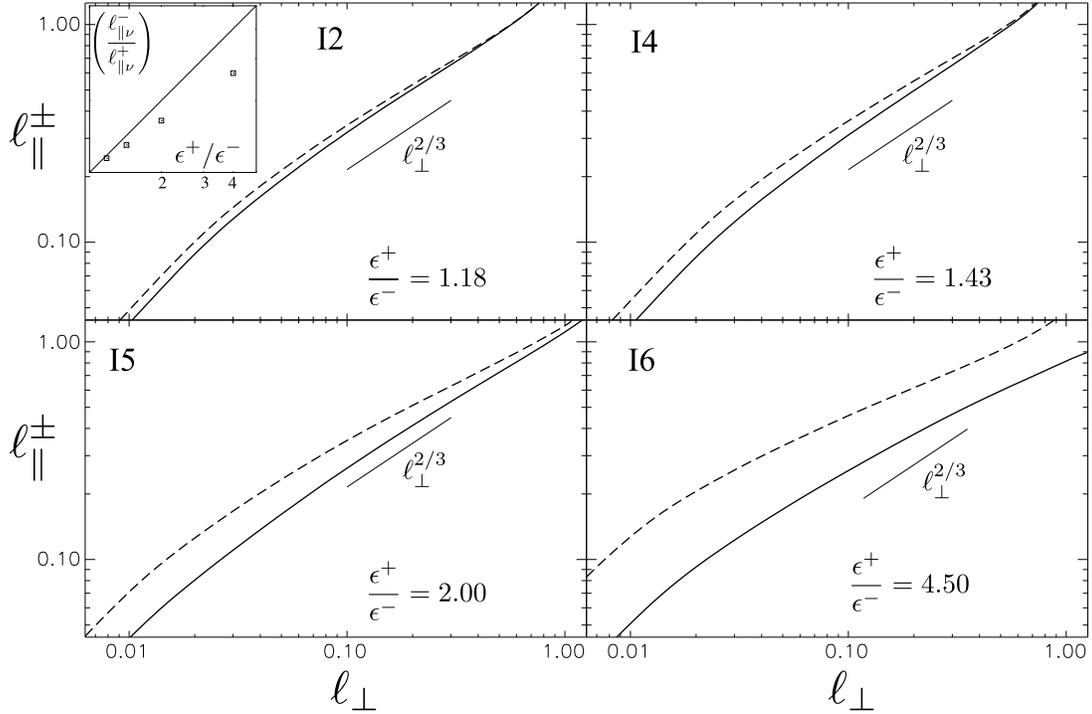}
\end{center}
\caption{Anisotropies for $w^+$ (solid) and $w^-$ (dashed), simulations I1-I6. 
The small upper inset shows the ratio of anisotropies on the smallest scales
vs. the prediction of BL08, $\epsilon^+/\epsilon^-$.}
\label{imb_anis}
\end{figure}

Fig.~\ref{imb_anis} shows anisotropies for I1-6 simulations.
All simulations were driven by the same anisotropies on the outer scale, which is unfavorable
for obtaining the asymptotic power law solutions of BL08. 
It is, however, favorable to the LGS07 model, which predicts the same $w^+$ and $w^-$ anisotropies for all scales. Therefore, these
simulations are a sensitive test between LGS07 and BL08 models, both of which are roughly consistent in terms of energy ratios and spectra for small imbalances. If the LGS07 model is correct, we would observe the same anisotropy on all scales, but this is not what is observed in Fig.~\ref{imb_anis}, where anisotropies start to diverge on smaller scales.
The ratio of anisotropies is roughly consistent with the BL08 prediction of $\epsilon^+/\epsilon^-$ for small imbalances and somewhat smaller for larger imbalances.

\section{MHD dynamo}
\label{dynamo}
One of the main questions of MHD dynamics is how conductive fluid generates its magnetic field, a process known broadly as ``dynamo''. Turbulent dynamo is known as ``large-scale/mean-field dynamo'' and ``small-scale/fluctuation dynamo'', in the first case magnetic fields are amplified on scales larger than the outer scale of turbulence in the seconds on smaller scales.
An example of the flow generating no dynamo is an axisymmetric situation \cite{Cowling1933},
the natural turbulence, however, possesses no exact symmetries and is expected to amplify
magnetic field by stretching, due to the particle separation in a turbulent flow.
For the large-scale dynamo, a ``twist-stretch-fold'' mechanism was
introduced in \cite{vainshtein1972}. 

If the turbulent flow possess \emph{statistical isotropy}, 
it can not generate a large-scale field, i.e., the field with scales larger than the outer scale of turbulence.
To generate the observed large-scale fields, such as fields in the disk galaxies, 
large-scale asymmetries of the system must break the statistical symmetry of turbulence.
Further analysis of the induction equation shows that the rotation (described by the pseudovector
of angular velocity) is insufficient to provide large-scale dynamo, stratification or shear
should provide a real vector, so that the pseudoscalar alpha-effect result in a large-scale turbulent EMF \cite{Vishniac2001,kapyla2009}.

One approach to large-scale dynamo is mean field theory, see, e.g.\cite{krause1980}, where
the magnetic and velocity fields are decomposed into the mean and fluctuating part. The equations for the mean field are closed using statistical or volume averaging over the fluctuating part. 
The traditional theories of mean field dynamo, however, often fail
due to issues related to magnetic helicity \cite{Vishniac2001,Vishniac2003}.

The studied of the large-scale dynamos is big and complex science due to the variety of ways, large-scale symmetries are broken in different astrophysical objects.
In this review, we propose the reader follow other reviews focused on large-scale dynamos,
e.g., \cite{brandenburg2005,charb2010} and instead concentrate on the small-scale or fluctuating dynamo as more universal, generic and crucial to understand the overall level of magnetization
in astrophysical environments. We also emphasize that proper understanding of turbulent
magnetic fields is crucial as it is subsequently slowly ordered and made large-scale by the large-scale dynamo process.

\subsection{Kinematic dynamo}
\label{kinematic}
If the magnetic energy is less than the kinetic energy of turbulent motions, the turbulence may generate the magnetic field, which is referred to as ``turbulent dynamo''. There are two distinct regimes: 1) the magnetic energy is much less than the kinetic energy of driving eddies at all scales down to the dissipation scale and 2) the kinetic and magnetic energies come to the equipartition at some scale. The first regime is called ``kinematic'' or ``linear'' dynamo, referring to induction equation (\ref{mhd0}) being linear with respect to the magnetic field.
If the magnetic field is so weak that it provides no back-reaction to velocity, the problem 
reduces to studying the solutions of the induction equation with a prescribed velocity field.
This kinematic regime was, historically, the most studied (see Kazantsev model outlined in \cite{kazantsev1968,kraichnan1967,kulsrud1992,vincenzi2002}). For the Kolmogorov-type turbulence, the fastest magnetic field amplification comes from the fastest turbulent eddies, i.e., the eddies at the dissipation scales. It would follow that the growth rate $\gamma=1/\tau_\eta$ (see Eq.~\ref{tau_eta}). The spectrum of the magnetic field will be $E(k)\sim k^{3/2}$ rising sharply to the magnetic dissipation scale. Kazantzev picture was not without drawbacks since it had relied on an artificial delta-correlated velocity field instead of a realistic turbulent velocity field.
This resulted in an overestimated $\gamma$ compared to reality. Precise numerical experiments with ${\rm Pr}_m={\rm Re}_m/{\rm Re}=1$ have found a prefactor $\gamma\tau_\eta=0.0326$ \cite{haugen2004,Schekochihin2004,B12a} which
is much smaller than unity. This small number, as had been suggested in \cite{B12a} is due to
turbulence being time-asymmetric. 

From kinematic models, it is not clear whether magnetic energy will continue to grow
after the end of the kinematic regime, however, keeping in mind $\gamma\sim {\rm Re}^{1/2}$ and very large astrophysical ${\rm Re}$, the kinematic growth is incredibly fast and occupies a tiny fraction
of the dynamical time of the system. For example, while the galaxy clusters form on timescales
of 15 billion years, the characteristic growth time of kinematic dynamo is less than
a million years \cite{BM16}, so we do not expect kinematic dynamo to operate in present time. 
The magnetic spectrum of the kinematic dynamo, with its positive spectral index,
is incompatible with observations in galaxy clusters \cite{Laing2008}, see Fig.~\ref{feretti}.
These observations indicate steep spectrum with negative power index at small scales.

Summarizing, the kinematic dynamo is inapplicable in most astrophysical environments, since the observed magnetization corresponds to Alfv\'en speed which is many orders of magnitude higher
than the Kolmogorov velocity $\epsilon^{1/4}\nu^{1/4}$.

\subsection{Nonlinear dynamo}
\label{dynamo:sec}
Before numerical simulations were commonplace, it was proposed that after saturation of kinematic dynamo the magnetic energy will stop growing. If we agree to this proposition and assume that the magnetic energy indeed saturates as soon as the dynamo becomes nonlinear, then the saturation level, in this case, will be of order $\rho v_\eta^2/2$, where $v_\eta$ is a Kolmogorov velocity scale. This is a factor of ${\rm Re}^{-1/2}$ smaller than the kinetic energy density and will be completely dynamically unimportant in high-Re astrophysical environments. In fact, observations indicate the opposite -- a sizable energy density of the magnetic field in large-scale systems, see Fig.~\ref{feretti}

An alternative had been proposed in the early work by Schl\"uter and Bierman \cite{schluter1950}, who suggested that dynamo will not stop and will continue to grow, saturating on each subsequent scale after a dynamical time. Recently small-scale dynamo underwent revival due
to the availability of direct numerical simulations. Simulations of the dynamo saturated state produced steep spectra and significant outer-scale fields. The saturated state was only weakly dependent on ${\rm Re}$ and ${\rm Pr}_m$ as long as ${\rm Re}$ was large, see, e.g., \cite{haugen2004}.

\begin{figure}[t]
\begin{center}
\includegraphics[width=0.8\columnwidth]{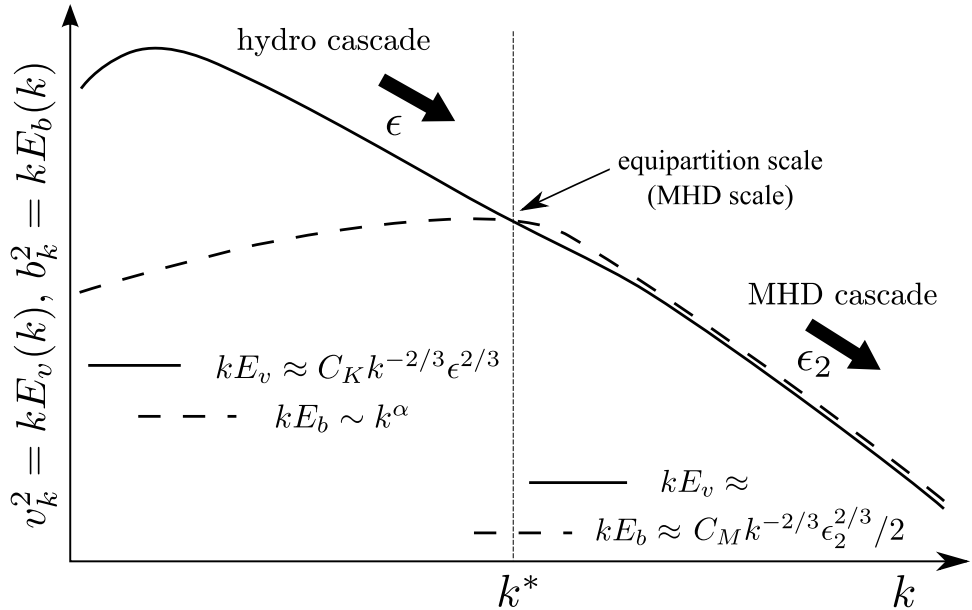}
\end{center}
\caption{A cartoon of kinetic and magnetic spectra in small-scale dynamo, at a particular moment of
time when equipartition wavenumber is $k^*$.}
\label{spect_dyn}
\end{figure}

Apart from the saturated state, the growth stage was suggested to have growth
of magnetic energy which is linear in time \cite{schekochihin2007,CVB09,ryu2008,BJL09,B12a}.
In \cite{B12a} the locality of the dynamo, which is necessary for the linear growth picture was argued analytically. Let us imagine that the magnetic and kinetic spectra at a particular moment of time are similar to what is presented in Fig.~\ref{spect_dyn}.
Magnetic and kinetic spectra cross at some ``equipartition'' scale $1/k^*$,
below which both spectra are steep, typically $k^{-5/3}$ due to the MHD cascade (Sec.~\ref{alfvenic}). A number of arguments suggest this assumption.
Firstly, we expect kinematic dynamo \cite{kulsrud1992} to proceed until the moment when
magnetic spectrum intersect the kinetic spectrum at the viscous scales (assuming 
${\rm Pr}_m=1$), which will correspond to the beginning of the nonlinear regime.
Secondly, this is supported by the wealth of numerics, e.g. \cite{BL09b,CVB09}. Thirdly, this is also somewhat supported by observations of magnetic fields in clusters
\cite{Laing2008}.

At large scales magnetic spectrum is shallow, $k^\alpha,\, \alpha>0$,
while kinetic spectrum is steep due to the hydrodynamic cascade. Most of the magnetic energy is contained at the scale of $1/k^*$. We designate $C_K$ and $C_M$ as Kolmogorov constants
of hydro and MHD respectively. The hydrodynamic cascade rate is $\epsilon$ and the MHD cascade rate as $\epsilon_2$.
Due to the conservation of energy in the inertial range, magnetic energy will grow at a rate of $\epsilon-\epsilon_2$.
We will designate $C_E=(\epsilon-\epsilon_2)/\epsilon$ as an ``efficiency of the small-scale dynamo'' and will argue that this is a true constant, since a) turbulent dynamics is local in scale in the inertial range; b) ideal MHD or Euler equations do not contain any scale explicitly.
Magnetic energy will grow linearly with time if $\epsilon={\rm const}$:

\begin{equation}
\frac{1}{8\pi}\frac{dB^2}{dt}=C_E\epsilon, \label{C_E}
\end{equation}

The equipartition scale $L_B=1/k^*$ will grow with time as $t^{3/2}$ \cite{BJL09}:
\begin{equation}
L_B=c_l v_A^3/\epsilon_{\rm turb}, \label{Lb}
\end{equation}
here we introduced dimensionless constant $c_l$.
Alternatively one can say that small-scale dynamo saturates at several dynamical times at scale $1/k^*$ and proceeds to a twice larger scale \cite{schluter1950,schekochihin2007}. If magnetic
energy grows approximately till equipartition \cite{haugen2004,CVB09}, the whole process
will take around several outer timescales of the system, or more quantitatively, $(C_K^{3/2}/C_E)(L/v_L)$.

It was demonstrated analytically in \cite{B12a} that as long as the kinetic spectrum is steep
(spectral slope between -1 and -3), the magnetic spectrum is steep below the equipartition scale and
magnetic spectrum is shallow (slope higher than 0) above the equipartition scale, the dynamo
is indeed local and the picture described above can indeed be rigorously argued.
So, besides the fact that local interactions dominate the kinetic cascade at large scales and the MHD cascade at small scales, we also know that dynamo is governed by local interactions. Assuming Kolmogorov phenomenology, it is also possible to estimate the upper limit on the wavevector interval
in which nonlinear dynamo operates, namely the interval $k^*[C_E^{3/2}C_K^{-9/4},C_E^{-3/2}C_M^{9/4}]$. Substituting
numerically known values $C_K=1.6$, $C_M=4.2$, $C_E=0.05$ we get the interval of
$k^*[0.004,2000]$, which not very restrictive as a practical upper limit for numerical simulations,
but undoubtedly essential for astrophysical situations where the inertial range is many orders
of magnitude in scale. Given the sufficiently large inertial range, $C_E$ a universal dimensionless constant of MHD dynamics, much like Kolmogorov constant and since it relates energy fluxes, not energies, it is also unaffected by intermittency.
Interestingly, magnetic energy dynamics at $k\ll k^*$ is likely dominated by nonlocal
interactions with $k^*$ but this part of the spectrum contains negligible
magnetic energy and the universality claim is unaffected by this nonlocality.
\begin{figure}[t]
\begin{center}
\includegraphics[width=0.7\columnwidth]{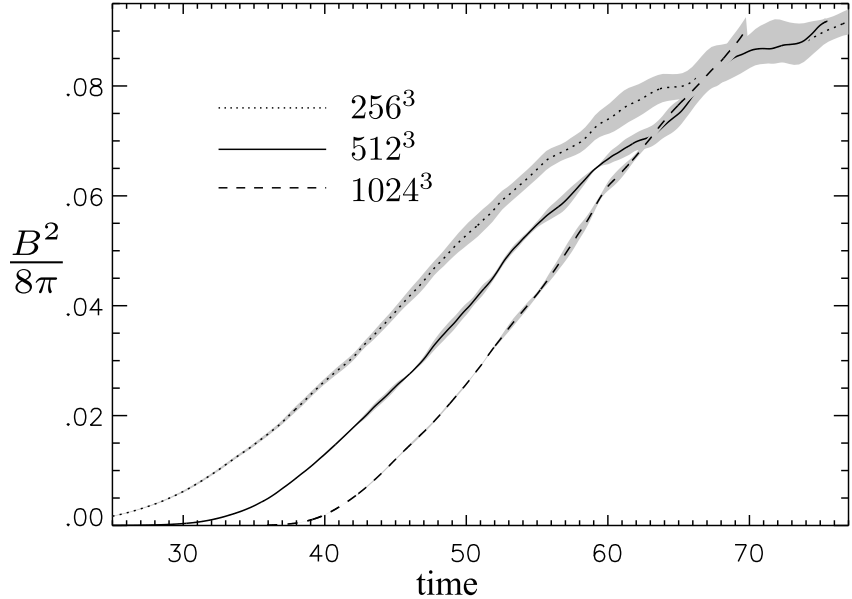}
\end{center}
\caption{Magnetic energy growth vs. time in code units, observed in simulations
with ${\rm Re}_m=1000$ ($\tau_\eta=0.091$ in code units), ${\rm Re}_m=2600$ ($\tau_\eta=0.057$) and ${\rm Re}_m=6600$ ($\tau_\eta=0.036$). We used sample averages which greatly
reduced fluctuations and allowed us to measure $C_E$ with sufficient precision.}
\label{dyn1}
\end{figure}

\subsection{Numerical results}
Numerical simulations of statistically homogeneous isotropic
small-scale dynamo in \cite{B12a} were performed by solving MHD equations with stochastic non-helical driving and explicit dissipation with $Pr_m=1$.
The results in Fig.~\ref{dyn1} is the statistical average over three different simulations. 
Growth is initially exponential and smoothly transition into the linear stage. Note that scatter is initially small, but grows with time, which is
consistent with the picture of the magnetic field that grows at the progressively larger scales, and has progressively less independent realizations in a single datacube.
The value of the dynamo efficiency that we measure $C_E=0.05$ is much smaller than unity.
One would expect this quantity of order unity because this is
a universal number, determined only by strong interaction on the equipartition scale.
If we refer to the ideal incompressible MHD equations, written in terms of Els\"asser variables,
$\partial_t{\bf w^\pm}+\hat S ({\bf w^\mp}\cdot\nabla){\bf w^\pm}=0$, dynamo could be understood as decorrelation of ${\bf w^\pm}$ which are originally equal to each other
in the hydrodynamic cascade. In our case, this decorrelation is happening at the equipartition scale $1/k^*$. Being time-dependent, it propagates upscale, while, ordinarily, energy cascade goes downscale. The small value of $C_E$ might be due to this. As opposed to picture with multiple reversals and dissipation due to microscopic diffusivity, typical for the kinematic case, in our picture we appeal to \emph{turbulent diffusion} which helps to create the large-scale field. Both stretching and diffusion depend on turbulence at the same designated scale $1/k^*$, so,
in the asymptotic regime of large Re, one of these processes must dominate.
As $C_E$ is small, we conclude that stretching and diffusion are close to canceling each other.
In \cite{B12a}, the interplay of stretching and mixing was studied by simulations of
kinematic dynamo forward and backward in time. Basically, forward in time
stretching is less efficient, while mixing is more efficient. This also tells a cautionary
tale that using artificial statistics of velocity, such as delta-correlated statistics,
may be grossly misleading when dealing with real turbulence.

\begin{figure}
\begin{center}
\includegraphics[width=0.54\columnwidth,valign=c]{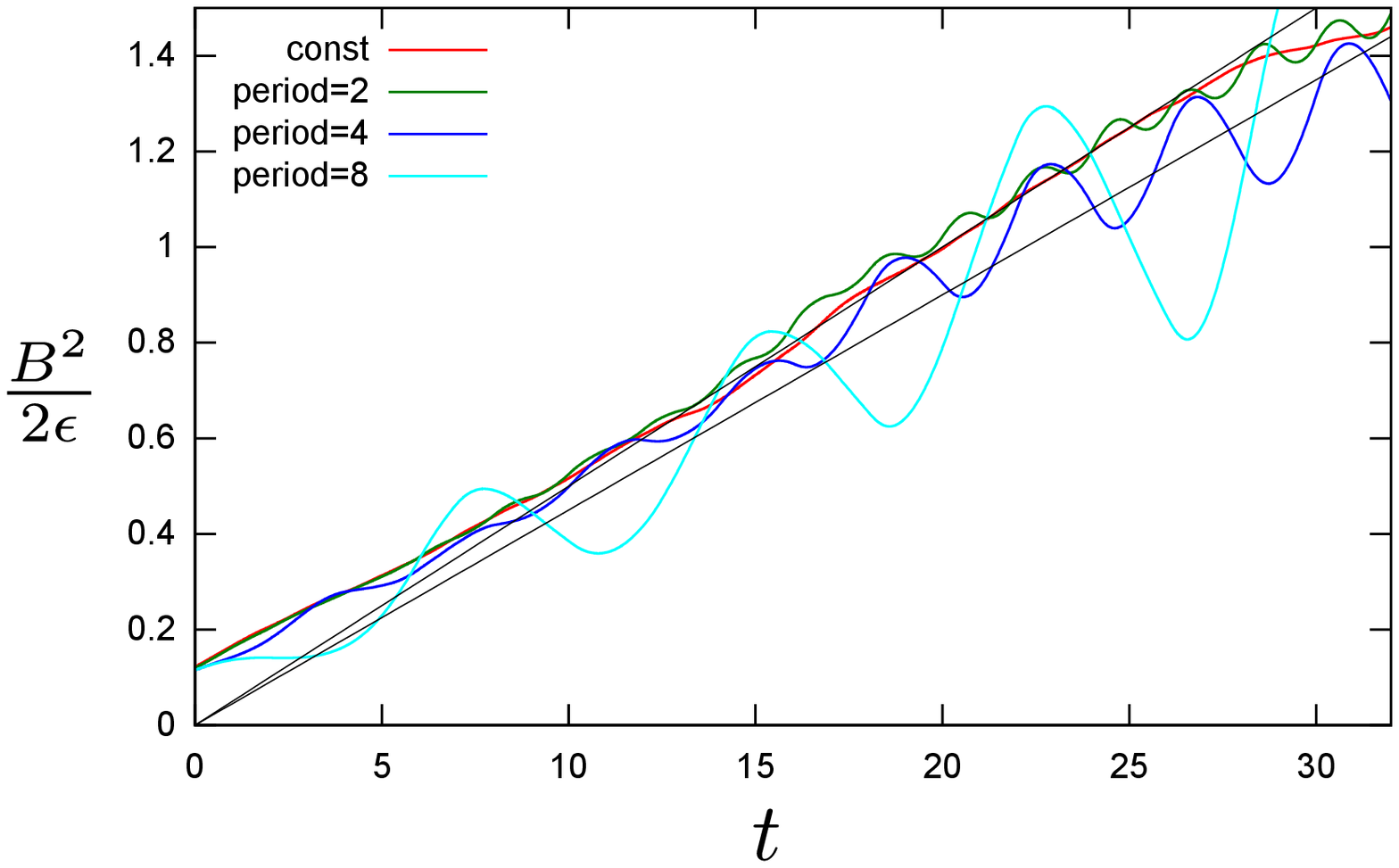}
\includegraphics[width=0.45\columnwidth,valign=c]{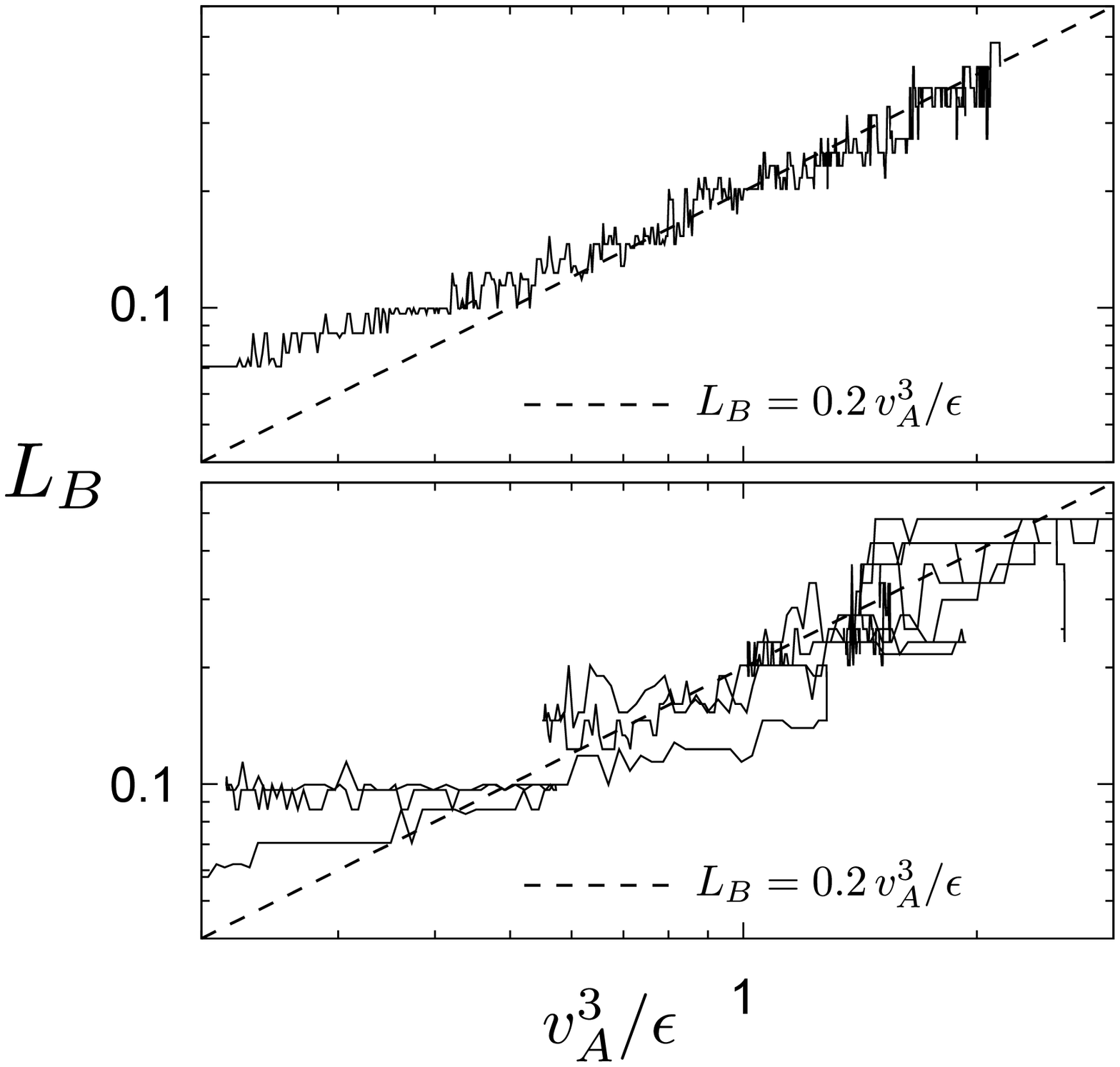}
\end{center}
\caption{Left: The magnetic energy in Alfv\'en units, divided by the energy driving rate, i.e. $B^2/2\epsilon$ versus time (both in numerical time units, approximately correlation time $\tau_c$). Different curves correspond to intermittent driving with different periods. Right: The relation between magnetic energy and the outer scale. The upper plot corresponds to the case with constant driving (from \cite{B12a}), while the lower plot correspond to intermittent driving with the period of $8 \tau_c$. In this relation we used the dissipation rate $\epsilon$ averaged over $2 \tau_c$. Outer scale was determined by the peak wavenumber of the spectrum. Best fit $c_l\approx0.18$ in Eq.~(\ref{Lb}). From \cite{BM16} }
\label{b_field}
\end{figure}

The arguments of the previous section can be applied even if the energy injection
rate is not stationary. On Fig.~\ref{b_field} we presented simulations with intermittent driving
checking relations (\ref{C_E},\ref{Lb}). 

\subsection{Caveats in simulating astrophysical dynamo}
As we alluded above in Section~\ref{kinematic} the timescales of initial growth of the magnetic field are extremely short compared to the dynamical timescale of the system -- these two are separated by the factor of ${\rm Re}^{1/2}$, which is very large in astrophysical
environments. Not so in numerical simulations, which are limited in terms of ${\rm Re}$.
This situation is further exacerbated by the small prefactor $0.0326$ in the growth rate of $\gamma=0.0326/\tau_\eta$. For example, simulation with ${\rm Re}=1600$ will result in characteristic growth time of $1/(0.0326 {\rm Re}^{1/2}) \approx 0.77$ of the outer timescale of the system, while in astrophysical reality this is a negligible fraction of the outer timescale\footnote{See, e.g., how the growth on Fig~\ref{dyn1} starts with rather high values of $t$}. The lesson from this is that if we take a very small
initial magnetic field while simulating rather dynamically young object, e.g., collapsing cloud or a forming galaxy cluster, this may artificially delay the onset of the nonlinear dynamo and grossly underestimate the magnetic field at the end of the evolution \cite{BM16}. This delay effect can also lead to the artificial dependence on the initial
field value or direction, which should not normally appear in nonlinear dynamo, which, as any turbulence, erases any traces of the initial condition. 
The solution to this issue could be injecting initial field based on the
amount of energy in the cascade, along the lines of Section~\ref{dynamo:sec}.

Another issue is that popular ILES codes lack any knowledge or prescription of the
microscales. While very often this is not an issue, because ILES code would simply absorb cascade energy into the thermal energy on the grid scales, but there is a qualitative difference between nature and an ILES simulation with zero initial field, which would produce zero magnetic field for all subsequent times. In nature, the small-scale non-ideal contributions to the induction equation, such as Biermann battery term, will always jump-start the dynamo and result in non-zero fields, also large ${\rm Re}$ in nature will
ensure that average magnetization will be close to the estimate given by nonlinear dynamo in Section~\ref{dynamo:sec}, i.e. the magnetic energy will be a certain fraction of the dissipated cascade energy. From MHD point of view, taking diffusivities to zero allows
us to have two types of solutions - completely unmagnetized and fairly strongly magnetized,
however in nature only magnetized solution will be realized\footnote{Assuming small-scale dynamo is indeed unstable, this requires certain minimal magnetic Reynolds number, but this
is normally satisfied in most astrophysical environments}. We conclude that in dynamo situations ILES code should sometimes use subgrid dynamo prescription. 

\subsection{Application to galaxy clusters}
Galaxy clusters constitute an interesting case of the small-scale dynamo. All
properties of the cluster continue to evolve during the cosmic time, its
mass also determining virial velocity which is always around the inflow velocity.
Clusters are heated by the major mergers, which is also the primary source
of energy for the intracluster turbulence. On average, around 0.4 of heating comes
from dissipation of solenoidal turbulence \cite{miniati2015}. At the same time,
as we saw in the previous section dynamo converts a fraction of 0.05 of this energy
into magnetic energy. Thus we may conclude that magnetization $\beta$, the ratio
of magnetic to thermal energy, should stay approximately constant through cosmic
time, $\sim 40$ for the past 10 Gyr. Similarly, the outer magnetic scale will
stay a constant fraction of the cluster virial radius, $\sim 1/200$.

\section{Compressible and Supersonic Turbulence}
\label{compressible}

The study of supersonic ISM turbulence is vital for
understanding the structure of molecular clouds and subsequent star formation.
In this respect, the studies of density scalings and thermal instability in DNS
have become commonplace. One way to look at it is assuming that in addition to the Alfven mode, additional compressible modes are excited as well. How this approach is valid in the regime
where Alfven waves couple strongly among themselves is still a matter of debate. Below
we will show a mode decomposition techniques, as well as study density perturbations
which are perturbed nonlinearly, by orders of magnitude in the supersonic regime.

Compressible turbulence is characterized by two dimensionless
numbers, sonic Mach number $M_s=\delta v/c_s$ and Alfvenic Mach number $M_A=\delta v/v_A$.
Plasma beta, the ratio of gas pressure to magnetic pressure can be expressed as $\beta=8\pi P_{\rm gas}/B^2=2\gamma M_A^2/M_s^2$. 
In a turbulent environment, these numbers tend to evolve towards ``preferred'' values, depending
on the physics. For example in the absence of cooling, supersonic turbulence will heat
itself to trans-sonic state $M_s \sim 1$ in one dynamical time. The persistent supersonic regime, therefore, is limited only to regions which have very
efficient atomic or molecular cooling, such as molecular clouds, where $M_s\lesssim 10$.

Likewise, the Alfvenic Mach number is limited by the dynamo process that we described in Section~\ref{dynamo}. Turbulence with a very weak initial field tends to generate
equipartition-strength fields, with $M_A \sim 1$, in around 20 dynamical times.
We expect large (but not very large) $M_A$ only in dynamically young systems, such as in collapsing molecular clouds.

In galaxy clusters, which are continuously heated up by accretion and turbulence, its mass, and, therefore, virial velocity, also continue to increase. This leads to the situation when both $\delta v$ and $c_s$ are determined by virial velocity and $M_s \sim 1$. Due to the continuous growth of the cluster mass, it never approaches equipartition, so $M_A \sim 7$.

\subsection{Decomposition into modes}
\def\vv{{\bf v}}
\def\BB{{\bf B}}
\def\ee{{\bf e}}
\def\kk{{\bf k}}

\begin{figure*}[t]
\begin{center}	
  \includegraphics[width=0.33\textwidth]{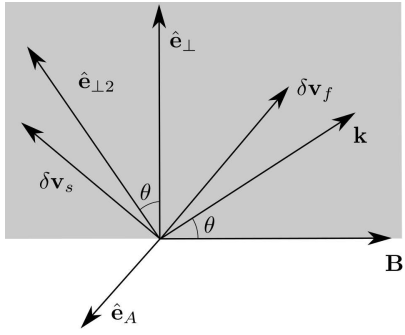}
\end{center}  
  \caption{Coordinate system for vector decomposition into MHD modes. Note that all vectors
  except $\ee_A$ are in the same plane.}
\label{coordinates}
\end{figure*}

Here we will use expressions for the phase wave speeds of the Alfven and slow/fast modes (\ref{u_A},\ref{u_fs}) from Section~\ref{mhd_eq} and the same notation for 
$\cos\theta= (\hat \kk \cdot \hat \BB)$. Our coordinate system will be defined
by the unit vector of the magnetic field $\hat \BB$, 
the unit vector of the Alfven
perturbation $\hat \ee_A = \kk \times \BB_0/|\kk \times \BB_0|$, 
and the third unit vector perpendicular
to the three, in the ${\bf (k, B)}$ plane: $\hat \ee_\perp = \hat \ee_A \times \hat \BB_0$, see Fig.~\ref{coordinates}.

The perturbations of the velocity for the Alfven mode are along $\hat \ee_A$, also
from induction equation $-\omega \delta {\bf B}= {\bf k\times(\delta v \times B_0)}$, however the sign of $\delta \BB$ depend on the sign of $\omega$, i.e. whether the wave is propagating along the field direction, or in the opposite direction. The waves where 
$\delta \BB \sim -\delta \vv$ the waves propagate along the field. This also
corresponds to $\delta w^+$ propagating opposite to the field and $\delta w^-$ along the field. 

The perturbations of the velocity for the slow and fast mode are along mutually
orthogonal vectors in the ${\bf (k, B)}$ plane:
\begin{equation}
 \delta \vv_{f,s} =  \hat \BB (u_{f,s}^2-v_A^2) \cos \theta + \hat \ee_\perp u_{f,s}^2 \sin \theta.
\end{equation}
This two vectors and $\hat \ee_A$ form an orthogonal coordinate system in wavenumber space.
The perturbation $\delta \BB$ for both modes is in the 
\begin{equation}
\delta \BB_{f,s}=\hat \BB_0 - \hat \kk \cos\theta
\end{equation}
 direction, however again the sign of the contribution for each mode will depend on the direction of the wave propagation. Introducing $\Delta \vv_{f,s}$ as perturbations where oppositely propagating wave
contribute with different signs, the contribution to $\delta \BB$ for each mode will be proportional
to $|\Delta \vv_{f,s}/u_{f,s}|$ (see, e.g., \cite{cho2003c}).

Alternatively, one can decompose in the coordinate system rotated around $\hat \ee_A$ by $\theta$, which is made of $\hat \kk$, $\hat \ee_A$ and $\hat \ee_{\perp2} = \hat \ee_A \times \hat \kk$, in which case the velocity and magnetic field vectors along fast and slow modes can be expressed as
\begin{eqnarray}
\delta \vv_{f,s} &=& {\bf \hat k}(u_{f,s}^2/v_A^2 - \cos^2 \theta) + \hat \ee_{\perp2} \cos \theta \sin \theta,  \\
\delta \BB_{f,s} &=& \hat \ee_{\perp2}.  
\end{eqnarray}
It was suggested in \cite{Cho2002a,cho2003c} that the Alfven mode have an independent
cascade (Sec.~\ref{alfvenic}), slow mode is passive to Alfven cascade and has the same spectra 
and anisotropy, and fast mode has an independent isotropic acoustic/wave turbulence cascade (Sec.~\ref{kolm_gen_sec}):
\begin{eqnarray}
   \mbox{ Alfv\'{e}n, Slow:~}  E^A(k)  \propto k^{-5/3}, 
                        ~~~k_{\|} \propto k_{\perp}^{2/3}.\\
   \mbox{ Fast:~~~}   E^f(k)  \propto k^{-3/2}, 
                        ~\mbox{isotropic spectrum}.   
\end{eqnarray}
This was broadly consistent with simulations in \cite{Cho2002a,cho2003c}, however,
steeper spectrum $\sim k^{-2}$ was reported in \cite{Kowal2010} for fast modes.
 \begin{figure*}
  \includegraphics[width=0.99\textwidth]{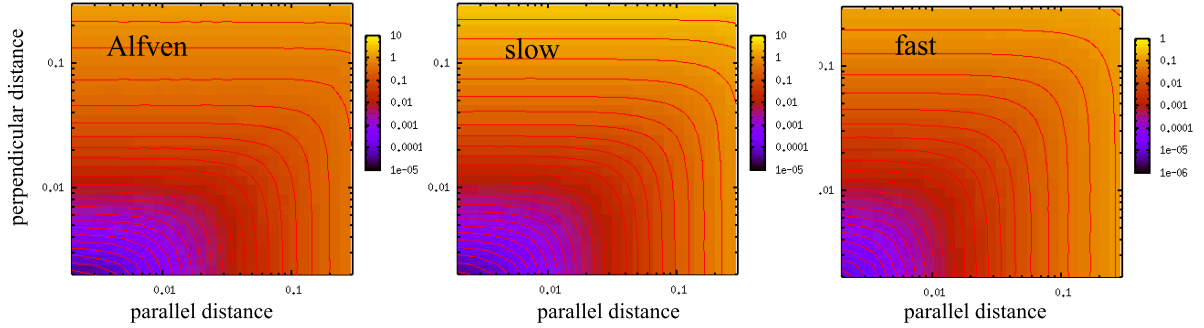}
  \caption{
  Anisotropy of the Alfv\'en, slow and fast modes as evidenced by the contours of the second order structure function. Here we used the SF decomposition method.
 The Alfv\'en and slow mode exhibit scale-dependent anisotropy, while the fast mode is almost isotropic.}
\label{sf_decomp}
\end{figure*}
Regarding the amplitude of each mode in realistic turbulence, it stronly depend on the way
the turbulence is driven. For a particular incompressible driving of \cite{Cho2002a,cho2003c} they
suggested a scaling relation 
\begin{equation}
  \frac{\delta E_{\rm fast}}{\delta E_{\rm Alf}}\approx \frac{\delta V_A V_A}{V_A^2+c_s^2},
\label{eq_high2}
\end{equation}
where $\delta E_{\rm fast}$ and $\delta E_{\rm Alf}$ are energy
of fast and Alfv\'en modes, respectively.

\subsection{Decomposition in real space}

Another way to decompose into modes is by using structure functions.
In this method, the separation vector $\vec l$ of the structure function plays the role of the wavenumber, because of the correspondence between one-dimensional structure function 
along the particular line and the power spectrum along the same line (Sec.~\ref{statistical}).
Fig.~\ref{sf_decomp} shows the contours of the structure function corresponding to each mode obtained
in datacubes from $M_s=10$ supersonic simulations used earlier in \cite{BLC05}.
The anisotropies of each mode show the same behavior as in the global decomposition method discussed
above. The advantages in using the SF decomposition method is that it is a local measurement,
so we can apply it even when the global average magnetic field is zero, e.g., it
has been applied to the decomposition of MHD turbulence
obtained in the high-resolution cosmological simulation of a galaxy cluster \cite{BXLS13}. 

\subsection{Density scalings}
\label{density}
\begin{figure}
\begin{center}	
  \includegraphics[width=0.5\textwidth]{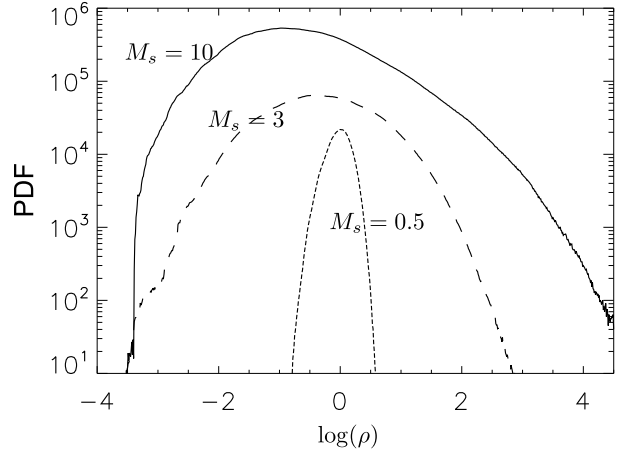}
\end{center}  
  \caption{Probability density function for a density in isothermal
  numerical simulations with $M_A\approx 1$ and various $M_s$. From \cite{BLC05}.}
\label{pdf_logrho}
\end{figure}
\begin{figure*}[t]
  \includegraphics[width=0.99\textwidth]{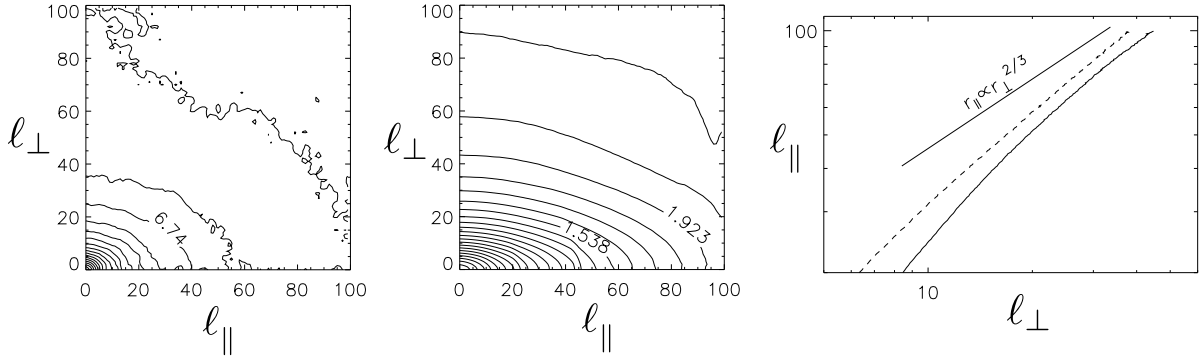}
  \caption{Contours of the structure function of density (left), log-density (center) and the anisotropy of log-density (right), solid line represents $M_s\sim 10$, dashed - $M_s\sim 3$. From \cite{BLC05}}
\label{log_density}
\end{figure*}

The properties of density in supersonic turbulence is interesting to astronomers due to its relation to star formation. The density is primarily perturbed by the slow mode; however, in supersonic case, these perturbations are not small. Instead, density varies by several orders of magnitude within the box (Fig.~\ref{pdf_logrho}).
At the same time, the statistics of density is very different from the statistics
of the slow mode velocity.
It is worth noting that supersonic isothermal hydrodynamics predicts the log-normal distribution of density \cite{Passot1998} due to the gauge symmetry of log-density in this case. This symmetry is broken in MHD. However the PDF still resemble
log-normal law (Fig.~\ref{log_density}). The paper \cite{BLC05} pointed out that in the low beta supersonic regime the perturbation of the slow mode velocity is almost along $\BB_0$, so the dynamics of the slow mode is quasi-one-dimensional.
This results in the generation of many slow shocks, which is indeed observed in
simulations. The perturbations of density, e.g., its log-normal PDF, are created by these
random slow shocks, similar to hydrodynamics. The Alf\'en mode only mixes these perturbations
by shearing motions, without affecting PDF. On the other hand,
the structure of density (SFs) is almost entirely determined by this shearing and is expected to have an anisotropic structure like velocity and magnetic field.
In other words, two distinct physical processes act simultaneously and affect different statistical measures of the turbulent density field. The random multiplication of density induced by shocks, affect
the PDFs, while the other, Alfv\'enic shearing, affects anisotropy and scaling of the structure function of the density. Revealing the structure created by shearing requires overcoming
the effect of high-density clumps which will dominate ${\rm SF}^2(\rho)$. So, instead, we should
use statistics of $\log(\rho)$, which have approximate Gaussian PDFs.
This exercise is shown in Fig.~\ref{log_density}. Indeed, the statistics of $\log(\rho)$ shows
anisotropy characteristic of Alfv\'en shearing. 

\section{Turbulence driven by magnetic field}
\label{mag_shear}
Turbulence in reconnection can appear as a result of instabilities, for example, resistive tearing \cite{biskamp1986}. The paper \cite{loureiro2007} demonstrated that the instability becomes faster
and not slower with decreasing resistivity above a critical Lundquist number around $10^4$.
Observing effects of the feedback of the release of magnetic energy in numerics is challenging because currently available 3D MHD numerics are limited by the Lundquist numbers of several of $10^4$.

Physically, periodic box simulations, like numerics in \cite{B17} correspond to early times in the current sheet disruption when the outflow did not develop. Importantly enough,
it did demonstrate fast (resistively-independent) reconnection rates, defined as mixing rates
of the fluid. Simulations with open boundaries in \cite{Kowal2015} have been performed for a sufficiently long time to allow for the establishment of the stationary state. They correspond to later times when the stationary inflow/outflow appears.
\begin{figure*}[t]
\includegraphics[width=0.43\textwidth]{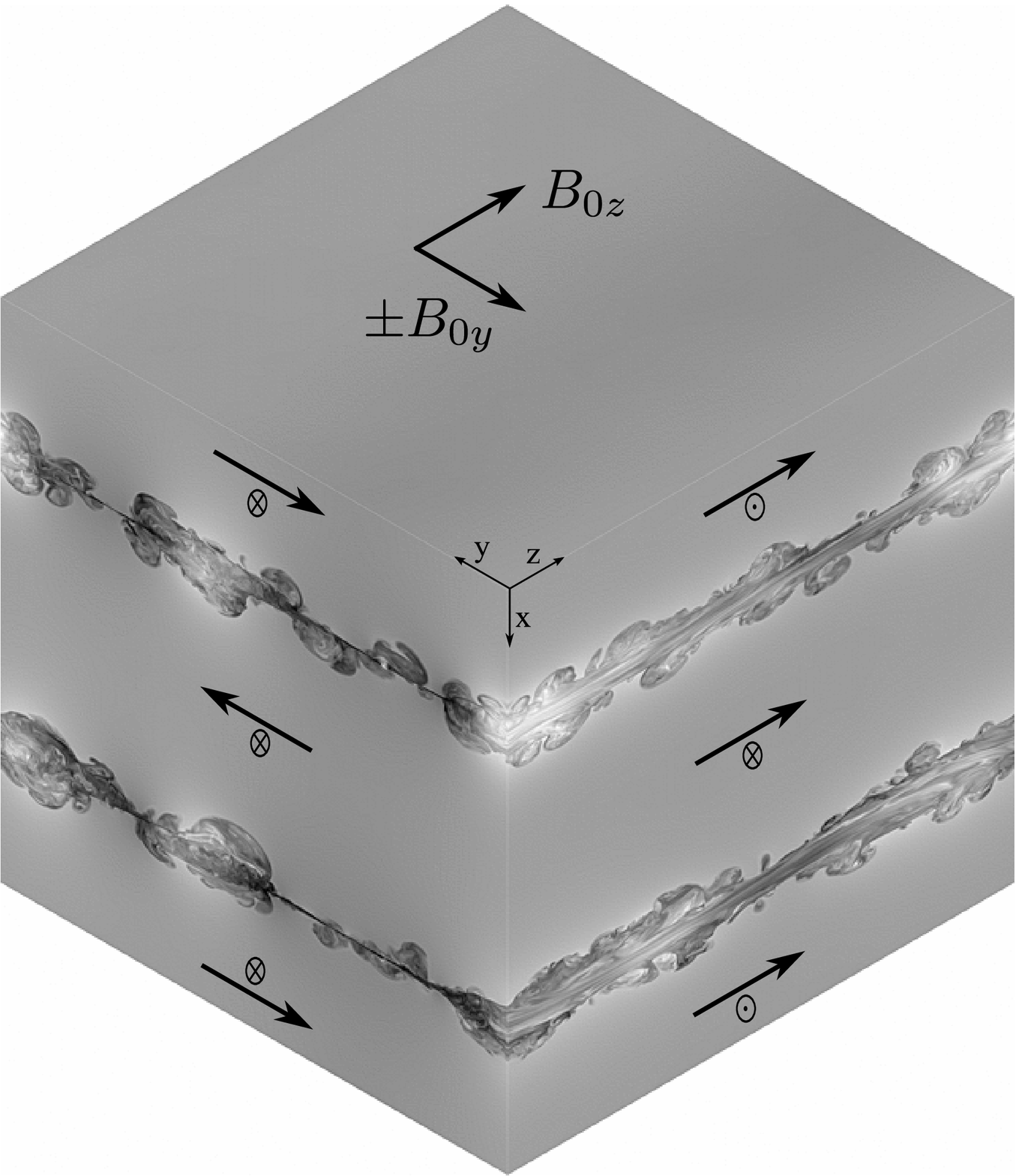}
\hfill
\includegraphics[width=0.52\textwidth]{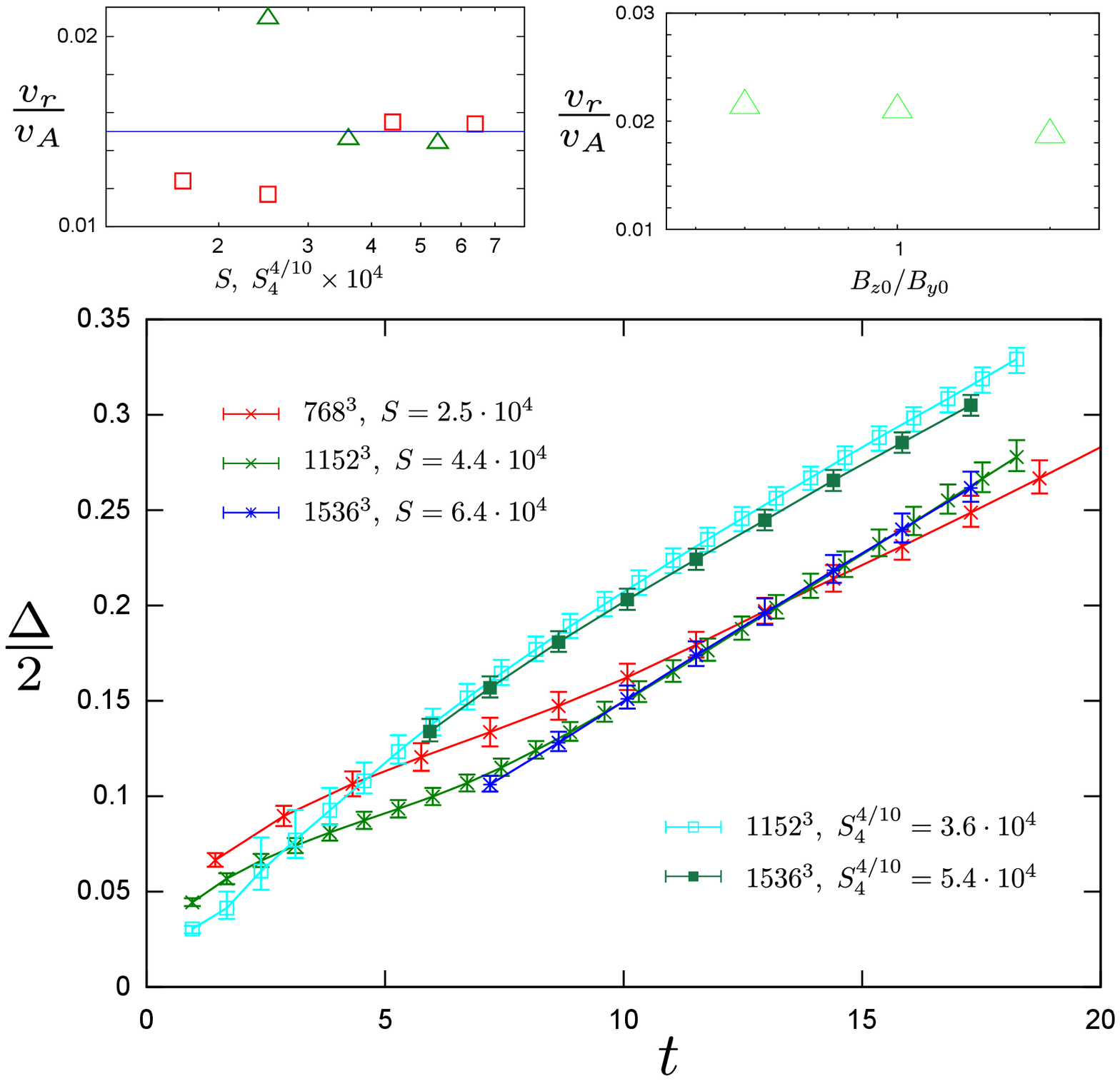}
\caption{Left: The setup of all-periodic reconnection with two current layers. The magnitude of
the magnetic field is shown in grayscale. Right: the evolution of the layer width $\Delta$ (bottom) and the reconnection (mixing) rate as a function of the Lundquist number $S$ (top left) and the ratio of $B_{z0}/B_{y0}$ (top right). From \cite{B17}.}
\label{width_reconn}
\end{figure*}

One of the simplest setups to study the development of turbulence in the thin current layer is a periodic setup with the mean field $B_{z0}$ threading the box, reconnecting field $\pm B_{y0}$ changing the sign in the $x$ direction, see Fig.~\ref{width_reconn}. We consider the incompressible case, in which case the only dimensionless parameters of the problem are the Lundquist number $S$ and the ratio $B_{y0}/B_{z0}$. We use the planar sheet in an attempt to simulate a zoomed-in portion of a very large and unstable Sweet-Parker current layer. Lundquist number is defined with the box size, as $S=v_{Ay} L/\eta$. We imagine that this box is a part of a bigger system with larger system size $L_{\rm global}$. We aim to simulate early times, $t<L_{\rm global}/V_A$, when the global outflow did yet develop. We also assume that the global Lundquist number, determined by the larger system, is asymptotically large so that we can ignore large-scale gradients.
The simulation end is determined by the development of structures with the size comparable to the box size, at which point our artificial periodic boundary starts influencing the result.
\begin{figure*}[t]
\includegraphics[width=0.5\textwidth]{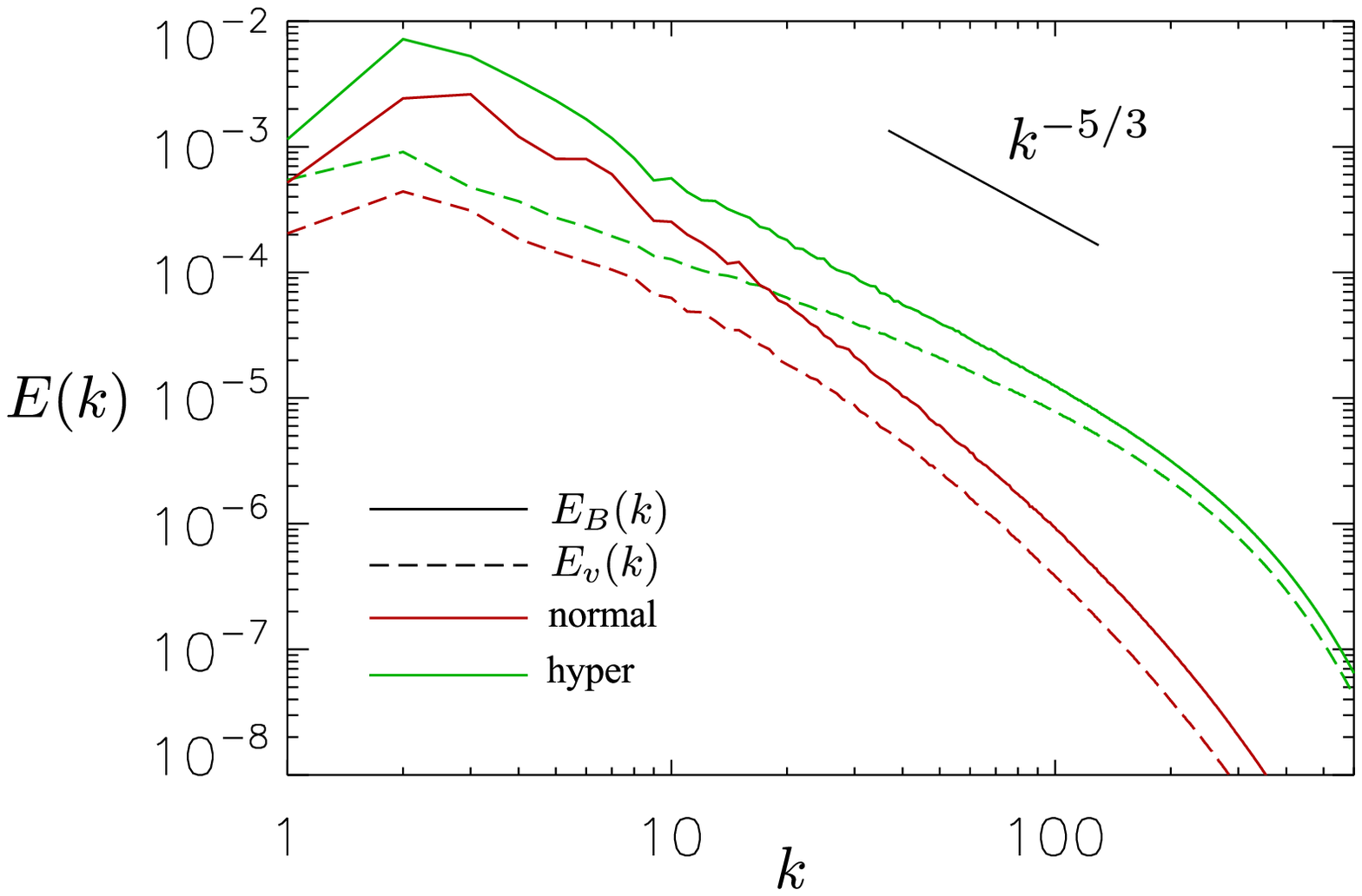}
\hfill
\includegraphics[width=0.45\textwidth]{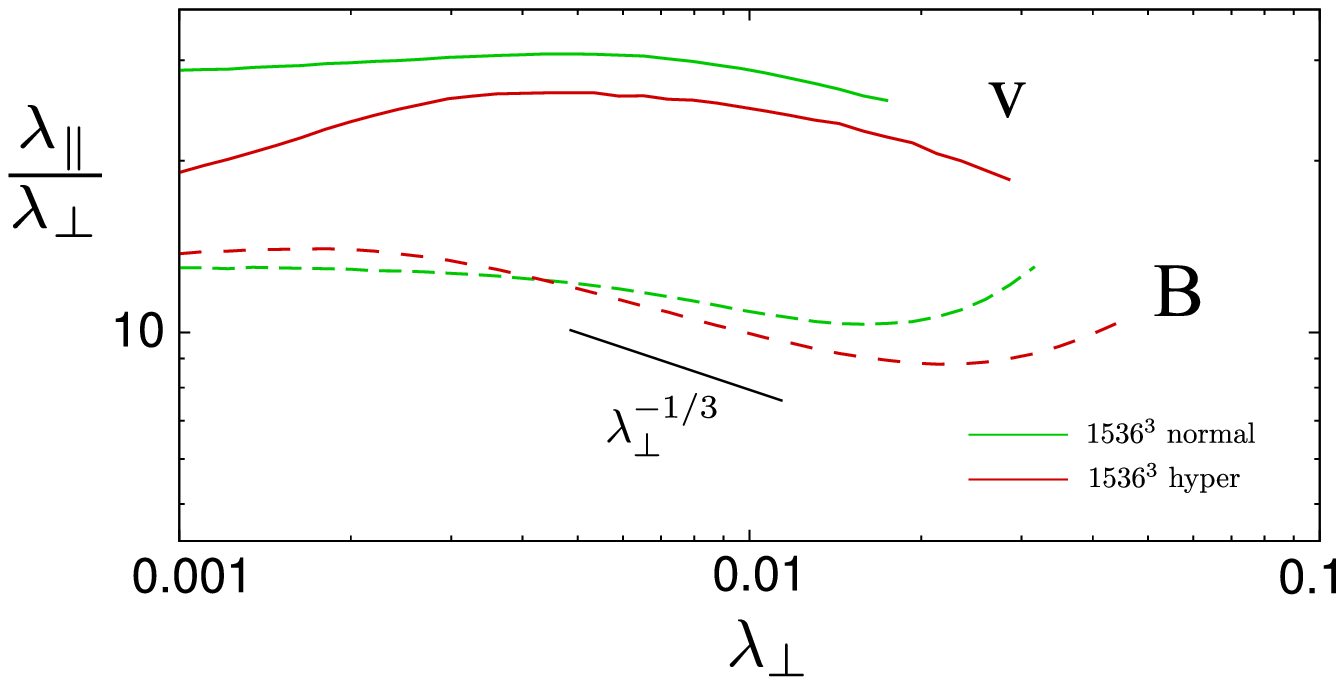}
\caption{Left: The y-z power spectra of velocity and magnetic perturbations of turbulence in the current layer. Right: Anisotropy from the ratio of parallel to perpendicular scales obtained from equating 2-order SFs (see Section 3.12). We used simulations with 2nd as well as 4th order diffusivities (hyper diffusivities) to evaluate the effect of the dissipation on the statistics
of turbulence in the layer. From \cite{B17}.}
\label{spec_reconn}
\end{figure*}
  
The free energy in the system is the energy density of the opposing fields $B_{y0}^2/8\pi$, which
is declining in the turbulent current layer due to dissipation. After $t\approx 0.3 L/v_{Az}$
the fraction of the dissipated energy $w_d$ becomes approximately constant, around $w_d\approx 0.4$. We calculate the reconnection rate as the speed of growth of the turbulent current layer width $\Delta$, i.e.
we define $V_r=d \Delta/dt$.  The evolution of $d$ and the inferred reconnection rate are shown in Fig.~\ref{width_reconn}. $V_r$ was around $0.015 v_{Ay}$ for high Lundquist numbers and is rather insensitive to the imposed mean field $B_{z0}$ (Fig.~\ref{width_reconn}). The dissipation rate per unit area of the current sheet can be calculated from $w_d$ and $v_r$ as 
\begin{equation}
\epsilon_S=2 w_d v_r (1/2) \rho v_{Ay}^2 \approx 0.006 \rho v_{Ay}^3,
\end{equation}
Note that we arrived at the expression not only for ``fast reconnection'' (independent on resistivity and viscosity) but also for ``fast dissipation''. This expression, modulo
numerical coefficient, can be obtained by dimensional analysis using only $\rho$ and $v_{Ay}$.

The field in the current layer can be analyzed statistically.
We show the spectrum for one time slice on Fig.~\ref{spec_reconn}.
The peak of the spectrum moves towards smaller wavenumbers, i.e., the outer scale of this turbulence is growing in time. This is unlike driven turbulence (Sec~\ref{alfvenic}) where this scale was determined by forcing and fixed. Another difference with driven turbulence is that magnetic spectrum is above kinetic on all scales, but closer to equipartition on smaller scales. This is similar to decaying MHD turbulence described, e.g., in \cite{biskamp2003}.
Qualitatively reconnection turbulence is very similar to decaying turbulence created by the initial random magnetic field. 

Scale-locality is an important component of turbulent reconnection. 
Our spontaneous reconnection numerics corroborate scale-locality, because
the spectral slope of perturbations is between -1 and -3.
In the real world, we expect the reconnection rate to be independent of
system size as long as ion Larmor radius $r_L$ and ion skip depth $d_i$ are both much smaller than
the layer width $\Delta$.
On the right-hand side, Fig.~\ref{spec_reconn} shows
anisotropy expressed as a ratio of parallel to perpendicular scale $\lambda_\|/\lambda_\perp$, obtained by a method we explain in Sec~\ref{alfvenic}. We can also estimate the interaction strength parameter $\xi=\delta v \lambda_\|/v_A \lambda_\perp$ and see that
for this case it is around unity, i.e., we are dealing with critically balanced strong turbulence. Note that the anisotropy of our turbulence, being around $k_\|/k_\perp \sim 1/20$ is very different from the tangent of the fastest growing oblique tearing mode, $k_\|/k_\perp = B_z/B_y = 1$. So turbulence forgets the properties of the oblique tearing that started it. From simulations with higher $B_z$ one also confirm Alfv\'en symmetry: increasing $B_z$ only increases parallel lengthscale, while keeping dynamics essentially unchanged (see \cite{B17}).

\section{Conclusion}
This review covers a set of topics that were chosen because they are either: a) basic topics
that are essential for the understanding of subsequent material or b) have seen rapid
progress recently, which is otherwise not covered in books or reviews. Due to this choice,
many things, especially astrophysical and space applications of MHD turbulence, has been
omitted or mentioned only in passing. This document, however, is a living review and will be
evolving, below we overview several topics that we expect to add or expand. For the
impatient, we mention older books and reviews that can be used to expand and deepen the
reader's knowledge of the topic. Most of these have been already mentioned in the course of
the review.

Mathematical tools to work with the statistical ensemble of turbulent realization can be
found in the monograph by Monin and Yaglom \cite{monin1975}. Overview of turbulence as a
nonlinear dynamical process can be found in the book of Frisch \cite{frisch1995}, Falkovich
(2011) \cite{falkovich2011}, Davidson (2015) \cite{davidson2015}. Comprehensive, although
older, book dedicated to MHD turbulence is by Biskamp (2003) \cite{biskamp2003}, a few
topics in MHD turbulence are also covered in Davidson (2013) \cite{davidson2013}. Older book
on mean-field dynamo is Krause and Raedler (1980) \cite{krause1980}, a more modern approach
to the same topic, primarily for solar dynamo applications is a living review by Charbonneau
(2010) \cite{charb2010}. A more broad review on dynamo is by Brandenburg and Subramanian
(2005) \cite{brandenburg2005}. In future editions, we plan to cover large-scale dynamo as
well. For an in-depth review of solar wind turbulence, see living review by Bruno and
Carbone \cite{bruno2005}. We plan to expand the section related to the solar wind and cover
energy flux \cite{macbride2008} as well as magnetic helicity measurements
\cite{brandenburg2011}. An interesting case of energy cascade with applications to
cosmological-scale magnetic fields and its dynamical evolution is a freely decaying
homogeneous MHD turbulence, see, e.g. \cite{brandenburg2015}, which we also plan to cover in
the future.

Several topics connecting astrophysical plasmas and MHD turbulence can be found in a book by
Kulsrud (2005) \cite{Kulsrud2005}. In the ISM, MHD turbulence coexists with cosmic rays.
Cosmic ray interaction with MHD turbulence is a fairly large topic, for an introduction to
cosmic rays as well as quasilinear scattering theory one can start with Schlickeiser (2002)
\cite{schlickeiser2002}. One particularly important application of mutual interaction
between cosmic rays and MHD turbulence which will be covered in future editions of this
review is the acceleration of cosmic rays in supernova remnants. In front of strong shocks,
MHD turbulence is self-generated by fast particles. This is supported by estimates of
diffusion coefficient $D$ of cosmic rays in supernova remnants. $D$ in front of the shock is
estimated to be many orders of magnitude smaller than $D$ in the ambient ISM, i.e. cosmic
rays create their own MHD turbulence and dynamo and scatter themselves.

Supersonic turbulence with applications to ISM and star formations is covered in Mac Low and
Klessen 2004 \cite{maclow2004} and McKee and Ostriker (2007) \cite{mckee2007}. The physics
of turbulent energy cascade in the supersonic case has been an open question for some time,
but recently we saw progress in deriving exact analytic relations in supersonic case
\cite{falkovich2010,galtier2011,banerjee2013}, as well as empirical findings and numerical
verification \cite{wagner2012,kritsuk2013}. The earlier phenomenological approach of
replacing statistics of velocity ${\bf u}$ with the statistics of $\rho^{1/3} {\bf u}$ in
the compressible case \cite{fleck1996,kritsuk2007} have found a firmer foundation
\cite{banerjee2014}. This has also been used to explain observed statistical correlations,
such as Larson's laws \cite{larson1981} in star-forming clouds \cite{kritsuk2013}.

Intermittency is the deviation from self-similarity of turbulence and is an important
property that reminds us of the richness of the field of nonlinear fluid dynamics. While
intermittency in hydrodynamics has been long studied as-is, the intermittency in different
variables in MHD turbulence may give us something to think about its dynamical origin. On
the other hand, extreme intermittency had been suggested to explain heating and molecular
synthesis in the ISM. This will be covered in more detail in the future editions of this
review. The numerical section will be expanded with mention of Lagrangian-Eulerian (moving
mesh) codes and recent progress in this area. In future editions, we also will pay more
attention to the connection between theory and observations. Big progress has been achieved
in the area of cosmological structure formation by massive ab-initio simulations including
$\Lambda$CDM initial conditions with $\Lambda$, dark matter and ordinary matter using grid
refinement down to the scales of galaxies (e.g. Illustris project). Some of these results
are relevant to understand magnetization in filaments and, possibly, clusters and will be
added later.
  




\def\apj{{\rm Astrophys. J.}}           
\def\apjl{{\rm ApJ }}          
\def\apjs{{\rm ApJ }}          
\def\grl{{\rm GRL }}
\def\aap{{\rm A\&A } }
\def\jgr{{\rm JGR}}
\def\mnras{{\rm MNRAS } }
\def\physrep{{\rm Phys. Rep. } }               
\def\prl{{\rm Phys. Rev. Lett.}} 
\def\pre{{\rm Phys. Rev. E}} 
\def\araa{{\rm Ann. Rep. A\&A } }
\def\prd{{\rm Phys. Rev. D}} 
\def\pra{{\rm Phys. Rev. A}} 
\def\ssr{{\rm SSR}}
\def\planss{{\rm Plan. Space Sciences}}
\def\apss{{\rm Astrophysics and Space Science}}
\def\nat{{\rm Nature}}
\def\jcap{{\rm JCAP}}
\def\memsai{{\rm MEMSAI}}
\def\lt{{<}}
\def\aapr{{\rm AAPR}}
\def\solphys{{\rm SolPhys}}

\bibliographystyle{unsrt}
\bibliography{all}

\end{document}